\documentclass[article,twocolumn,onecolappendix]{aastex61}

  \usepackage{amsmath}
  \usepackage{tabularx}
  \usepackage{color}
  \usepackage{graphicx}
  \usepackage{rotating}
  \usepackage{subfigure}

\hypersetup{linkcolor=red,citecolor=blue,filecolor=cyan,urlcolor=magenta}
\received{October , 2017}
\revised{\today}
\accepted{2017}
\submitjournal{ApJ}

\begin{document}

\title{Optical spectroscopic survey of a sample of Unidentified \textit{Fermi} objects. \\}

\correspondingauthor{Simona Paiano}
\email{simona.paiano@oapd.inaf.it}

%\author[0000-0002-0786-7307]{Greg J. Schwarz}
%\affil{American Astronomical Society \\
%2000 Florida Ave., NW, Suite 300 \\
%Washington, DC 20009-1231, USA}

\author{Simona Paiano}
\affiliation{INAF, Osservatorio Astronomico di Padova, Vicolo dell'Osservatorio 5 I-35122 Padova - ITALY}
\affiliation{INFN, Sezione di Padova, via Marzolo 8, I-35131 Padova - ITALY}

\author{Renato Falomo}
\affiliation{INAF, Osservatorio Astronomico di Padova, Vicolo dell'Osservatorio 5 I-35122 Padova - ITALY}

\author{Alberto Franceschini}
\affiliation{Dipartimento di Fisica e Astronomia, Universita' di Padova, Vicolo dell'Osservatorio 3, I-35 Padova - ITALY}
%\collaboration{(LaTeX collaboration)}

\author{Aldo Treves}
\affiliation{Universit\`a degli Studi dell'Insubria, Via Valleggio 11 I-22100 Como - ITALY}

\author{Riccardo Scarpa}
\affiliation{Instituto de Astrofisica de Canarias, C/O Via Lactea, s/n E38205 - La Laguna (Tenerife) - SPAIN}
\affiliation{Universidad de La Laguna, Dpto. Astrofsica, s/n E-38206 La Laguna (Tenerife) - SPAIN}



\begin{abstract}
We present optical spectroscopy secured at the 10m Gran Telescopio Canarias of the counterparts of 20 extragalactic $\gamma$-ray sources detected by the \textit{Fermi} satellite. 
The observations allow us to investigate the nature of these sources and to determine their redshift. 
We find that all  optical counterparts have a spectrum that is consistent with a BL Lac object nature. 
We are able to determine the redshift for 11 objects and set spectroscopic redshift limits for five targets.
Only for four sources the optical spectrum is found featureless. 
In the latter cases we can set lower limits on the redshift based on the assumption that they are hosted by a  typical massive elliptical galaxy whose spectrum is diluted by the non thermal continuum.
The observations allow us to unveil the nature of these gamma-ray sources and provide a sanity check of a tool to discover the counterparts of $\gamma$-ray emitters/blazars based on their multiwavelength emission.

\end{abstract}




\section{Introduction} \label{sec:intro}

High energy observations from space, like those performed by the EGRET \citep{thompson1993}, the AGILE \citep{tavani2009}, and the \textit{Fermi} \citep{atwood2009} missions, among others, offer us the prime tool for selecting the peculiar class of Active Galactic Nuclei known as blazars. 
In essence, blazars derive their extreme properties from relativistic flux amplification due to bulk motions of the non-thermal emitting plasma towards the observer. 
This combines with the tendency of such plasmas to ''inverse-Comptonize" lower energy photons in the medium, in producing very intense fluxes of photons in the high energy regime, hence offering a rather unique method for their selection.
The recent release of the 8-years Fermi survey catalog \citep[3FGL,][]{acero2015} includes 3033 detected sources, a large fraction of which are either identified or suspected blazars. 
Indeed, more than 1000 of the 3FGL objects are already confirmed blazars.

One of the clear advantages of such high energy observations, compared to other analyses based on broad-band colors (e.g. radio to optical to X-ray, etc.), is to provide us with a simple flux-limited and all-sky complete selection method, with good statistics.
The penalty to pay for this comes from the poor angular resolution (error box with radius of the order of 5 - 10 arcminutes). 
Although the Fermi-LAT observatory is greatly improving it by the continuous signal addition  over the years, the task remains challenging, if we consider that one third of the 3FGL lastly-released source catalog are still unassociated to counterparts in other frequency band. 
Preliminary searches of these unassociated/unidentified gamma-ray sources (henceforth UGS) by various authors \citep{mirabal2012x, massaro2012x, dabrusco2013x, acero2013x, doert2014x, landi2015, paiano2017ufo} found clear indications that the large majority of them are AGNs of the blazar class. 

This UGS source population then represents a very important and new component of the high-energy sky, and may hide new classes of AGNs or even new unknown high-energy phenomena.
It is then of utmost interest to perform a full investigation of these sources.

Of course, achieving a good level of completeness in the gamma-ray source identification is of primary importance for any statistical analyses of the samples. 
For instance, this is needed for understanding the physical differences between the two main blazar classes, the BL Lac objects (BLL) and the Flat-Spectrum Radio Quasars (FSRQ), to check the origin for such different spectral properties (BLLs showing very weak or absent emission lines in the optical, instead FSRQs with prominent lines), as well as the distributions of the two populations in space-time and their cosmological evolution \citep{ghisellini2017}.

The progress in the identification of sources from the various Fermi releases has been quite slow during the years \citep[e.g.][]{massaro2016}: the effort for source identification has been essentially balanced by the increasing depth of the catalogs with time, and the higher required sensitivity for the follow-up. 
While refined techniques have been implemented for preliminary associations of the sources via multi-wavelength analyses based on WISE data \citep{dabrusco2013x, massaro2016wise}, or broad-band studies of the Spectral Energy Distributions (SED) \citep{paiano2017ufo}, or via complementary radio data \citep{petrov2013, nori2014, schinzel2015x, giroletti2016}, the fundamental bottleneck in this process is set by the required optical spectroscopic characterization of the sources.
Even assuming that a substantial part of the missing objects are observed spectroscopically to a sufficient depth, the spectral analysis and interpretation might turn out overwhelmingly difficult.
To account for this, the Fermi collaboration has defined a new class of sources, the blazar candidates of uncertain type (BCU\footnote{The 3LAC sources classified as blazar candidates of uncertain type are categorized into three sub-types: the BCU-I sources where the counterpart has a published optical spectrum but is not sensitive enough for a classification as an FSRQ or a BL Lac; the BCU-II objects with the counterpart lacking an optical spectrum, but a reliable evaluation of the SED synchrotron-peak position is possible; the BCU-III sources for which the counterpart is lacking both an optical spectrum and an estimated synchrotron-peak position but shows blazar-like broad-band emission and a flat radio spectrum.}), that is sources showing SEDs characteristic of the blazar population, but without spectroscopic confirmation of their nature. 
These are very numerous in the 3FGL and 3LAC\footnote{The 3LAC catalog is the third catalog of AGN detected by the Fermi LAT \citep{3lac}} catalog classification and represent the second largest population among the $\gamma$-ray AGN emitters.

From this point of view, the two blazar populations, BLLs and FSRQs,  behave very differently one from the other. 
The emission lines in FSRQs are relatively easy to catch based on even moderate quality spectra, also allowing a relatively easy and secure redshift measurement. 
This is good because FSRQ are typically high redshift objects and of high luminosity.
Instead, this becomes very often prohibitive for the BLL class, whose weakness of the emission lines makes their identification, and particularly the redshift measurement very challenging. 
So far, 4-meter or lower class telescopes have been used for such observations \citep{sbarufatti2005b, sbarufatti2006b, sbarufatti2006a, shaw2013a, massaro2014, paggi2014,  landoni2015, ricci2015, marchesini2016, alvarez2016A, alvarez2016B}, that contributed to classify a large number of the optical counterparts , but only a small fraction of them were able to derive a redshift for BLLs.

With all this in mind, we started in 2015 an observational program, aimed to significantly improve the spectroscopic study of blazars, and the BLLs in particular, using observations at 10-meter class telescope, the Gran Telescopio de Canarias.

In previous works, we provided high signal-to-noise (S/N) ratio optical spectra for a sample of 15 TeV BLL and 7 TeV candidates with unknown or uncertain redshift \citep{paiano2017tev} and of 10 BLLs detected by Fermi satellite for which previous observations suggested that are at relatively high redshift \citep{paiano2017fgl}.

The present paper is intended to be the first in a series devoted to the UGSs detected by the \textit{Fermi} satellite.  
While these are rarely very faint sources in the optical (typical magnitudes ranging from 17 to about 20th mag), the weakness of the absorption and/or emission features require high S/N spectra to detect and identify them. 

We believe that, with such an approach, the elusive nature and mysterious cosmological significance of the BLL population could be effectively addressed.
Is their low-redshift confinement simply due to a selection bias, favoring the line-emitting FSRQ at higher redshifts? 
Which are the statistical properties of the population, like the luminosity functions, and how do they compare to those of the host galaxies?
Our strategy is to provide fundamental spectroscopic data to address these issues.

In Section \ref{sec:sample} we describes our sample and its main properties.  
In Section \ref{sec:data} we present the data collection and the reduction procedure.  
In Section \ref{sec:results} we show the optical spectra of each object, underlying their main features, and discuss their redshift.  
In Section \ref{sec:notes} we give detailed notes on individual objects and finally in Section \ref{sec:discu} we summarize and discuss the results.

In this work we assume the following cosmological parameters: H$_0=$ 70 km s$^{-1}$ Mpc$^{-1}$, $\Omega_{\Lambda}$=0.7, and $\Omega_{m}$=0.3.

\section{Sample selection} \label{sec:sample}

About 30\% of the $\gamma$-ray sources reported in the 3FGL catalog are unassociated to sources at other frequencies. 
These may be due either to the lack of high energy observations at other frequencies or to the presence of multiple sources in the large $\gamma$-ray error box.  
The active campaign provided by the \textit{Swift} satellite \citep{swift_mission} allows us to detect X-ray sources within the error box of the $\gamma$-ray emission. 
We use these data to identify new X-ray sources, as candidates for the $\gamma$-ray emission. 

We selected a sample of UGSs from the 3FGL catalog following these criteria:

\begin{itemize}

\item The source is not associated in the 2FGL or 3FGL and in other gamma-ray catalogues (i.e. the EGRET, AGILE).

\item Target coordinates outside the Galactic plane $| b | ~>$~20$^{\circ}$; this avoids the very crowded and confused region of the sky where the diffuse background is stronger and difficult to model. 
This also favors the selection of extragalactic sources.

\item The objects should be well observable from the La Palma site ($\delta > -20^{\circ}$).

\item Presence of at least one X-ray source detected within the UGS error box.  In case that more than one X-ray source is detected within the Fermi error box, we select the one that is coincident with a radio source.

\end{itemize}

The total number of the UGSs in the 3FGL catalog that satisfy only the first three criteria is 238 and, as of today, for $\sim$180 objects there are \textit{Swift} pointings with an integration time greater than 2000 sec. 
Only for 60 of them, the observations revealed a detection of an X-ray source  inside the 3FGL error box.

\begin{figure*}[htbp]
  \centering
  \begin{minipage}[c]{.4\textwidth}
    \includegraphics[width=1.\textwidth]{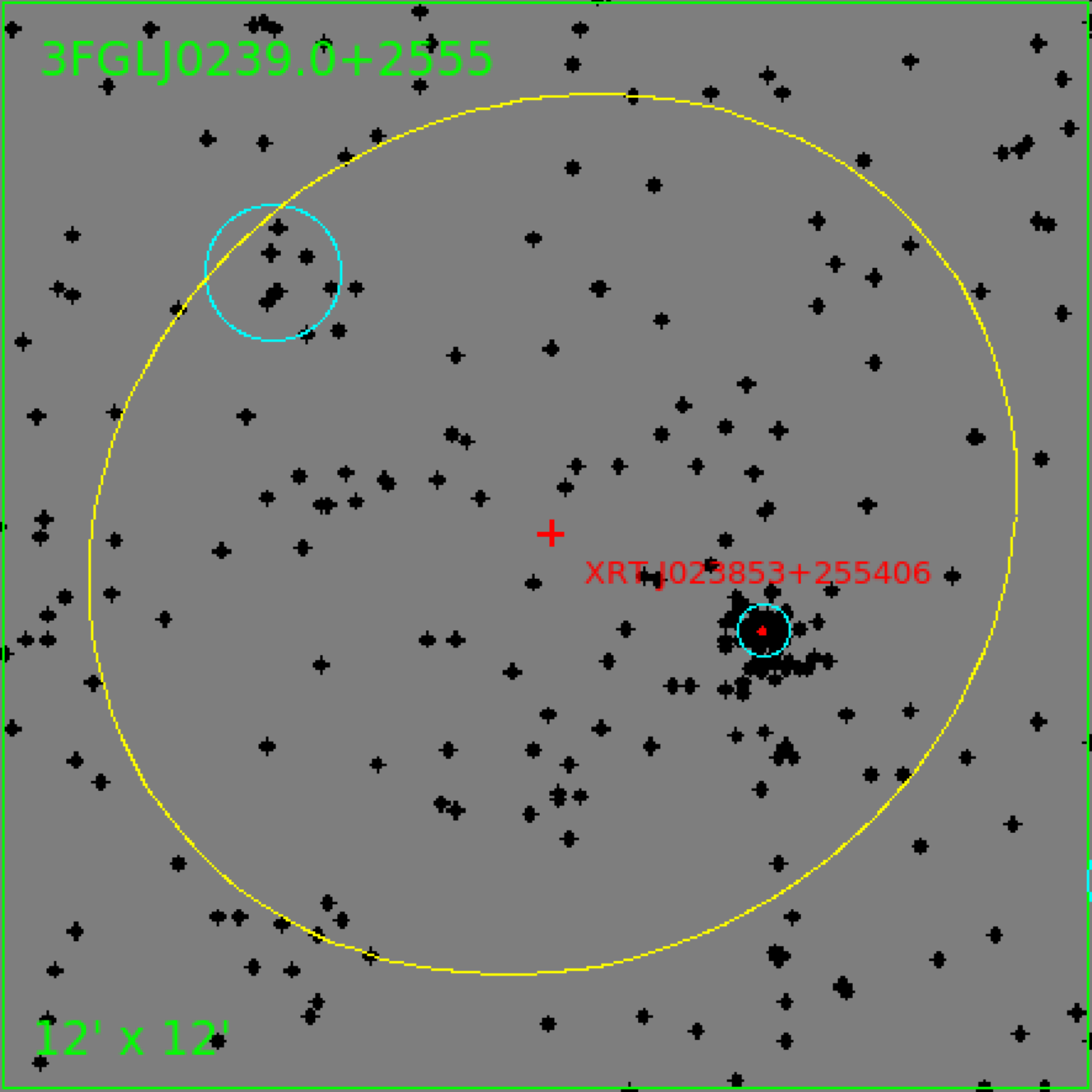}
  \end{minipage}%
  \hspace{3mm}%
  \begin{minipage}[c]{.4\textwidth}
    \includegraphics[width=1.\textwidth]{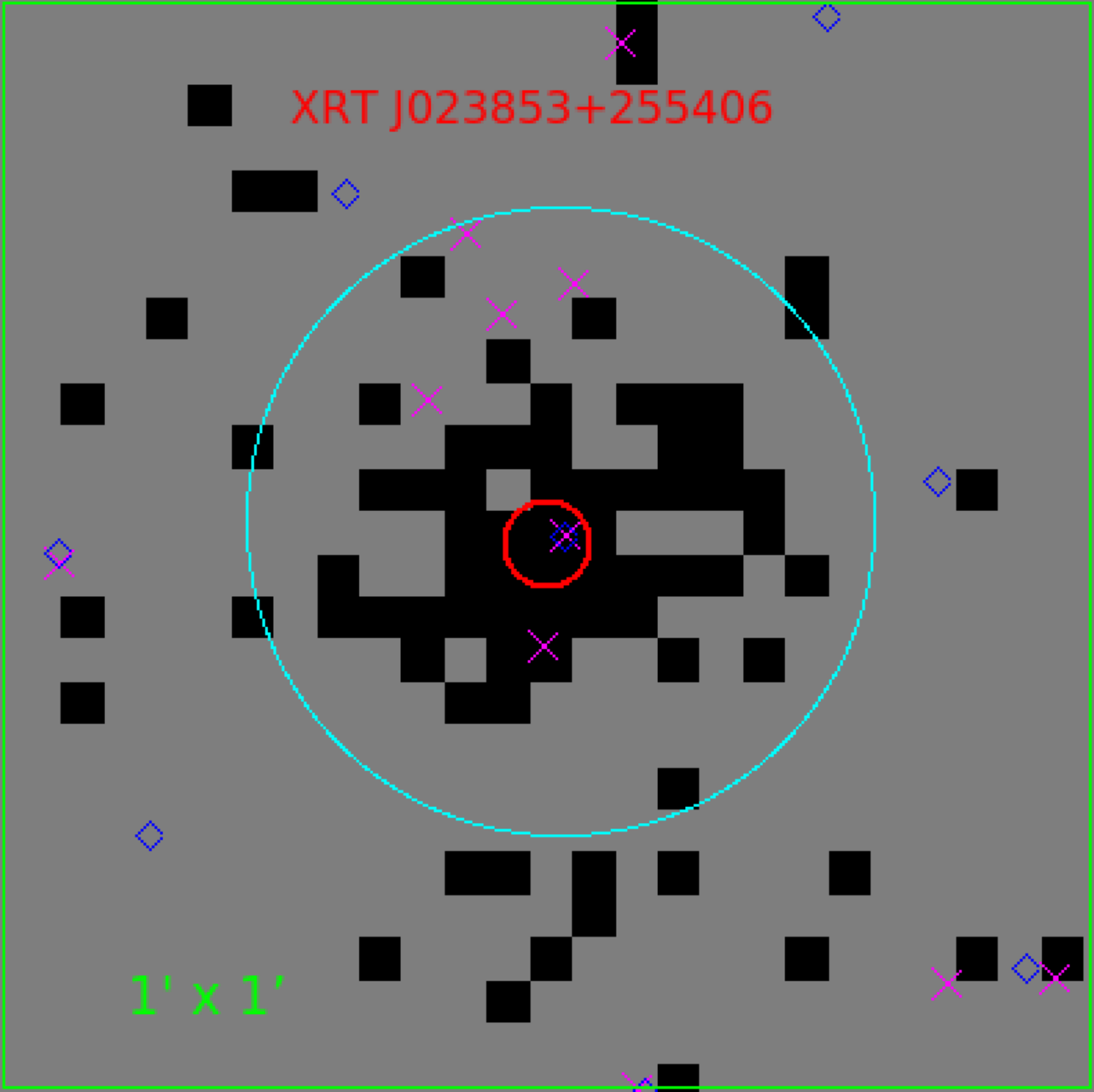}
  \end{minipage}
  \begin{minipage}[c]{.6\textwidth}
  \centering
   \includegraphics[width=0.7\textwidth, angle=-90]{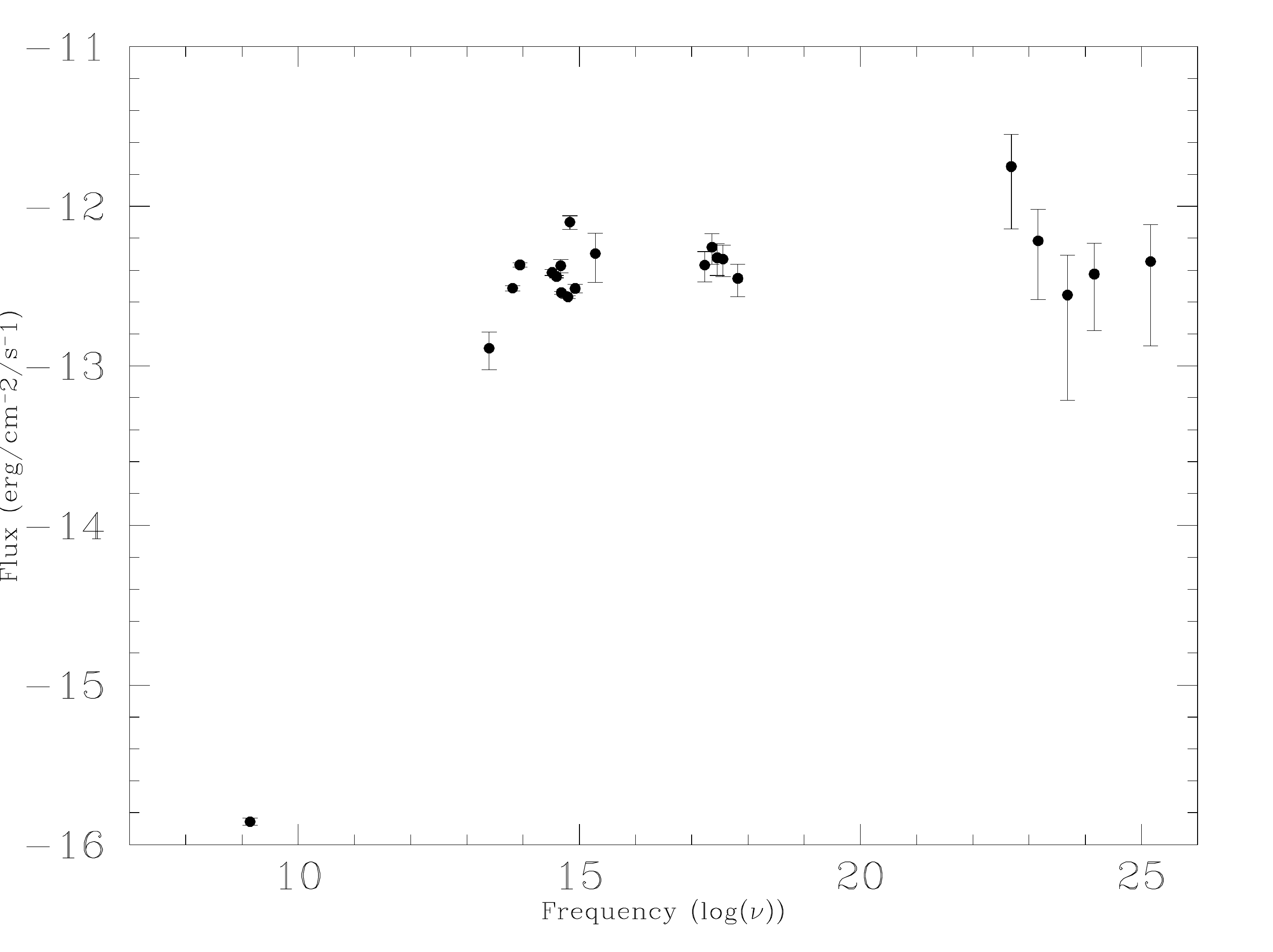}
  \end{minipage}
 \caption{Upper left panel: Swift/XRT images of 3FGL J0239.0+2555 created using the online data analysis tool of UK Swift Science Data Centre. The red cross is the position of 3FGL J0239.0+2555 as reported in the 3FGL catalog and the yellow ellipse is the 95\% error region of the 3FGL catalog. The XRT source detected in this work, XRT J023853+255406, is displayed as a red circle. The cyan circles show the error circles of the NVSS radio sources. Upper right panel: close-up of the XRT J023853+255406 sky map. The blue and magenta marks are the positions of WISE and SDSS objects. Bottom panel: broad-band SED of 3FGL J0239.0+2555 created using the SED Builder tool of the ASI ASDC Data Centre. We combine NVSS radio data, IR WISE data, SDSS optical data, the HE $\gamma$-ray data from the 3FGL catalog. The X-ray flux is derived from the analysis of the our \textit{Swift} data analysis.}
 \label{fig:0239}
\end{figure*}

Following previous works \citep{stephen2010, takahashi2012, takeuchi2013, acero2013, landi2015}, we search for X-ray emission inside the 3FGL error box (typically of a few arcmin). 
If an X-ray source is detected (with detection error circles $\sim$ 5 arcsec ), the next step is to search for likely counterparts of these X-ray sources in radio, infrared, and optical, in order to determine a positional association (an example is given in Fig.~\ref{fig:0239}). 
To further constrain the association we also build its multiwavelength SED by combining the available measured data-fluxes at different wavelengths and to select blazar-like objects \citep[for details see][(PFS)]{paiano2017ufo}. 

The obvious next step to single out the counterpart of the UGS is to obtain spectroscopic observations of candidate optical counterpart.
   
In this paper, we present the results for one third of the full sample (see Tab.~\ref{tab:table1}).
Note that the sample includes 6 sources classified as blazar candidates of uncertain type (BCU) and 2 BLL that were unassociated in the 2FGL. For these sources, no optical spectra were available or their redshift was very uncertain.

\section{Observations and data reduction} \label{sec:data}

The observations were obtained in service mode between December 2015 and May 2016  at the GTC using the low resolution spectrograph OSIRIS \citep{cepa2003}. 
The instrument was configured with the grism R500B\footnote{http://www.gtc.iac.es/instruments/osiris/osiris.php}, in order to cover the spectral range 3600-8400~$\textrm{\AA}$, and with a slit width~$=$~1.2''. 

The observation strategy and the data reduction followed the same procedure as reported in the \citet{paiano2017tev}.
For each source, three individual exposures were obtained that were then combined into a single average image, in order to perform optimal cleaning of cosmic rays and of CCD cosmetic defects. 
Detailed information on the observations are given in Tab. \ref{tab:table2}.

%\software{IRAF (Tody 1986, Tody 1993)}
Data reduction was carried out following standard IRAF\footnote{IRAF (Image Reduction and Analysis Facility) is distributed by the National Optical Astronomy Observatories, which are operated by the Association of Universities for Research in Astronomy, Inc., under cooperative agreement with the National Science Foundation.} procedures for long slit spectroscopy with bias subtraction, flat fielding, and bad pixel correction. 
Individual spectra were cleaned of cosmic-ray contamination using the L.A. Cosmic algorithm \citep{lacos}.

Wavelength calibration was performed using the spectra of Hg, Ar, Ne, and Xe lamps. Spectra were corrected for atmospheric extinction using the mean La Palma site extinction table\footnote{https://www.ing.iac.es/Astronomy/observing/manuals/}. 
For each source, during the same observation night,  we observed a spectro-photometric standard star in order to perform the relative flux calibration on every spectrum. 
The absolute flux calibration was possible thanks to the availability of a direct image of the source obtained as part of target acquisition. The average spectra were then calibrated to have the flux at 4750~$\textrm{\AA}$ equal to the photometry found for the targets (see Tab. \ref{tab:table2}).

Finally each spectrum has been dereddened for the Galaxy contribution, applying the extinction law by \citet{cardelli1989} and assuming the $E(B-V)$ values taken from the NASA/IPAC Infrared Science Archive \footnote{https://irsa.ipac.caltech.edu/applications/DUST/}.

\section{Results} \label{sec:results}

The flux-calibrated optical spectra of the optical counterparts of 20 $\gamma$-ray sources are presented in Fig. \ref{fig:spectra} and can be accessed at the website http://www.oapd.inaf.it/zbllac/.
In order to emphasize weak emission and/or absorption features, we show also the normalized spectrum. 
This was obtained by dividing the observed calibrated spectrum by a power law fit ($F_{\lambda}~\propto~\lambda^{\alpha}$) of the spectral continuum, excluding the telluric absorption bands.  
%($F_{\nu}~\propto~\nu^{\alpha}$)
These normalized spectra were used to evaluate the S/N in a number of spectral regions (see Tab. \ref{tab:table3}).

All spectra were carefully inspected to find emission and absorption features.  
When a possible feature was found, we determined its reliability by checking that it was present on the three individual exposures (see Sec. \ref{sec:data} for details).  
We were able to detect spectral lines for 16 targets: in particular we found absorption lines due to the host galaxy: Ca~II (3934~$\textrm{\AA}$, 3968~$\textrm{\AA}$),  G-band (4305~$\textrm{\AA}$), Mg~I (5175~$\textrm{\AA}$) and Na~I (5893~$\textrm{\AA}$) for 11 objects.
We observed the emission line due to [O~II] (3727~$\textrm{\AA}$) and [O~III] (5007~$\textrm{\AA}$) from 3FGLJ0049.0+4224, 3FGLJ0305.2-1607, 3FGLJ1049.7+1548 and 3FGLJ1704.1+1234 and the absorption lines attributed to Mg~II (2800~$\textrm{\AA}$) intervening systems in the spectrum of 3FGLJ0338.5+1303, 3FGLJ0644.6+6035, 3FGL J1129.0+3758, 3FGLJ1511.8-0513 and 3FGLJ2115.2+1215, that allow us to derive a spectroscopic lower limit of the redshift.
Details in Fig. \ref{fig:spectraCU} and Tab. \ref{tab:line}. 

For four targets, the observed spectra are completely featureless. 
Based on the assumption that all BLLs are hosted by a massive elliptical galaxy, one may look for faint absorption features from the starlight \citep{sbarufatti2006a}, provided that the SNR and the spectral resolution are sufficiently high.
Following the scheme outlined in \citet{paiano2017tev}, in these cases it is possible to set a lower limit to the redshift based on the minimum Equivalent Width (EW) that can be measured in the spectrum (see Tab.~\ref{tab:table3}).

\section{Notes for individual sources } \label{sec:notes}
\label{individual}

\begin{itemize}
\item[] \textbf{3FGL J0049.0+4224}: %OB0012-34
The analysis of \textit{Swift}-XRT data reveals one X-ray object (F=5.7$\times 10^{-13}$ erg cm$^{-2}$ s$^{-1}$ ) in the 3FGL error box, that is spatially coincident with the optical source SDSS J004859+422351 (g = 19.9 ) and the radio source NVSS J004859+422350. 
Our optical spectrum is characterized by a power law (PL) emission ($\alpha$ = -0.16) with a signature of diluted galaxy starlight typical of BLL.  
We clearly detect stellar absorption features identified as Ca II (3934~$\textrm{\AA}$, 3968~$\textrm{\AA}$),  G-band (4305~$\textrm{\AA}$), and  Mg I (~$\textrm{\AA}$5175), and two weak emission lines due to [O II] (3727~$\textrm{\AA}$) and [O III] (5007~$\textrm{\AA}$) at the redshift z~=~0.302.
This confirms the blazar nature of this $\gamma$-ray source (see Fig. \ref{fig:spectra} and Fig. \ref{fig:spectraCU}).

\item[] \textbf{3FGL J0102.1+0943}: %OB0002-34  
This source is proposed to be associated to the optical source SDSS~J010217+094409 as an high-synchrotron-peaked sources peak blazar at a redshift of 0.4~-~0.5 \citep{paiano2017ufo}.
Our optical spectrum is dominated by a PL emission ($\alpha$~=~-0.10 ). 
We find a weak signature of a Ca~II break at $\sim$5700~$\textrm{\AA}$ (see Fig. \ref{fig:spectraCU}) yielding a tentative redshift of 0.42.

\item[] \textbf{3FGLJ0239.0+2555}: %OB0013-34
Through the XRT data analysis, we find only one X-ray source within the 3FGL error box with a flux at the same level of the $\gamma$-ray emission flux (see Fig.\ref{fig:0239}).  
%with a F=1.4$\times 10^{-12}$ erg cm$^{-2}$ s$^{-1}
We propose the spatially coincident object SDSS~J023853+255407 (g = 20.2) as the likely optical counterpart for this source (see Fig. \ref{fig:opt_skymap}).  
Our optical spectrum for this source is clearly dominated by a PL emission ($\alpha$= -0.39 ) and a doublet absorption features (6231~$\textrm{\AA}$, 6286~$\textrm{\AA}$) is detected. 
If identified as Ca II (3934~$\textrm{\AA}$, 3968~$\textrm{\AA}$), the redshift is 0.584.
Note that the red component of this doublet is partially contaminated by the telluric band at 6280~$\textrm{\AA}$.

\item[] \textbf{3FGLJ0305.2-1607}: %OB0003-34 
This $\gamma$-ray emitter is associated with the radio source PKS 0302-16 and classified as BCU-II in the 3LAC catalog, but no optical spectrum for this target is available in literature.
The GTC optical spectrum clearly exhibits absorption features of the overall stellar population superimposed onto the non-thermal emission.
In particular we detect absorption lines of Ca~II (3934~$\textrm{\AA}$, 3968~$\textrm{\AA}$),  G-band (4305~$\textrm{\AA}$), Mg~I (5157~$\textrm{\AA}$) and Na~I (5893~$\textrm{\AA}$) at z=0.311, and moreover an emission line due to [O II]  (L=6.2$\times$10$^{40}$ erg/s) indicative of modest star formation \citep{gilbank2010}.

\item[] \textbf{3FGLJ0338.5+1303}: %OB0004  
In the 3LAC catalog, the source is associated to the radio loud active galaxy RX J0338.1+1302 and classified as BCU-II.
The optical spectrum of this object exhibits a featureless continuum except for a clear absorption line at 3867~$\textrm{\AA}$ with an EW~=~3.0~$\textrm{\AA}$. 
If attributed to Mg II (2800~$\textrm{\AA}$), we can set a spectroscopic redshift lower limit of 0.382.
A featureless spectrum for this object was also reported by \citet{marchesini2016}. Note that the absorption line is also present in their published spectrum but it was not noted and identified.

\item[] \textbf{3FGL J0409.8-0358 }: %OB0006-34
The optical counterpart of this source was identified with a BLL by \citet{massaro2015b} on the basis of the featureless optical spectrum.
We obtain a much better S/N spectrum and we confirm that it is lineless. 
We can set a lower limit of the redshift of z~$>$~1.1, following the procedure reported in \citet{paiano2017tev}.

\item[] \textbf{3FGLJ0644.6+6035}: %OB0007-34
This object is an UGS in the 3FGL. 
The optical counterpart of this $\gamma$-ray source was proposed to be WISE J064459.38+603131.7 and classified as a blazar at z~$=$~0.358 \citep{paggi2014}.
Based on the new 3FGL error box  and the analysis of the \textit{Swift} data, we propose the X-ray source XRT J064435+603850 as the likely counterpart for the \textit{Fermi} emitter.
Our optical spectrum confirms the blazar nature of the candidate and a prominent intervening absorption system of Mg II (2800~$\textrm{\AA}$) is detected at 4425 $\textrm{\AA}$, setting a redshift lower limit of the object at z~$>$~0.581.

\item[] \textbf{3FGLJ0937.9-1435}: %OB0011-34
We found an X-ray emission within the 3FGL error box (XRTJ093754-143350) that is coincident with the optical source USNOB0754-0223141.
The same association was also proposed using IR objects from the WISE survey \citep{massaro2013b}
The optical spectrum shows a characteristic non thermal emission with signature of host galaxy. 
We detect Ca II (3934~$\textrm{\AA}$, 3968~$\textrm{\AA}$), G-band (4305~$\textrm{\AA}$), Mg I (5157~$\textrm{\AA}$) and Ca+Fe (5269~$\textrm{\AA}$) at z=0.287.

\item[] \textbf{3FGLJ0952.8+0711}: %OB0015-34
Based on XRT data, the most plausible optical counterpart is the source SDSS J095249.57+071329.9 (g=18.9).
No optical spectra are found in the literature.  
Our optical spectrum is clearly dominated by non-thermal power law emission and we are able to detect faint absorption features of Ca II (3934~$\textrm{\AA}$, 3968~$\textrm{\AA}$) and G-band (4305~$\textrm{\AA}$) at z=0.573 .
It is worth noting that from the SDSS data in the environment of this source there are two galaxies (projected distance $<$ 300 kpc at z~$=$~0.573) that exhibit the same redshift of the target.

\item[] \textbf{3FGLJ1049.7+1548}: %OB0014-34
In the 3FGL this object is classified as UGS. However, the source was associated to the optical counterpart SDSS J104939.35+154837.6 by \citet{paggi2014} who propose a redshift z~=~0.327 based on Ca~II (3934~$\textrm{\AA}$, 3968~$\textrm{\AA}$) absorption lines. 
On the other hand a better quality optical spectrum was obtained by the SDSS Boss survey suggesting a redshift z~=~1.452.
Given the large inconsistency we obtain an high SNR spectrum for this object that confirms the redshift of z~=~0.3271 (see Fig. \ref{fig:spectra} and Fig. \ref{fig:spectraCU}). 
In addition to Ca II absorption doublet, we detect a possible faint [OII] emission. 
The source was found $\sim$~1~mag brighter than observed by SDSS.

\item[] \textbf{3FGL J1129.0+3758}: %OB0009-34
This source is classified as UGS in the 3FGL catalog with a flux of F$_{(1-100 GeV)}$~=~6.99$\times$10$^{-10}$ ph~cm$^{-2}$~s$^{-1}$.
\citet{paiano2017ufo} proposed as optical counterpart the source SDSSJ112903+375656 with g~=~20.3, classifying as a blazar at high redshift (z~$\sim$~1.4~-~1.8). This object is located at $\sim$5 arcsec West of a g~=~14. star.
In its optical spectrum, there are many clear absorption features in the spectral range between 4000~$\textrm{\AA}$ and $<6000~\AA$. The strongest one, at 6189~$\textrm{\AA}$, is a doublet that is consistent with an intervening absorption system due to Mg~II (2800 $\textrm{\AA}$) at redshift z~=~1.211. This is clearly present also in the SDSS spectrum. 
For the other features (see Tab. \ref{tab:table4}), no clear identification is found. 
It is worth to note that some of them are close to stellar lines as H$_{\gamma}$,  Mg~I (5157~$\textrm{\AA}$) and Ca+Fe (5269~$\textrm{\AA}$), but the level of contamination of our spectrum by the presence of the bright star is negligible, since the slit intersects only marginally the stellar flux.
In addition to the spectral lines, also the continuum appears somewhat unusual: at $\lambda>$~5000$\textrm{\AA}$ the continuum emission is rather flat, while at shorter wavelength a rise of the flux is noted, suggesting a thermal component. The SDSS spectrum shows a similar shape.

\item[] \textbf{3FGLJ1222.7+7952}: %OB0006-64
The spectrum of the optical counterpart \citep{massaro2015b} for this $\gamma$-ray source failed to detect any spectral features and the redshift remained undetermined.
Our better SNR spectrum clearly shows the Ca II (3934~$\textrm{\AA}$, 3968~$\textrm{\AA}$) and other lines (see Tab. \ref{tab:table4}) characteristic of stellar population at z~=~0.375.

\item[] \textbf{3FGLJ1340.6-0408}: %OB0002-64
An optical spectrum of the counterpart of this $\gamma$-ray object was obtained by \citet{ricci2015} who found it featureless.
In our new spectrum we detect  a weak absorption doublet  (EW~=~0.4A$\textrm{\AA}$) at $\sim$~4830$\textrm{\AA}$ and other weak absorptions. These lines are all consistent with a redshift of z=0.223 (see Fig. \ref{fig:spectra} and Fig \ref{fig:spectraCU}.)

\item[] \textbf{3FGLJ1411.4-0724}: %OB0010-64
We obtain the optical spectrum of the X-ray counterpart source XRTJ141133-072253, likely associated to the $\gamma$-ray emitter (see Fig. \ref{fig:opt_skymap}), with a moderate S/N ratio.
No emission or absorption lines are detected, however the BLL nature of the source is confirmed.

\item[] \textbf{3FGL1511.8-0513} %OB0001-64
The source is classified as a BCU-III and associated to the radio source NVSSJ151148-051345 in the 3LAC catalog.
The optical spectrum of \citet{alvarez2016A} was found featureless.
We obtain a high S/N ratio ($\sim$200) spectrum that exhibits a marked non thermal featureless continuum form 4100~$\textrm{\AA}$ to 8500~$\textrm{\AA}$.
We detect a single absorption at 4053 $\textrm{\AA}$ with EW~$=$~2.1~$\textrm{\AA}$ that, if interpreted as Mg II (2800~$\textrm{\AA}$), sets a spectroscopic lower limit of the redshift of z$>$0.45.

\item[] \textbf{3FGLJ1704.4-0528} %OB0003-64
From the \textit{Swift}-XRT data analysis we found only one bright X-ray source within the 3FGL error box, coincident with the optical source USNOB0845-0308445 (B = 19.5) (see Fig. \ref{fig:opt_skymap}).
We obtained a S/N $\sim$~100 optical spectrum that is featureless and confirms the BLL nature for this object. 
Based on the assumption of the typical BLL host galaxy, we can set a lower limit of the redshift of $>$0.7.

\item[] \textbf{3FGLJ1704.1+1234} %OB0005-64
On the basis of the X-ray imaging, in the 3FGL error box of this $\gamma$-ray emitter we found an X-ray source XRTJ170409+123421 and we propose that it is associated to the optical source SDSSJ170409+123421 (g=19.1).
The same association was also proposed by \citet{alvarez2016B} who also obtained an optical spectrum, deriving a redshift of z~$=$~0.45. 
Our better S/N spectrum confirms this redshift and, in addition to absorptions due to  Ca II (3934~$\textrm{\AA}$, 3968~$\textrm{\AA}$), G-band (4305~$\textrm{\AA}$) and Mg I (5157~$\textrm{\AA}$), we also detect the [O II] (3727~$\textrm{\AA}$) and [O III] (5007~$\textrm{\AA}$) emission features at the redshift z~$=$~0.452.

\item[] \textbf{3FGLJ2115.2+1215} %OB0009-64
From the XRT analysis, we found the X-ray object XRTJ211522+121801 coincident with the radio source NVSSJ211522+121802 and the optical source SDSSJ211522+121802.
The source is relatively bright (g~$=$~17.2) and the optical spectrum exhibits a non thermal continuum. 
We detect a prominent absorption feature at 4191 $\textrm{\AA}$ (EW~$=$~5.0) that, if interpreted as Mg II (2800~$\textrm{\AA}$), implies a spectroscopic redshift lower limit of z$>$0.497.
In addition, we detect two other fainter absorptions at 4270~$\textrm{\AA}$ and 4571~$\textrm{\AA}$. 
They are likely two faint absorption systems that could be interpreted as intervening Mg II at z~$=$~0.525 and z~$=$~0.633.

\item[] \textbf{3FGLJ2246.2+1547} %OB0001-34
This $\gamma$-ray emitter is associated to the radio source NVSSJ224604+154437 and classified as BCU-II in the 3LAC catalog. No optical spectra were previously found in the literature.
Our spectrum of the optical counterpart confirms the BLL nature of the source and exhibits a pure featureless spectrum.
We can set a lower limit of the redshift of z$>$~0.7 (see Tab. \ref{tab:table3}).

\item[] \textbf{3FGLJ2346.7+0705} %OB0008-34
In the 3LAC catalog, this source is associated to the radio source TXS~2344+068 and classified as BCU-II.
There is a quasi-featureless optical spectrum provided by SDSS survey that, depending on the Data Release, proposed two different redshift values (z~$=$~0.171 and z~$=$~5.06). 
In our spectrum we detect weak absorptions of Ca II (3934~$\textrm{\AA}$, 3968~$\textrm{\AA}$), G-band (4305~$\textrm{\AA}$) and Mg I (5157~$\textrm{\AA}$), corresponding to a redshift of 0.171, superimposed to the non-thermal continuum.
\end{itemize}

\section{Discussion and conclusions} \label{sec:discu}

We secured optical spectra of the counterparts of 20 UGSs detected by the \textit{Fermi} satellite with the aim to investigate the nature of these sources and to determine their redshift. 
For all these objects the optical spectrum was either not previously known, or secured with modest S/N, not allowing clear spectral features to be detected \citep[see e.g.][ and references therein]{massaro2016}.
This allows us to classify the optical counterparts of the UGS sample and to derive their redshift or lower limits of it.

The optical spectrum of all these objects is characterized by a typical power law arising from the non thermal emission. For 11 sources, we found absorption and emission lines, allowing us  to determine their redshift, while for 5 targets we set spectroscopic lower limits by the detection of absorption lines from intervening systems.
Only for 4 sources the optical spectrum is entirely featureless in spite of the good S/N. For these objects, we can set lower limits of the redshift based on the minimum equivalent width of absorption features expected from their host galaxy as discussed in \cite{paiano2017tev}.

The measured redshifts are in the range between 0.2 and 0.6 for most of the sources in the sample.
One object (3FGL J1129.0+3758) is found at z~$>~$1.2, while four sources, with featureless spectra, are likely at z$>$~0.6. 
Fourteen out of 20 objects are detected by \textit{Fermi} satellite at energies above 10 GeV \citep[see the Third Catalog of Hard Fermi-LAT Sources (3FHL), ][]{ajello2017}. Therefore, they are candidates to be good TeV targets for the Cherenkov telescopes. Given their redshift (of $\sim$0.4 average), these offer a good opportunity to increase the number of blazars at $z~>~0.2$ detected at VHE energies (E$~>~$100 GeV) suitable to probe the UV-optical EBL attenuation in their TeV spectra.

The optical spectra and redshifts reported allow us to test a recently published tool for blazar recognition and classification \citep[PFS, see details in ][]{paiano2017ufo}. This, based on analyses of the Spectral Energy Distributions and luminosities of blazars, also provides us with rough estimates of the redshift (see Table \ref{tab:table5}). 
Now, considering the 16 objects of our sample with spectroscopic redshift measurements or lower limits, for about 50\% the estimated redshift by PFS turns out to be in agreement within the 2 sigma uncertainties with the redshift measurement, and for the remaining sources, 30\% of them are in agreement  within 3 sigma. 
Altogether, these results support the fair effectiveness of the PFS blazar recognition tool to unveil and roughly characterize blazars among the numerous population of the UGSs detected by the \textit{Fermi} satellite.

Finally, we note that, although the optical spectra were obtained with a large aperture telescope and modern instrumentation, for 4 sources their spectra are still featureless, thus preventing a redshift measurements. 
This will likely remain unknown until the advent of the next generation of extremely large telescopes, as E-ELT\footnote{http://www.eso.org/sci/facilities/eelt/} and TMT\footnote{http://www.tmt.org/}.

%\newpage
%\section{TABELLE}
%\input{/home/paiano/Dropbox/GTC_TeV/tabelle/Tabelle_Latex/tev_bllac_table1.tex}
\begin{table*}
\caption{THE SAMPLE  }\label{tab:table1}
%\centering
%\begin{tabular}{llcccccccll}
\begin{tabular}{lllllllll}
\hline 
3FGL Name  &    Optical Counterpart     &  RA      &    DEC     &  Class  & mag & E(B-V)  & $z$  & Reference \\
\hline
3FGLJ0049.0+4224 & SDSSJ004859+422351 & 00:48:59.1 & +42:23:51.0  & ugs   & 19.9 &  0.07 & ?     &   \\
3FGLJ0102.1+0943 & SDSSJ010217+094409 & 01:02:17.1 & +09:44:09.5  & ugs   & 18.8 &  0.03 & ?     &   \\
3FGLJ0239.0+2555 & SDSSJ023853+255407 & 02:38:53.8 & +25:54:07.1  & ugs   & 20.2 & 0.13 & ?     &     \\ 
3FGLJ0305.2-1607  & SDSS J030515-160816 & 03:05:15.0 & -16:08:16.6   & bcuII & 18.9 & 0.04 & ?     &    \\
3FGLJ0338.5+1303 & USNOB1030-0045117   & 03:38:29.2 & +13:02:15.7  & bcuII & 19.9 & 0.30 &  ?    & \footnotesize{\citet{marchesini2016}} \\
3FGLJ0409.8-0358  & NVSSJ040946-040003  & 04:09:46.5 & -04:00:03.5   & bll     & 19.4  & 0.07 &  ?    &  \footnotesize{\citet{massaro2015b}} \\
3FGLJ0644.6+6035 & USNOB1506-0162421   & 06:44:35.7 & +60:38:51.3  & ugs   &  19.9 & 0.07 &  0.358 ?  & \footnotesize{\citet{paggi2014}} \\
3FGLJ0937.9-1435  & USNOB0754-0223141   & 09:37:54.7 & -14:33:50.4   &  ugs  & 18.8 &  0.05 &   ?   &      \\
3FGLJ0952.8+0711 & SDSSJ095249+071329  & 09:52:49.5 & +07:13:29.9  & ugs  &  18.9 &  0.04 &   ?   &      \\
3FGLJ1049.7+1548 & SDSSJ104939+154837  & 10:49:39.3 & +15:48:37.6  & ugs  &  18.1 &  0.02 &   0.327  &       \footnotesize{\citet{paggi2014}} \\
3FGLJ1129.0+3758	 & SDSSJ112903+375656  & 11:29:03.2 & +37:56:56.7  &  ugs  &  20.3 &  0.02  &  4.09 ?  &     SDSS\\ 
3FGLJ1222.7+7952 & USNOB1698-0045483   & 12:23:58.1 & +79:53:28.6   & ugs   &  20.1 &  0.10 &  ?  &  \footnotesize{\citet{massaro2015b} } \\
3FGLJ1340.6-0408  & NVSSJ134042-041006   & 13:40:42.0 & -04:10:07.0   &  bcuII &  18.2 &  0.03 &   ?   &    \footnotesize{\citet{ricci2015}}  \\
3FGLJ1411.4-0724  & USNOB0826-0334743    & 14:11:33.3 & -07:22:53.3   &  ugs    &  19.5 &  0.03 &    ? &        \\
3FGLJ1511.8-0513  & NVSSJ151148-051345    & 15:11:48.5 & -05:13:46.7   &  bcuIII &  18.4 &  0.08 &   ?   &  \footnotesize{\citet{alvarez2016A}}  \\
3FGLJ1704.4-0528  & USNOB0845-0308445    & 17:04:33.8 & -05:28:41.1   &  ugs    &   19.5 & 0.46   & ?   &       \\
3FGLJ1704.1+1234 & SDSSJ170409+123421  & 17:04:09.6 & +12:34:21.3  &  ugs   &   19.1 &  0.06   &  0.45  &   \footnotesize{\citet{alvarez2016B}} \\
3FGLJ2115.2+1215 & SDSSJ211522+121802   & 21:15:22.0 & +12:18:02.8  &  ugs   &   18.2 &  0.03 &   ?   &      \\
3FGLJ2246.2+1547 & NVSSJ224604+154437  & 22:46:04.9 & +15:44:35.5  &  bcuII &   19.5  & 0.07  &   ?  &      \\
3FGLJ2346.7+0705 & TXS2344+068                 & 23:46:39.8 & +07:05:06.8  & bcuII  &   17.3 &  0.18 &   0.17   &   SDSS\\
\hline
\end{tabular}
\tablenotetext{}{
\raggedright
\footnotesize \texttt{Col.1}: Name of the target; \texttt{Col.2}: Optical counterpart of the target; {Col.3 - 4 }: Right ascension and declination of the optical counterpart; \texttt{Col.5}: Classification in the 3LAC catalog; \texttt{Col.6}: g magnitude from SDSS-DR13 (for the SDSS sources) or B-band magnitudes taken from USNOB1.0 catalog; \texttt{Col.7}: $E(B-V)$ taken from the NASA/IPAC Infrared Science Archive (https://irsa.ipac.caltech.edu/applications/DUST/); \texttt{Col.8}: Redshift; \texttt{Col.9}: Reference for the redshift.}
\tablenotetext{}{
\raggedright
 } 
\end{table*}

\newpage
\begin{table*}
\caption{LOG OF THE OBSERVATIONS }\label{tab:table2}
\centering
\begin{tabular}{lllll}
\hline
\hline
Obejct          & t$_{Exp}$ (s)  &       Date , Time (UT)            & Seeing ('') & g  \\
\hline
3FGLJ0049.0+4224 & 3600 & 2015 Dec 09, 21:40:16 & 1.5 & 19.7 \\
3FGLJ0102.1+0943 & 1800 & 2015 Dec 09, 22:55:55   & 2.0 & 19.3  \\
3FGLJ0239.0+2555 & 3600 & 2015 Dec 08, 23:35:25 & 2.0 & 20.1 \\
3FGLJ0305.2-1607  & 3600 & 2015 Dec 03, 00:03:53  & 0.9 & 18.9 \\
3FGLJ0338.5+1303 & 3600 & 2015 Dec 19, 20:42:03 & 1.4 & 18.4 \\
3FGLJ0409.8-0358  & 4200 & 2015 Dec 09, 00:02:11 & 1.9 & 19.7 \\
3FGLJ0644.6+6035 & 3600 & 2015 Dec 09, 05:02:43 & 1.7 & 19.1 \\
3FGLJ0937.9-1435  & 3000 & 2015 Dec 23, 03:35:35 & 1.5 & 18.6 \\
3FGLJ0952.8+0711 & 3000 & 2015 Dec 08, 04:52:08 & 1.8 &19.3 \\
3FGLJ1049.7+1548 & 1200 & 2015 Dec 09, 06:25:25 & 1.3 & 16.9 \\
3FGL J1129.0+3758	& 4200 & 2015 Dec 18, 04:48:16 & 1.7 & 20.8 \\
3FGLJ1222.7+7952 & 4500 & 2016 May 16, 21:27:58 & 1.8 & 19.3 \\
3FGLJ1340.6-0408  & 1500  & 2016 Apr 23, 02:23:13 & 1.0 & 17.2 \\
3FGLJ1411.4-0724  & 4500 & 2016  Apr 23, 03:05:20& 1.5 & 17.9 \\
3FGLJ1511.8-0513  & 3600 & 2016 Mar 24, 03:13:07 & 1.4 & 17.5 \\
3FGLJ1704.4-0528  & 4500 & 2016 May 16, 00:02:18 & 1.8 & 18.6 \\
3FGLJ1704.1+1234 & 3600 & 2016 Mar 24, 04:47:46 & 1.2 & 18.9 \\
3FGLJ2115.2+1215  & 2100 &	2016 Jun 04, 04:04:04 & 0.7 & 17.2 \\
3FGLJ2246.2+1547  & 3600 & 2015 Dec 08,  22:07:30 & 3.0 & 18.8 \\
3FGLJ2346.7+0705  & 750 & 2015 Dec 09, 21:12:51  & 1.5 & 17.3 \\
\hline
\end{tabular}
\tablenotetext{}{
\raggedright
\footnotesize \texttt{Col.1}: Name of the target; \texttt{Col.2}: Total integration time; \texttt{Col.3}: Date of observation; \texttt{Col.4}: Seeing during the observation; \texttt{Col.5}: g mag measured from the acquisition image.}
\end{table*}

\newpage
\begin{table*}
\caption{PROPERTIES OF THE OPTICAL SPECTRA }\label{tab:table3}
\centering
\begin{tabular}{lcccl}
\hline
OBJECT           & $\alpha$  &   SNR       &   EW$_{min}$            &  z                             \\
\hline
3FGLJ0049.0+4224        &  -0.16    &   40       &  0.68 - 0.80   	&  0.302$^{e,g}$         \\  
3FGLJ0102.1+0943        &  -0.10   &    50       &  0.65 - 0.90   	&  0.42$^g$:                \\  
3FGLJ0239.0+2555        &   -0.39   &   60       &  0.45 - 0.80   	&  0.584$^g$:              \\ 
3FGLJ0305.2-1607         &   -0.25   &   80       &  0.45 - 0.55   	&  0.312$^{e,g}$          \\ 
3FGLJ0338.5+1303        &    -1.10  &   200     &  0.20 - 0.30      &  $>$~0.382$^a$        \\ 
3FGLJ0409.8-0358         &   -0.50   &  110      &  0.35 - 0.75      &  $>$0.7$^h$              \\ 
3FGLJ0644.6+6035        &   -0.17  &   100      &  0.25 - 0.50      &  $>$~0.581$^a$        \\ 
3FGLJ0937.9-1435         &   -0.22   &  100      &  0.40 - 0.50      &  0.287$^g$                \\ 
3FGLJ0952.8+0711        &   -0.62   &   65       &  0.40 - 1.15      &  0.574$^g$                \\ 
3FGLJ1049.7+1548        &   -0.80   &  300      &  0.08 - 0.15      &  0.326$^{e,g}$          \\ 
3FGL J1129.0+3758	      &	    -0.23  &   120     &  0.20 - 0.50      &  $>$~1.211$^a$         \\
3FGLJ1222.7+7952        &  +0.91    &  30       &  0.65 - 1.30   	&   0.375$^g$              \\ 
3FGLJ1340.6-0408         &   -0.75   &  130      &  0.20 - 0.40    	&  0.223$^g$              \\ 
3FGLJ1411.4-0724         &   -0.23   &   45       &  0.50 - 1.50       & $>$~0.72$^h$        \\ 
3FGLJ1511.8-0513         &   -1.36   &  200      &  0.15 - 0.25    	&  $>$~0.45$^a$        \\
3FGLJ1704.4-0528         &   -0.96   &  80        &  0.40 - 0.55       &	$>$0.7$^h$        \\
3FGLJ1704.1+1234        &  -0.41    &  95        &  0.25 - 0.50       &  0.452$^{e,g}$         \\
3FGLJ2115.2+1215        &   -0.71   &  155      & 0.17 - 0.45     	&  $>$~0.497$^{a}$*   \\ 
3FGLJ2246.2+1547        &   -0.49   &  80        &  0.45 - 1.20      &  $>$~0.7$^h$           \\ 
3FGLJ2346.7+0705        &  -0.55    & 160       &  0.20 - 0.35       &	 0.171$^g$          \\
\hline
\end{tabular}
\tablenotetext{}{
\raggedright
\footnotesize \texttt{Col.1}: Name of the target; \texttt{Col.2}: Optical spectral index derived from a power law fit in the range 3800-8400; \texttt{Col.3}: S/N of the spectrum; \texttt{Col.4}: Range of the minimum equivalent width (EW$_{min}$) derived from different regions of the spectrum; \texttt{Col.5}: Spectroscopic redshift: the superscript letters are: \textit{e} = emission line, \textit{g} = galaxy absorption line, \textit{a}= intervening absorption assuming Mg~II (2800$\textrm{\AA}$) identification, \textit{h}= lower limit derived on the lack of detection of host galaxy absorption lines assuming assuming a BLL host galaxy with M(R) = -22.9.\\
(:) This marker indicates that the redshift is tentative, (*) For this source we found other two absorption line systems. (See details text). }
\end{table*}

\setcounter{table}{3}  
\begin{table*}
\caption{MEASUREMENTS OF THE SPECTRAL LINES}\label{tab:table4}
\centering
\begin{tabular}{lrlcl}
\hline
Object name          &  $\lambda$    &    EW   &     Line ID    &   z   \\
%                          &  $\textrm{\AA}$             &     $\textrm{\AA}$           &                &               \\
\hline
3FGLJ0049.0+4224 &  4852  & 1.4 &  [OII] 3727        & 0.302 \\
                                 &  5122  & 1.6 &  Ca II 3934       & 0.302 \\
                                 &  5167  & 2.0 &  Ca II 3968       & 0.302 \\
                                 &  5601  & 0.6 &  G-band 4305   & 0.301\\
                                 &  6517  & 1.1 &  [OIII] 5007       & 0.302 \\
                                 &  6734  & 1.4 &  Mg I 5175        & 0.301\\
3FGLJ0102.1+0943  &  $\sim$5650  & -	&  Ca II break & $\sim$0.42:\\
3FGLJ0239.0+2555  &  6231 & 2.7	& Ca II 3934       & 0.584: \\
                                  &  6286 & 2.0*	& Ca II 3968       & 0.584:*\\
3FGLJ0305.2-1607   &  4890  & 1.3 & [OII] 3727       & 0.312 \\
                                  &  5160 &  1.3 & Ca II 3934      & 0.312 \\
                                  &  5206 &  1.2 & Ca II 3968      & 0.312 \\
                                  &  5647 &  1.4 & G-band 4305  & 0.312\\
                                  % &  6378.9 & 2.09 & Hbeta     & 0.312\\
                                  &  6785 & 1.4  & Mg I 5175       & 0.311 \\
                                  &  7729 & 0.6  &  Na I 5893      & 0.312\\
3FGLJ0338.5+1303  &  3870 & 3.0  & Mg II  2800   & $>$~0.382$^a$\\
3FGLJ0644.6+6035   & 4425  & 5.0   & Mg II 2800 & $>$~0.581$^a$ \\
3FGLJ0937.9-1435    & 5061  & 0.9	 & Ca II 3934   & 0.287\\
                                   & 5106  & 0.7   & Ca II 3968   & 0.287 \\
                                   & 6660  & 0.8   & Mg I 5175    & 0.287 \\
                                   & 6781  & 0.4   & Ca+Fe 5269    & 0.287 \\
3FGLJ0952.8+0711   & 6192  & 0.8	 & Ca II 3934      & 0.574\\
                                   & 6246  & 0.5   & Ca II 3968      & 0.574 \\
                                   & 6770  & 1.0   & G-band 4305  & 0.573\\
3FGLJ1049.7+1548   & 4939  & 0.2   & [OII]  3727       & 0.325 \\
                                   & 5216  & 0.2  	 & Ca II 3924       & 0.326 \\
                                   & 5263  & 0.2   & Ca II 3968       & 0.326 \\
3FGL J1129.0+3758	  & 4336  & 2.0 & ?  &  - \\
                                   & 5182  & 0.7 & ?  &   - \\
                                   & 5268  & 2.2 & ?  &   - \\
                                   & 5718  & 0.6 & ?  &   - \\
                                   & 5749  & 2.1 & ?  &   - \\
                                   & 6189  & 9.1 & Mg II 2800  & $>$~1.211$^a$\\
3FGLJ1222.7+7952    & 5409 & 1.8 & Ca II 3924       & 0.375\\
                                    & 5457 & 2.5 & Ca II 3868       & 0.375\\
                                    & 5919 & 2.1 & G-band 4305   & 0.375\\
%                                    &  7115.7 &	2.31 & Mg I   & 0.375\\
 \hline
\end{tabular}
\tablenotetext{}{
\raggedright
\footnotesize (:) This marker indicates that the redshift is tentative, (*): Asterix indicates that the line is partially contaminated by the telluric band.}
\end{table*}

\setcounter{table}{3}                                   
\begin{table*}
\caption{MEASUREMENTS OF THE SPECTRAL LINES \textit{(continued)}} \label{tab:line}
\centering
\begin{tabular}{lllll}
\hline
Object name          &  $\lambda$    &    EW   &     Line ID    &   z   \\
%                          &  $\textrm{\AA}$             &     $\textrm{\AA}$           &                &               \\
\hline                                
3FGLJ1340.6-0408     &  4810  &  0.4 & Ca II 3924       & 0.223\\
                                    & 4852   &  0.4 & Ca II 3968       & 0.223\\
                                    &  5260  &  0.5 & G-band 4305  & 0.222\\
                                    %&  5951 ?    &	0.78 & Hbeta   & 0.222\\
3FGLJ1511.8-0513     &  4053  &   2.1 &  Mg II 2800       & $>$~0.45$^a$\\
3FGLJ1704.1+1234   &  5411   &   1.7  & [O II] 3727	         & 0.452 \\
                                   &  5712  &    1.0  & Ca II	3924         &  0.452 \\
                                   &  5763  &    1.1  & Ca II	3968         & 0.452  \\
                                   &  6250  &    1.6   & G-band 4305    & 0.452 \\
                                  %   &  *7199 &   1.1   & [O III] XXXX	& 0.452 \\
                                   &  7269  & 3.0  &  [O III]  5007         &  0.452  \\
                                   &  7515  & 0.9  &  Mg I 5175             & 0.452 \\
3FGLJ2115.2+1215   &  4191 &  5.0	& Mg II 2800  & $>$~0.497$^a$ \\
                                  &  4270 &  0.9	& Mg II 2800   & $>$~0.525$^a$\\
                                  &  4571 &  0.9	& Mg II 2800   & $>$~0.633$^a$\\
3FGLJ2346.7+0705  &  4606  & 0.3  &  Ca II  3924      & 0.171   \\
                                  &  4647  & 0.5  &  Ca II  3968      & 0.171   \\
                                  &  5040  & 0.5  &  G-band	 4305   & 0.171  \\
                                  &  6060  & 1.1  &  Mg I 5175 	   & 0.171   \\
                                  &  6170  & 0.6	&  Ca+Fe 5269    &  0.171  \\
\hline
\end{tabular}
\tablenotetext{}{
\raggedright
\footnotesize \texttt{Col.1}: Name of the target; \texttt{Col.2}: Barycenter of the detected line; \texttt{Col.3}: Measured equivalent width; \texttt{Col.4}: Line identification (\textit{a}= intervening absorption assuming Mg~II (2800$\textrm{\AA}$) identification); \texttt{Col.5}: Spectroscopic redshift.\\}
\end{table*}

\newpage
\begin{table*}
\caption{SUMMARY OF THE PROPOSED BLAZAR CLASSIFICATION BY THE BROAD-BAND SED TOOL (PFS).}\label{tab:table5}
\centering
\begin{tabular}{lllllll}
\hline
3FGL name & Counterpart name & 3FGL SED    & 3FHL & AGN class & Redshift  & Classification and Redshift \\
                    &                               & classification &           & proposed   & from PFS &   from spectroscopy            \\
\hline
3FGLJ0049.0+4224   & SDSSJ004859+422351  & UGS                  &  y & HSP  &  0.4 - 0.6  & BLL, 0.302  \\
3FGLJ0102.1+0943   & SDSSJ010217+094409  & UGS                  &  n & HSP  &  0.4 - 0.5  & BLL, 0.42        \\
3FGLJ0239.0+2555   & SDSSJ023853+255407  & UGS                  &  n & HSP  &  0.3 - 0.5  & BLL, 0.584       \\
3FGLJ0305.2-1607    & PKS0302-16                   & BCU-II / HSP     &  y &	 HSP &  0.5 - 0.6  & BLL, 0.312  \\
3FGLJ0338.5+1303   & RXJ0338.4+1302            & BCU-II / HSP     &  y &	 HSP &  0.3 - 0.6  & BLL, $>$0.382          \\
3FGLJ0409.8-0358    & NVSSJ040946-040003	   & BLL / ISP          &   y & ISP	 &  0.2 - 0.7  & BLL,  $>$0.7        \\
3FGLJ0644.6+6035   & USNOB1506-0162421    & UGS                 &   y  & HSP & 0.2 - 0.5   & BLL, $>$0.581          \\
3FGLJ0937.9-1435    & USNOB0754-0223141	   & UGS                 &   y  & HSP & 0.3 - 0.4   & BLL, 0.287   \\
3FGLJ0952.8+0711   &  SDSSJ095249+071329  & UGS                 &   n & HSP & 0.4 -  0.5   & BLL, 0.574           \\
3FGLJ1049.7+1548   & SDSSJ104939+154837   & UGS                 &   n & HSP & 0.3 - 04     & BLL, 0.326  \\
3FGL J1129.0+3758	  & SDSSJ112903+375656   &  UGS               &   n & LSP & 1.4 - 1.8     & BLL , $>$1.211      \\
3FGLJ1222.7+7952   & USNOB1698-0045483	   & UGS                  &   y & HSP & 0.3 - 0.5    & BLL, 0.375    \\
3FGLJ1340.6-0408    & NVSSJ134042-041006	   & BCU-II / HSP     &   n & HSP & 0.3 - 0.4   & BLL, 0.223     \\
3FGLJ1411.4-0724    & USNOB0826-0334743      & UGS                &    y  &  HSP - ISP & 0.3 - 0.6  & BLL , $>$0.72\\
3FGLJ1511.8-0513    & NVSSJ151148-051345	    & BCU-III            & y  &	HSP & 0.1 - 0.2    & BLL, $>$0.45  \\
3FGLJ1704.4-0528   &  USNOB0845-0308445	    &  UGS               &  y &   HSP & 0.3 - 0.4    & BLL, $>$0.7\\
3FGLJ1704.1+1234 &  SDSSJ170409+123421    &   UGS               &  y  &   HSP & 0.2 - 0.4    & BLL, 0.452 \\
3FGLJ2115.2+1215   & SDSSJ211522+121802	    &  UGS                &  y  &   HSP & 0.4 - 0.6   & BLL, $>$0.497 \\
3FGLJ2246.2+1547   & NVSSJ224604+154437    &  BCU-II / ISP    & y  &	 ISP   & 0.3 - 0.8   &  BLL, $>$0.7\\
3FGLJ2346.7+0705   & TXS2344+068      	    &  BCU-II / ISP     &  y & ISP - HSP& $\sim$~0.2   & BLL, 0.171\\
\hline
\end{tabular}
\tablenotetext{}{
\raggedright
\texttt{Col.1}: the 3FGL name; \texttt{Col.2}: name of the counterpart; \texttt{Col.3}: \textit{Fermi} SED classification as reported in the 3FGL and 3LAC catalog (UGS = unassociated gamma-ray source, BCU = active galaxy of uncertain type, BLL = BL Lac object type, HSP = the high-synchrotron-peaked sources peak ($\nu_{(syn-peak)}>10^{15}$ Hz), ISP = the intermediate-synchrotron-peaked sources ($\nu_{(syn-peak)}$ between 10$^{14}$ and 10$^{15}$), and LSP = the low-synchrotron-peaked sources ( $\nu_{(syn-peak)} < 10^{14}$ Hz) ); \texttt{Col.4}: Presence of the target in the third catalog of hard \textit{Fermi}-LAT sources \citep[3FHL,][]{3fhl}; \texttt{Col.5 and Col.6}:  classification type and the redshift range (within 2 sigma), found by running the blazar recognition tool \citep[PFS,][]{paiano2017ufo}; \texttt{Col.7}:  classification and spectroscopic redshift found in this work. }
\end{table*}

\newpage

\setcounter{figure}{1}
\begin{figure*}%[htbp]
\includegraphics[width=0.4\textwidth, angle=-90]{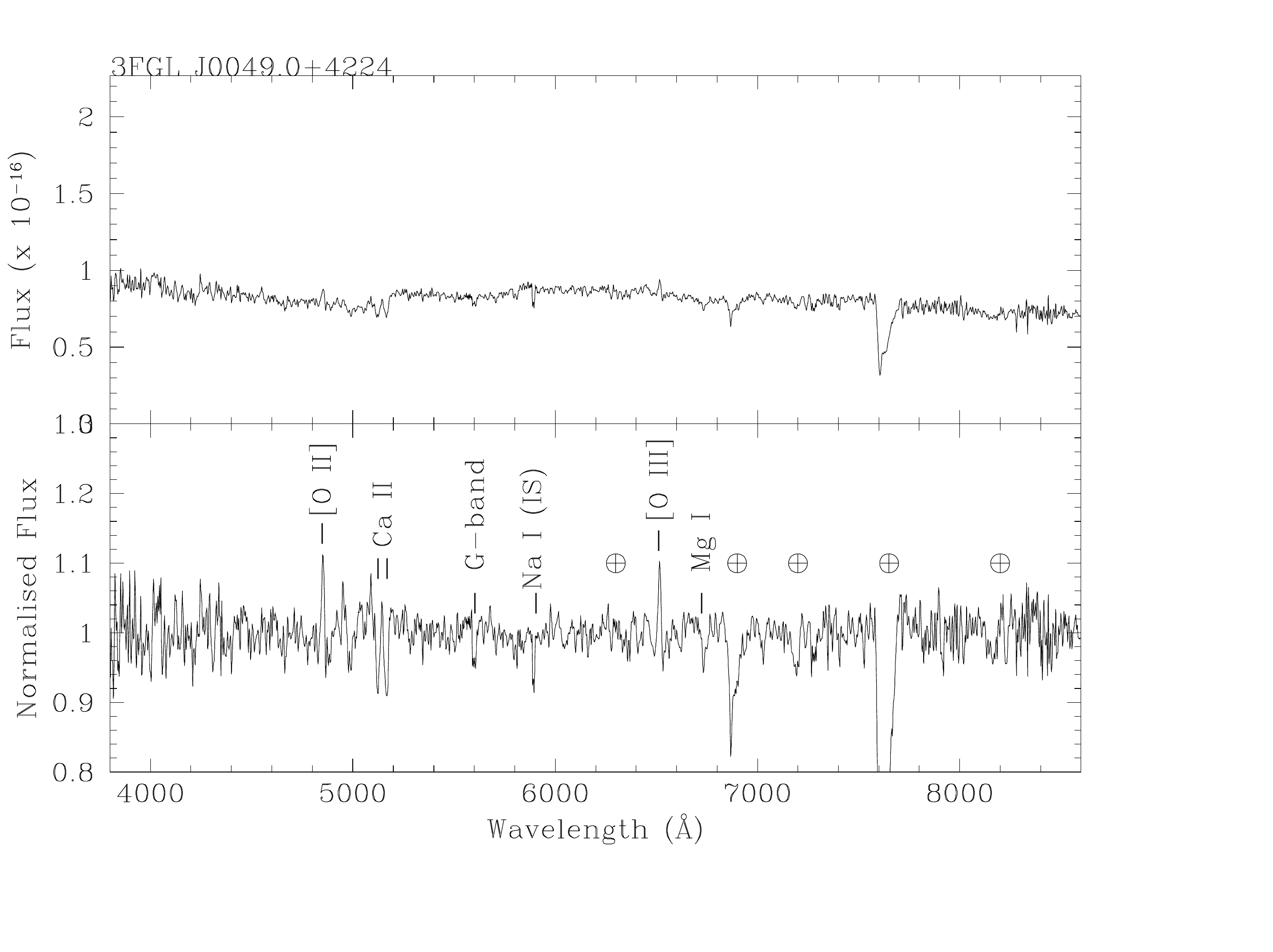}
\includegraphics[width=0.4\textwidth, angle=-90]{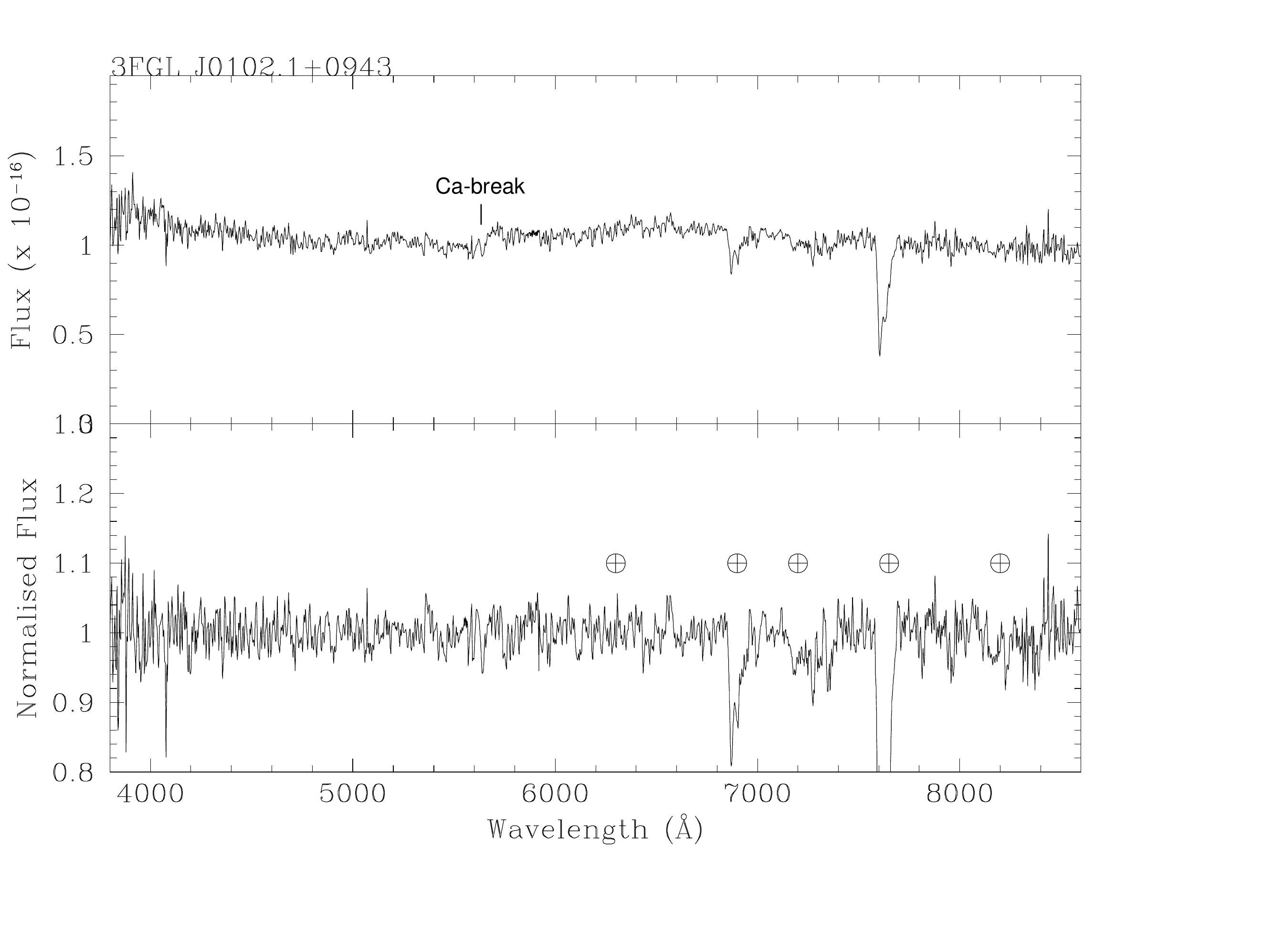}
\includegraphics[width=0.4\textwidth, angle=-90]{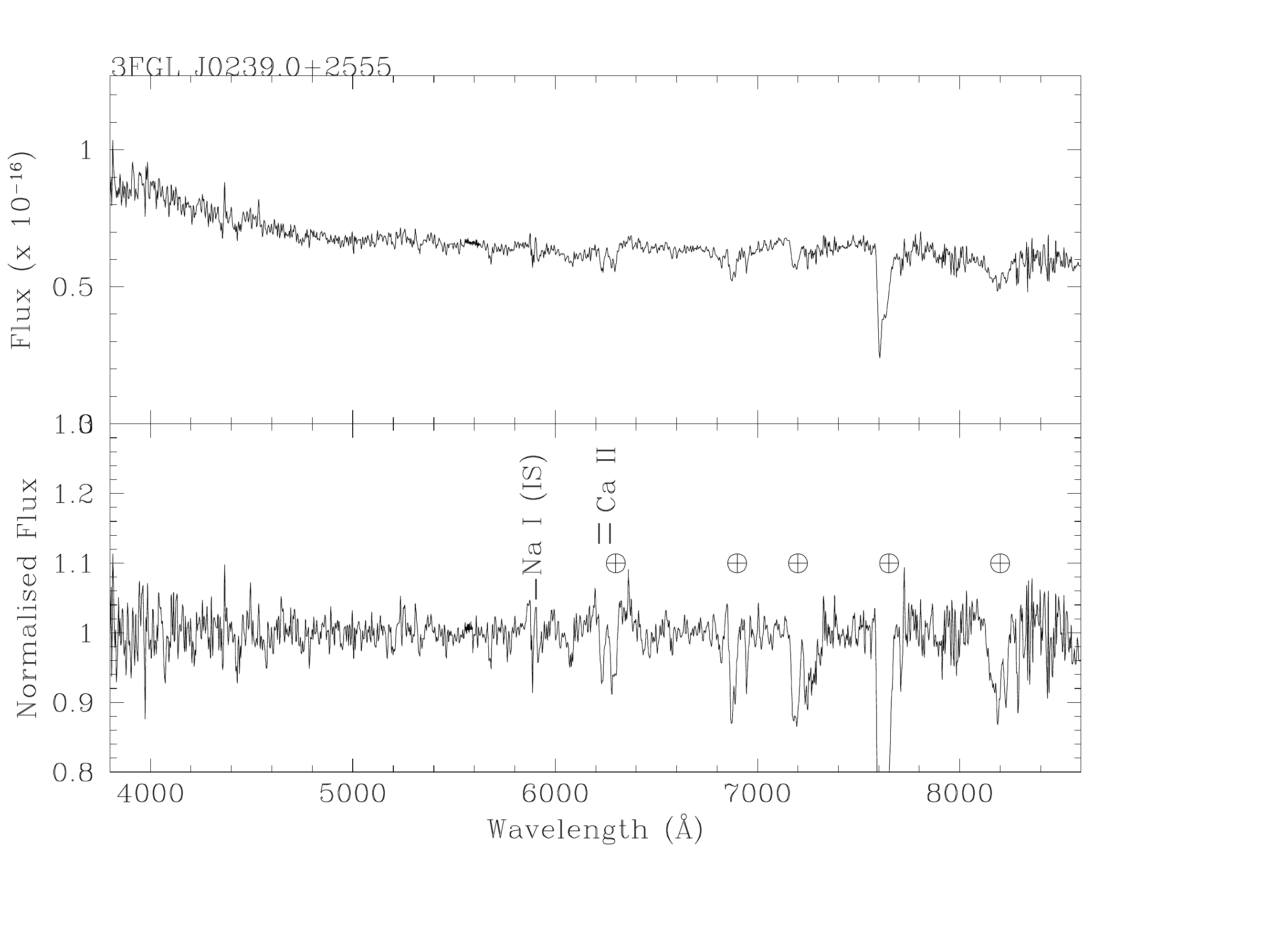}
\includegraphics[width=0.4\textwidth, angle=-90]{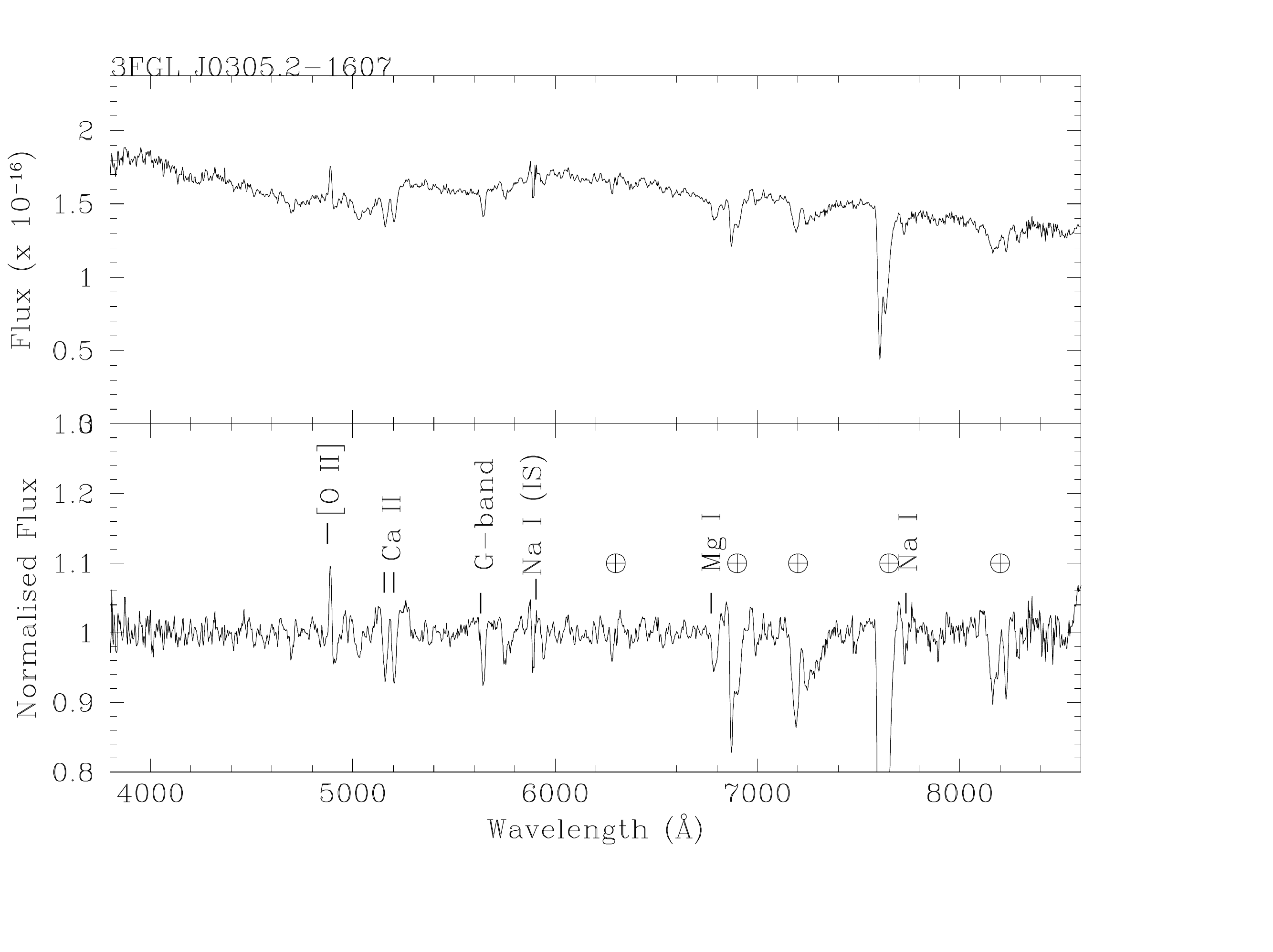}
\includegraphics[width=0.4\textwidth, angle=-90]{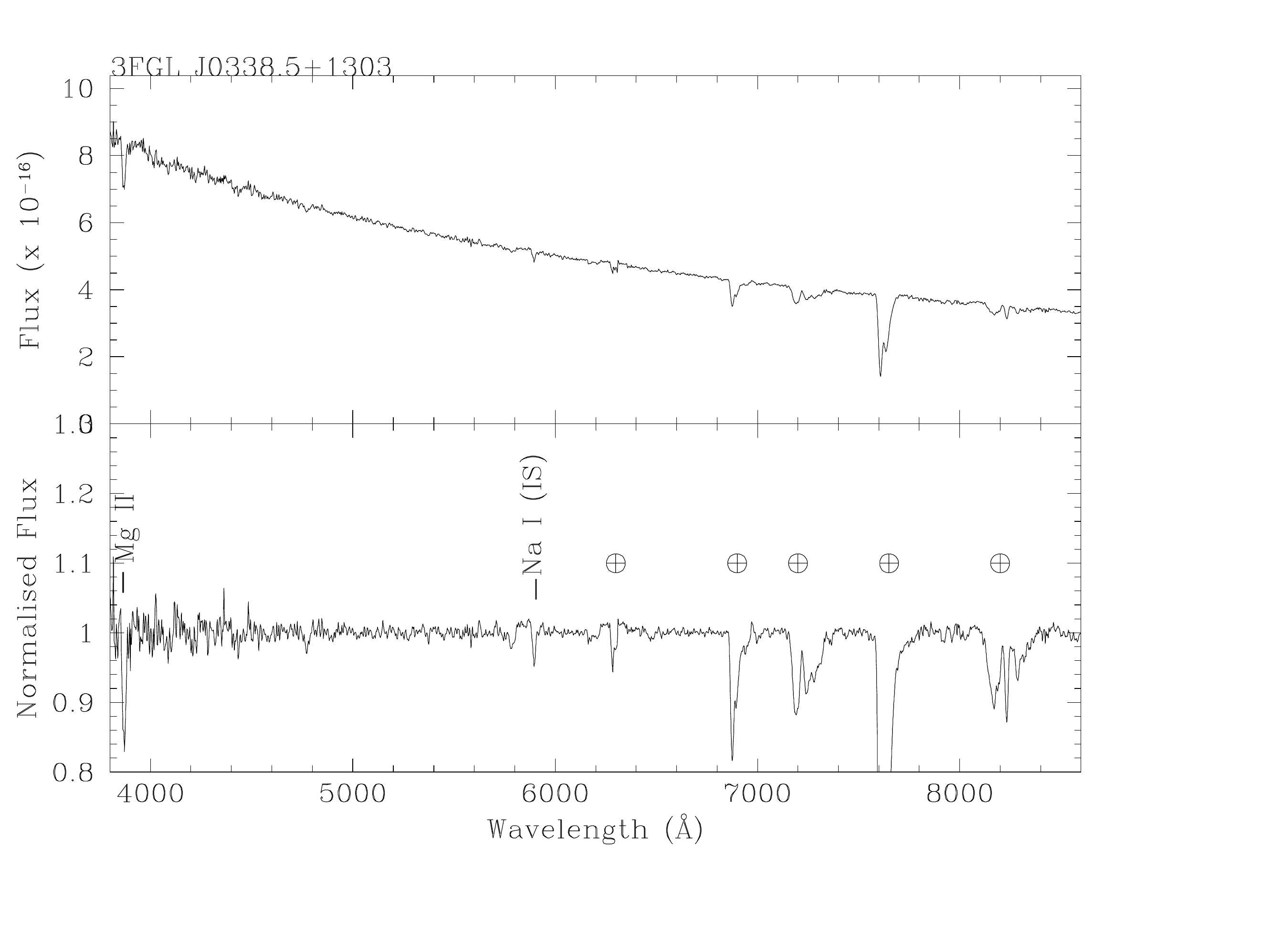}
\includegraphics[width=0.4\textwidth, angle=-90]{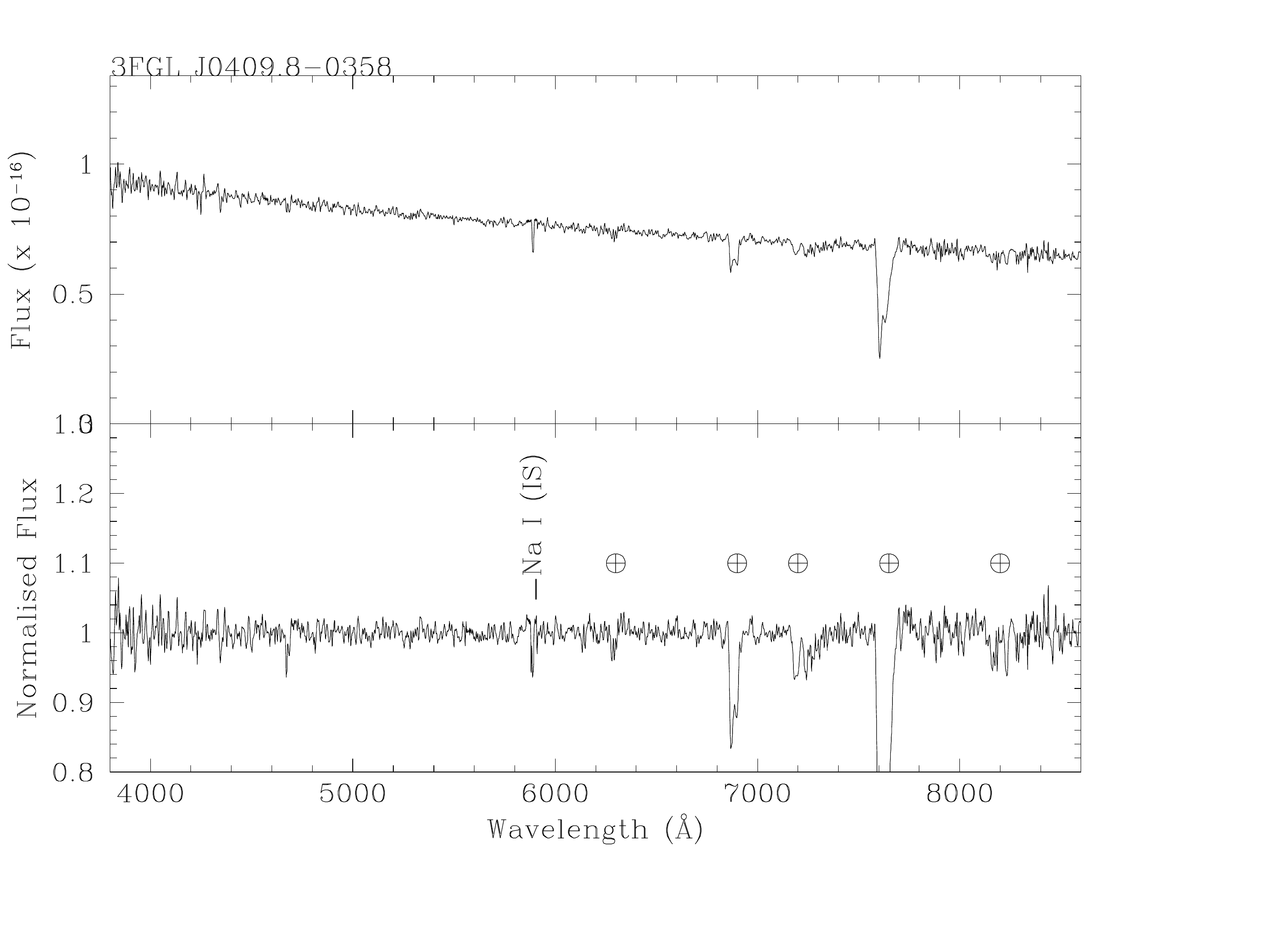}
\caption{Spectra of the UGSs obtained at GTC. \textit{Top panel}: Flux calibrated and dereddered spectra. \textit{Bottom panel}: Normalized spectra. The main telluric bands are indicated by $\oplus$, the absorption features from interstellar medium of our galaxies are labelled as IS (Inter-Stellar)}
\label{fig:spectra}
\end{figure*}%[htbp]

\setcounter{figure}{1}
\begin{figure*}%[htbp]
\includegraphics[width=0.4\textwidth, angle=-90]{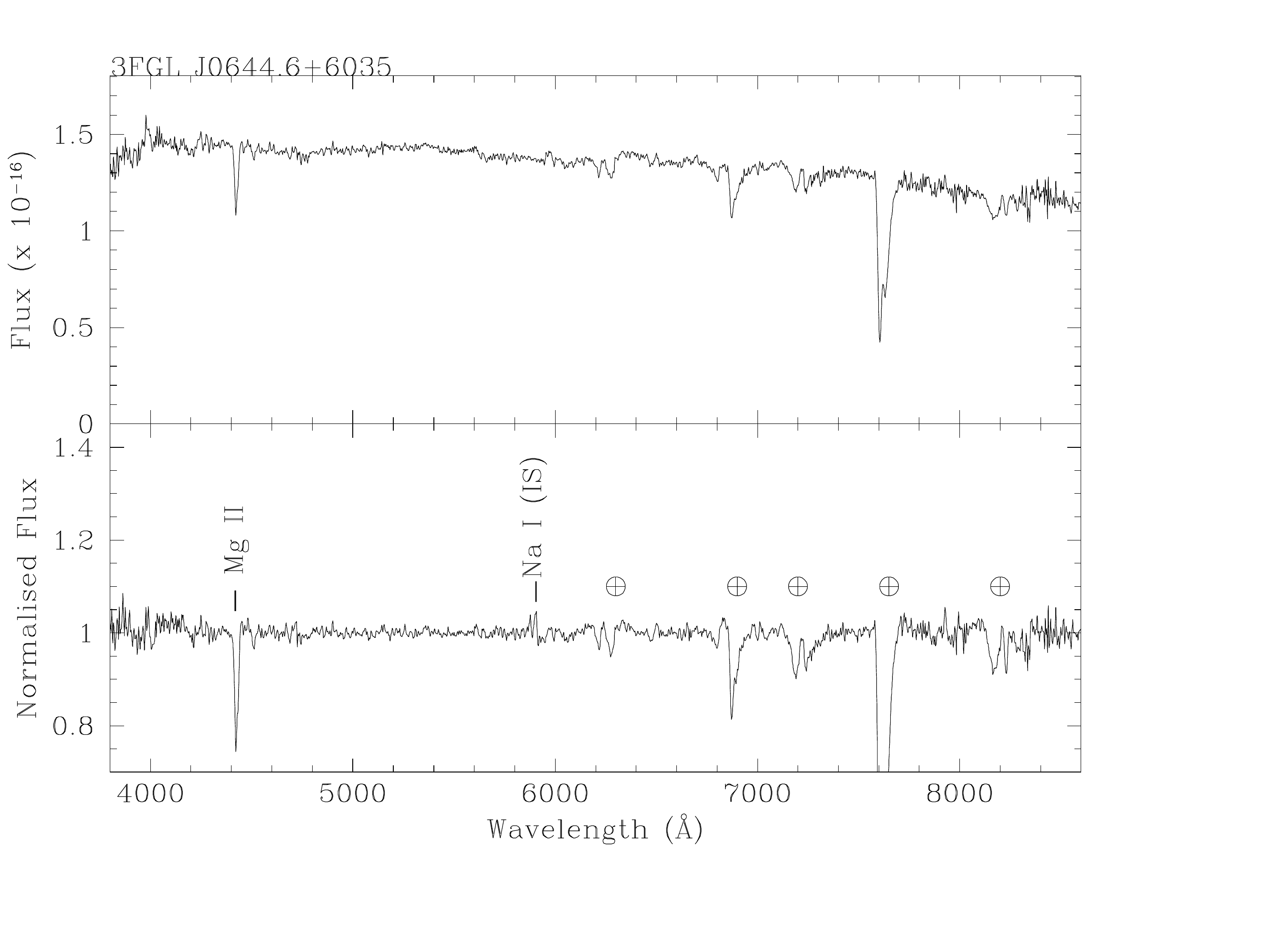}
\includegraphics[width=0.4\textwidth, angle=-90]{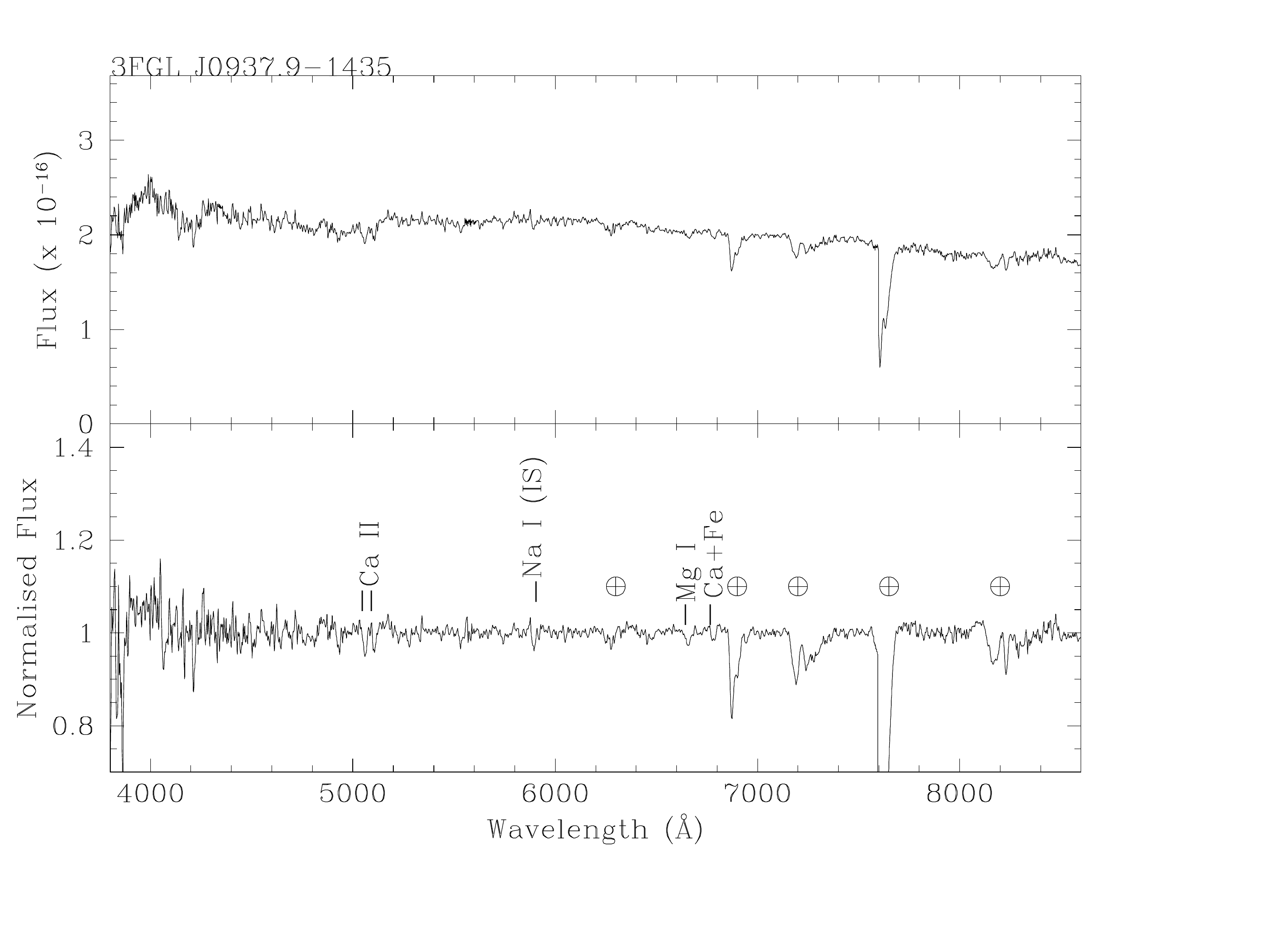}
\includegraphics[width=0.4\textwidth, angle=-90]{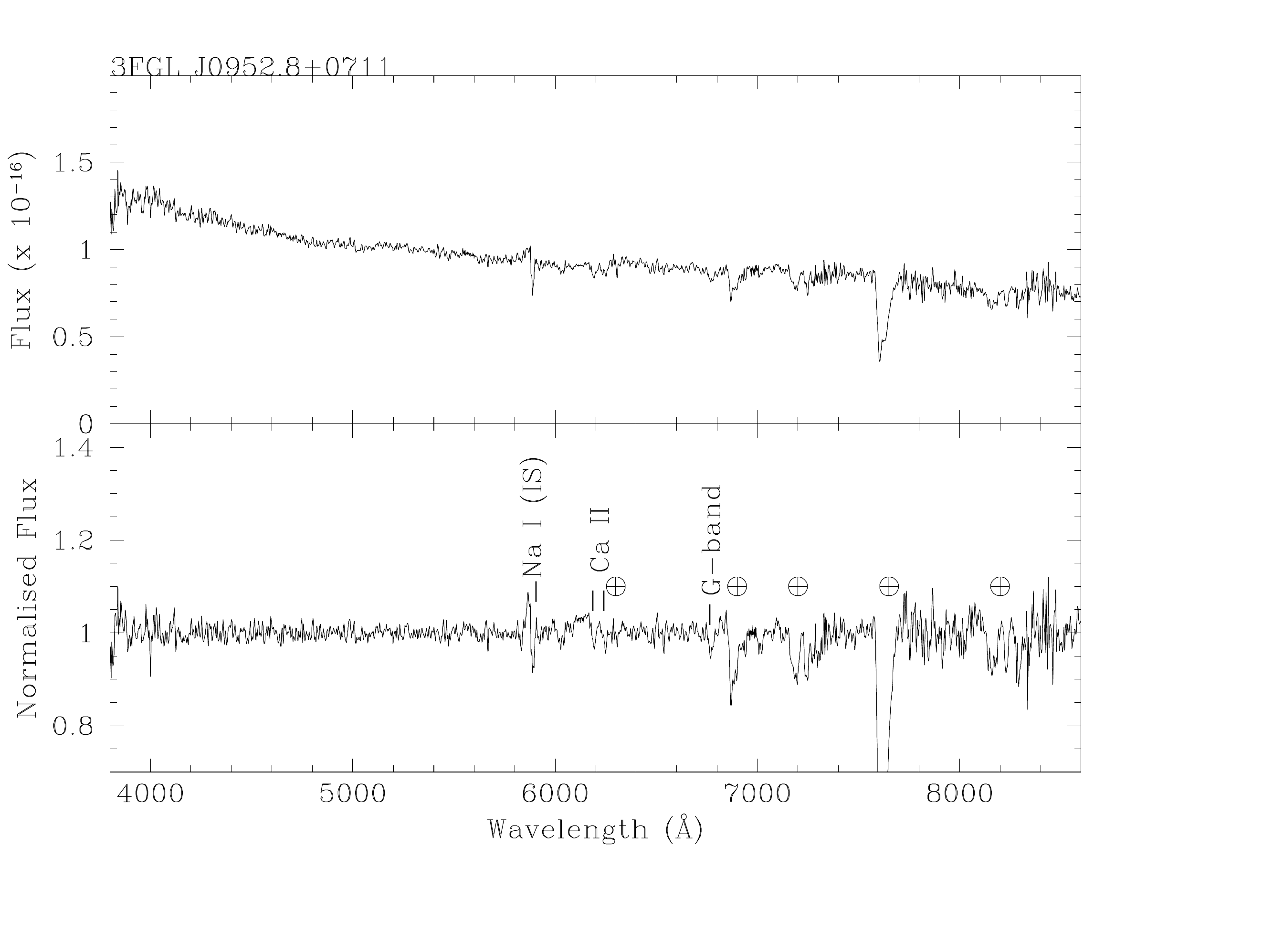}
\includegraphics[width=0.4\textwidth, angle=-90]{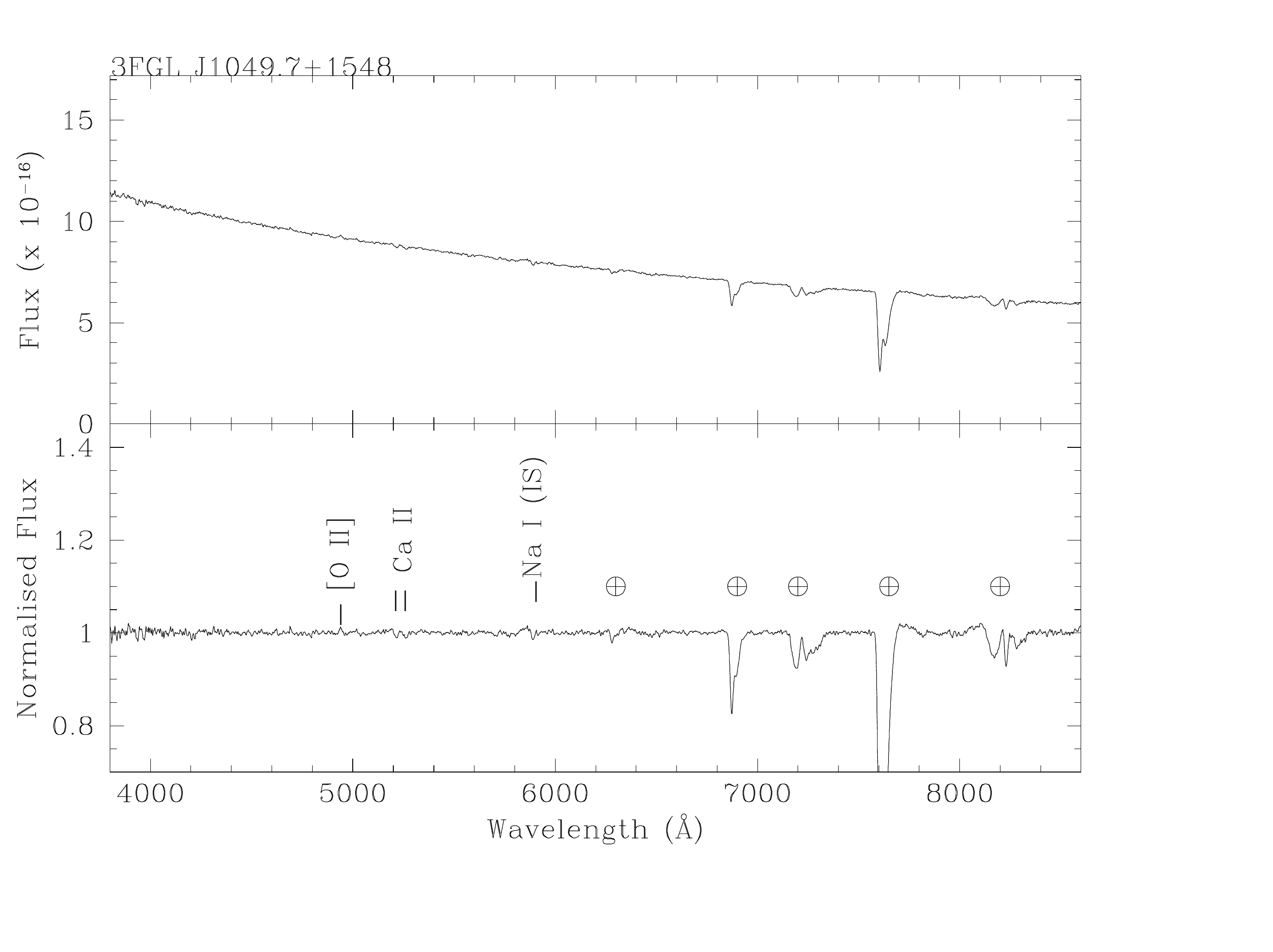}
\includegraphics[width=0.4\textwidth, angle=-90]{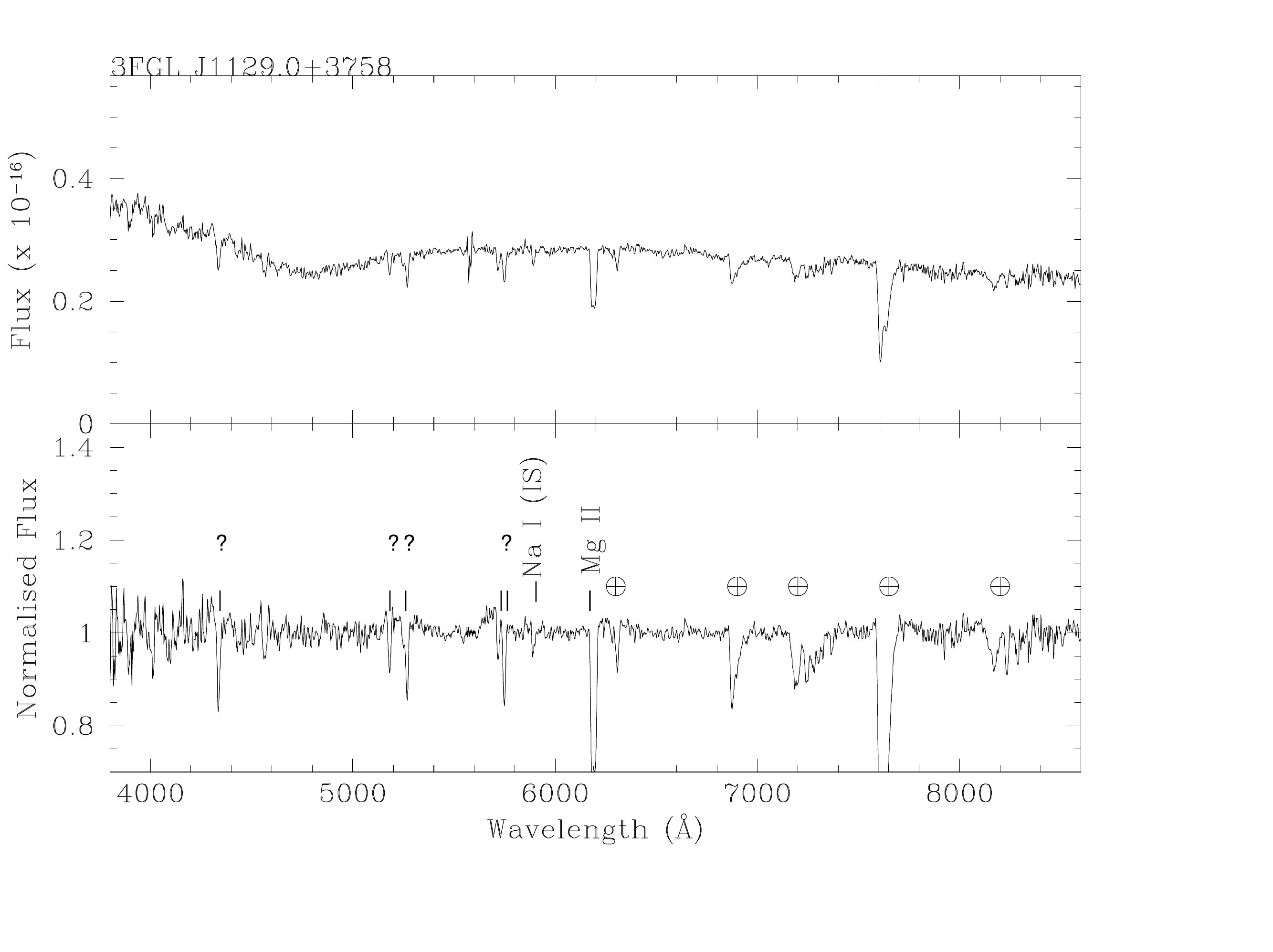}
\includegraphics[width=0.4\textwidth, angle=-90]{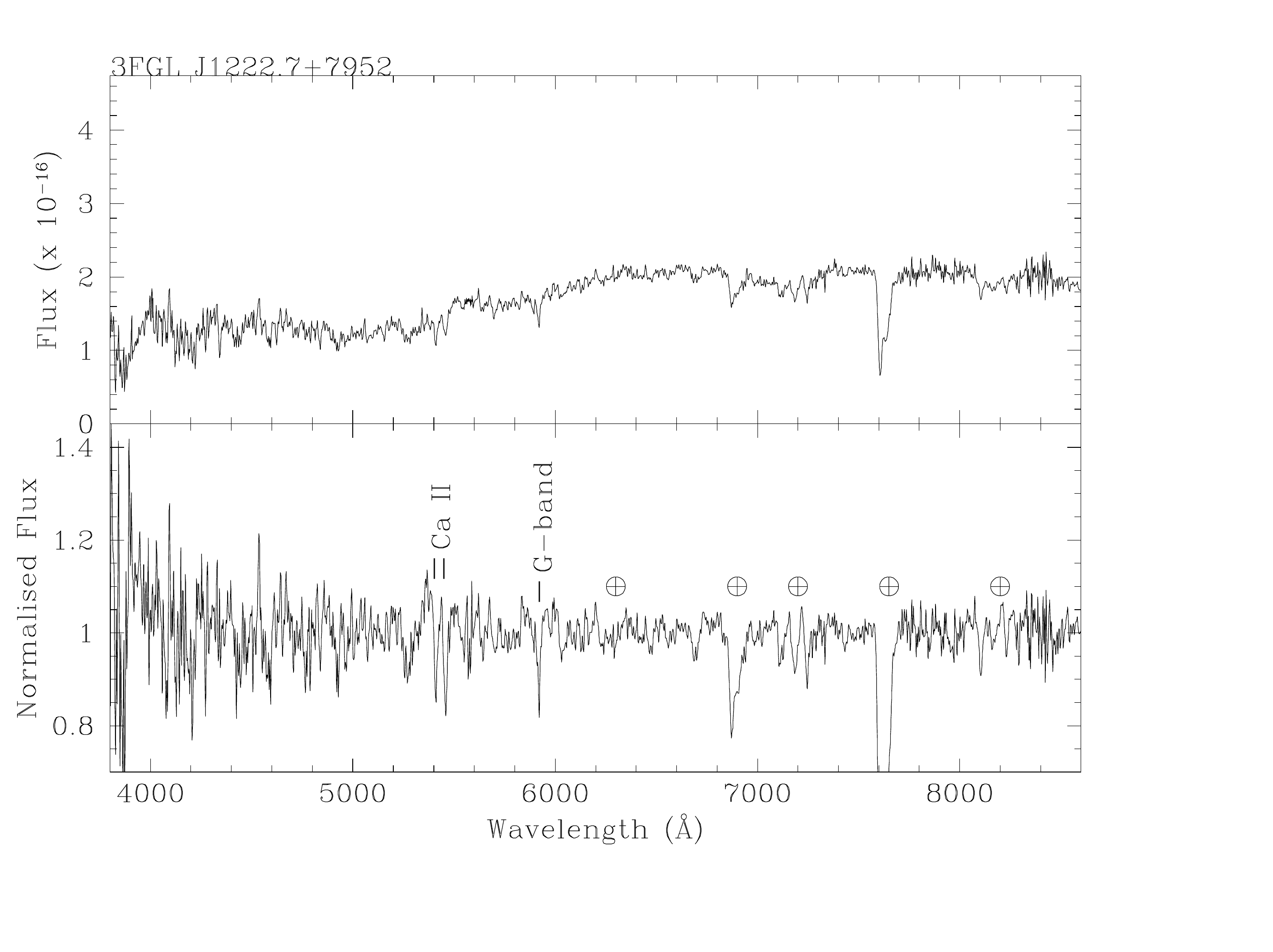}
\caption{Continued from Fig. \ref{fig:spectra}.}
\end{figure*}%[htbp]

\setcounter{figure}{1}
\begin{figure*}%[htbp]
\includegraphics[width=0.4\textwidth, angle=-90]{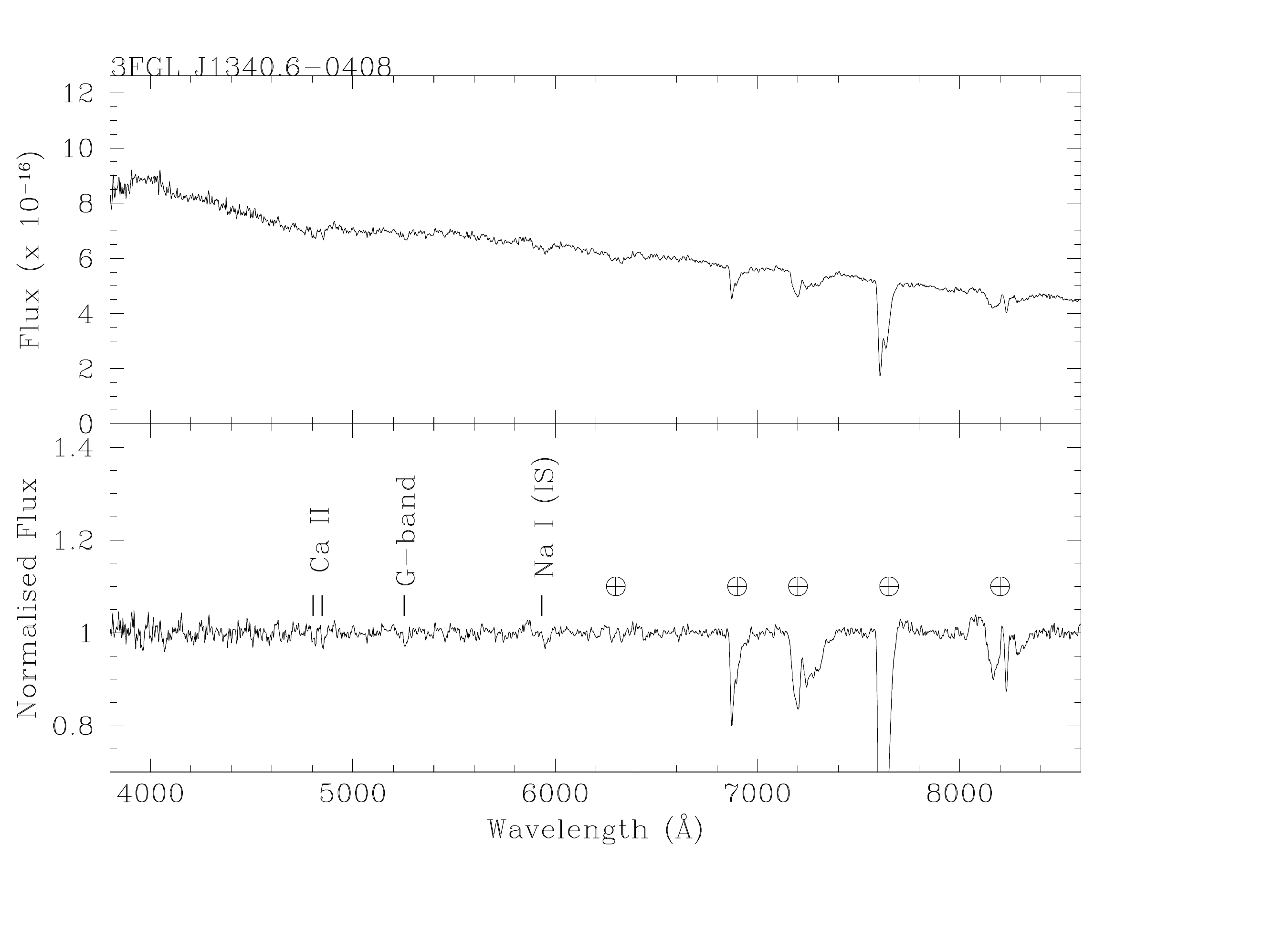}
\includegraphics[width=0.4\textwidth, angle=-90]{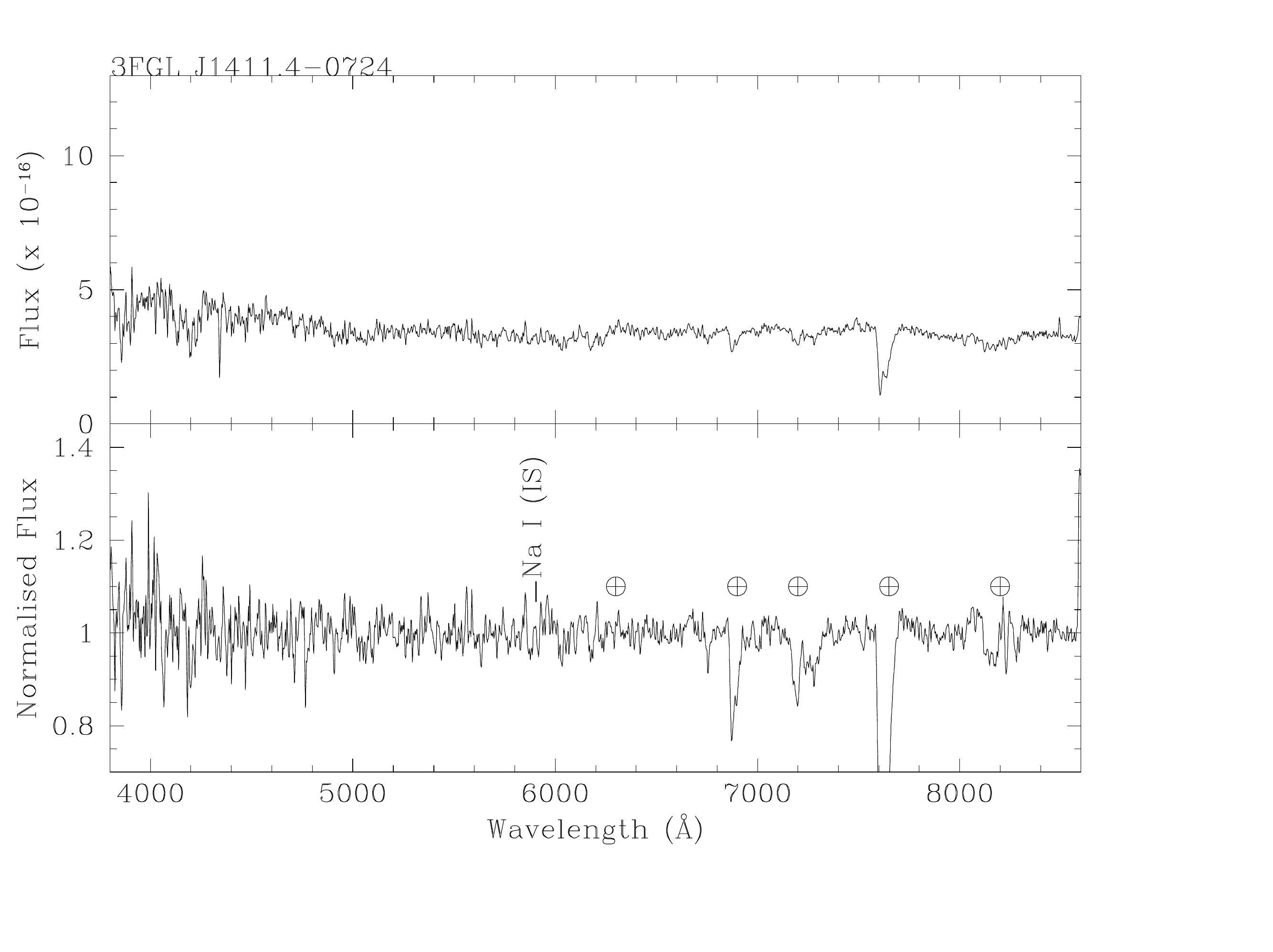}
\includegraphics[width=0.4\textwidth, angle=-90]{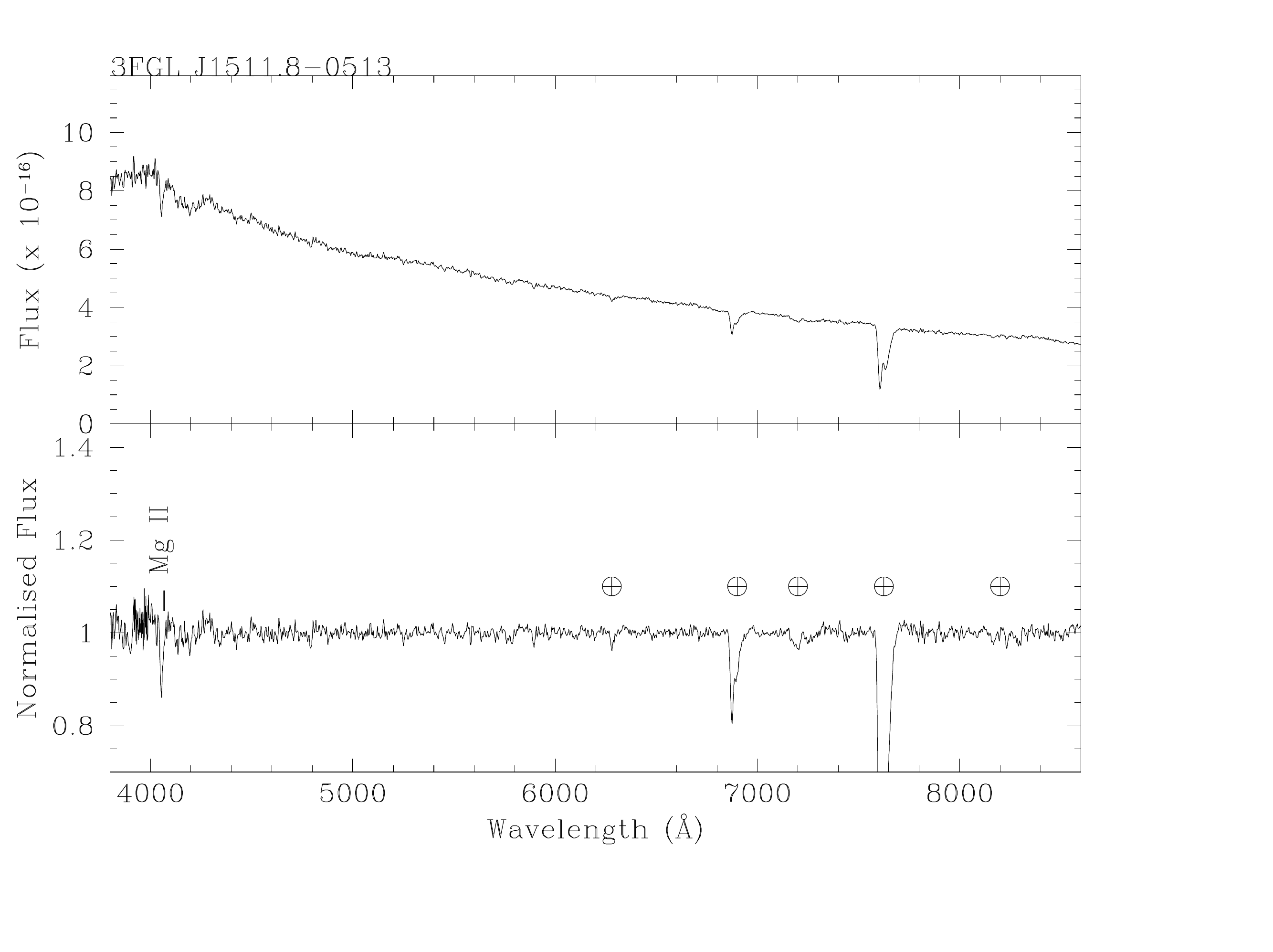}
\includegraphics[width=0.4\textwidth, angle=-90]{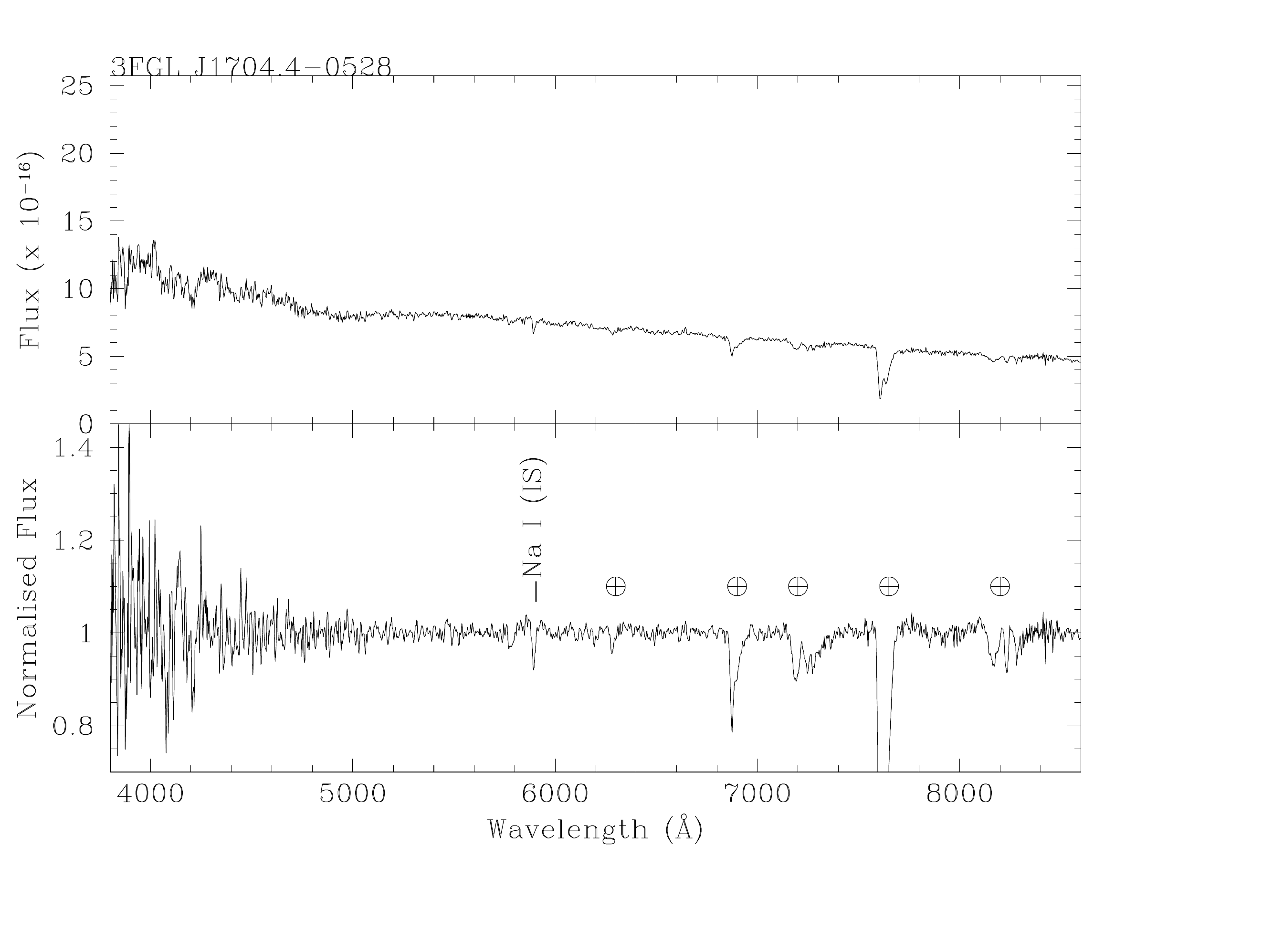}
\includegraphics[width=0.4\textwidth, angle=-90]{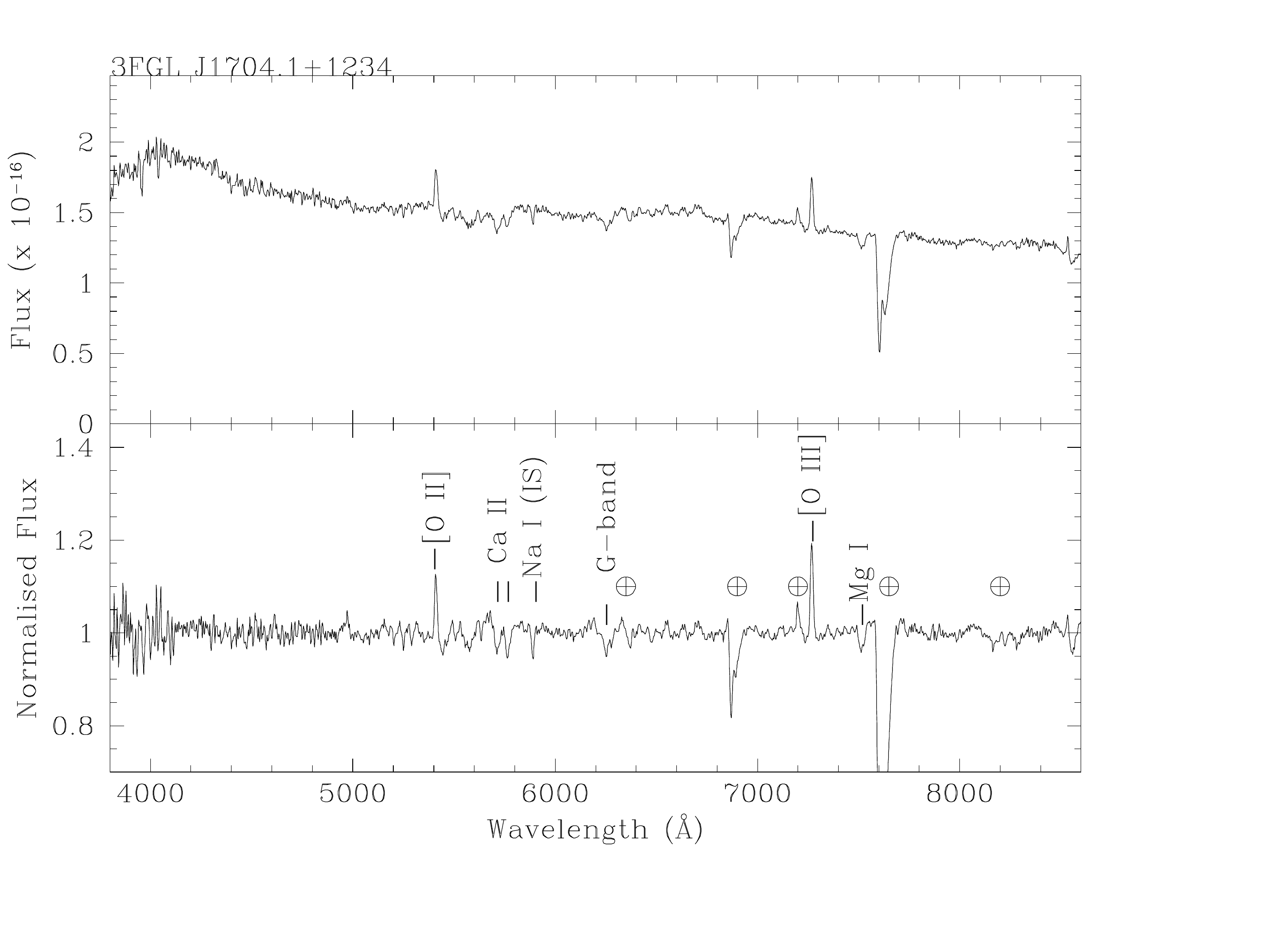}
\includegraphics[width=0.4\textwidth, angle=-90]{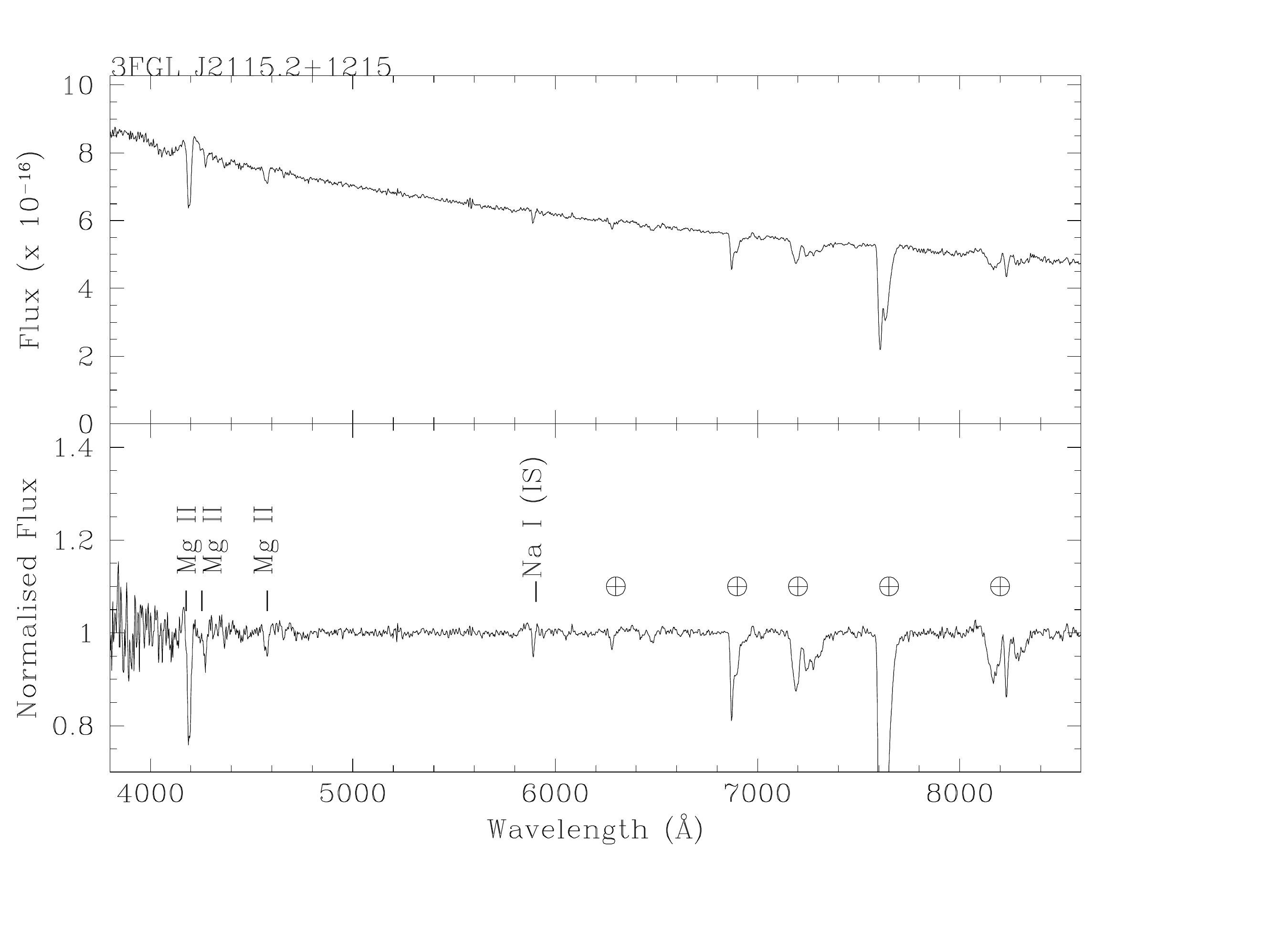}
\caption{Continued from Fig. \ref{fig:spectra}.}
\end{figure*}%[htbp]

\setcounter{figure}{1}
\begin{figure*}%[htbp]
\includegraphics[width=0.4\textwidth, angle=-90]{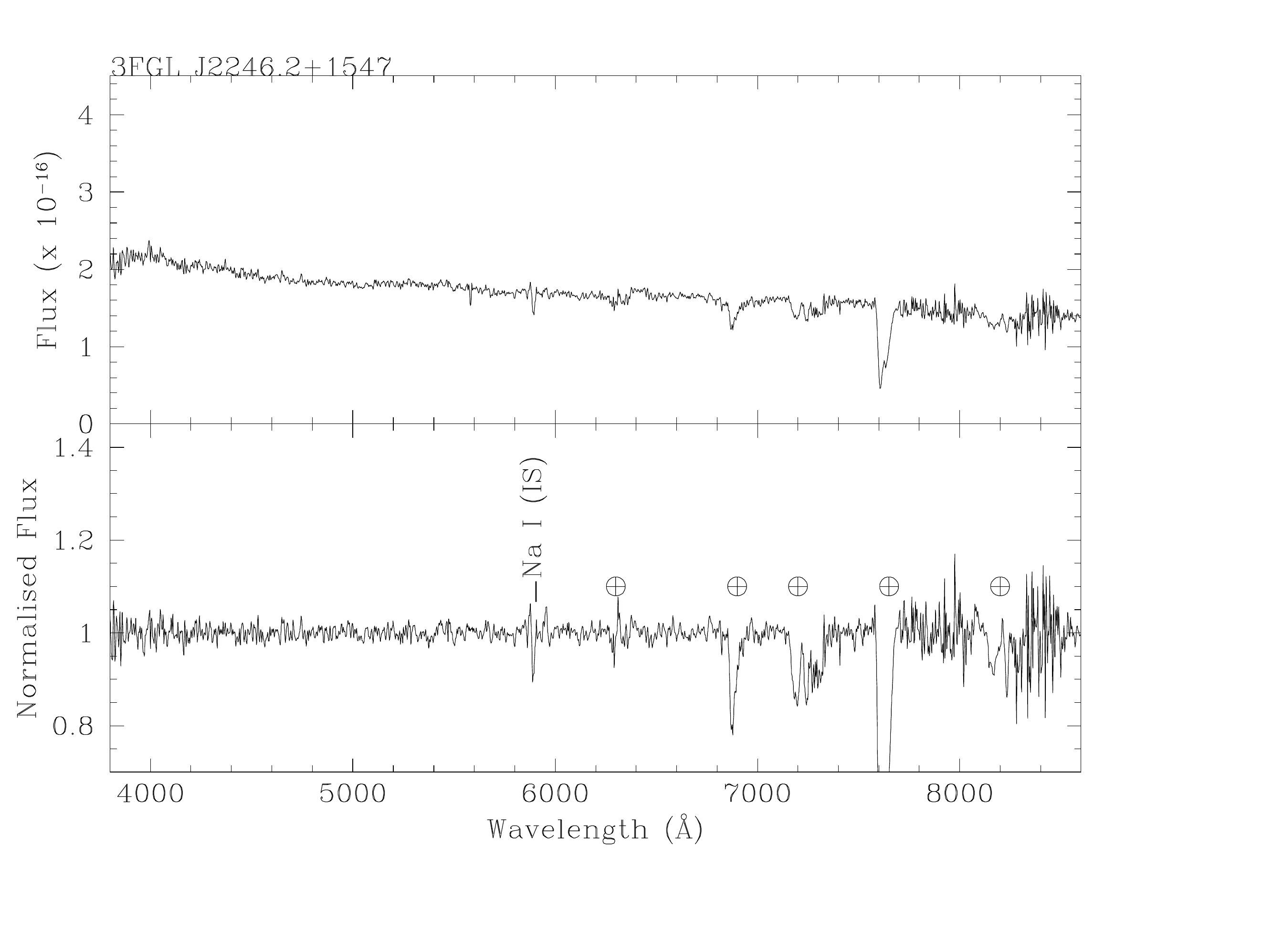}
\includegraphics[width=0.4\textwidth, angle=-90]{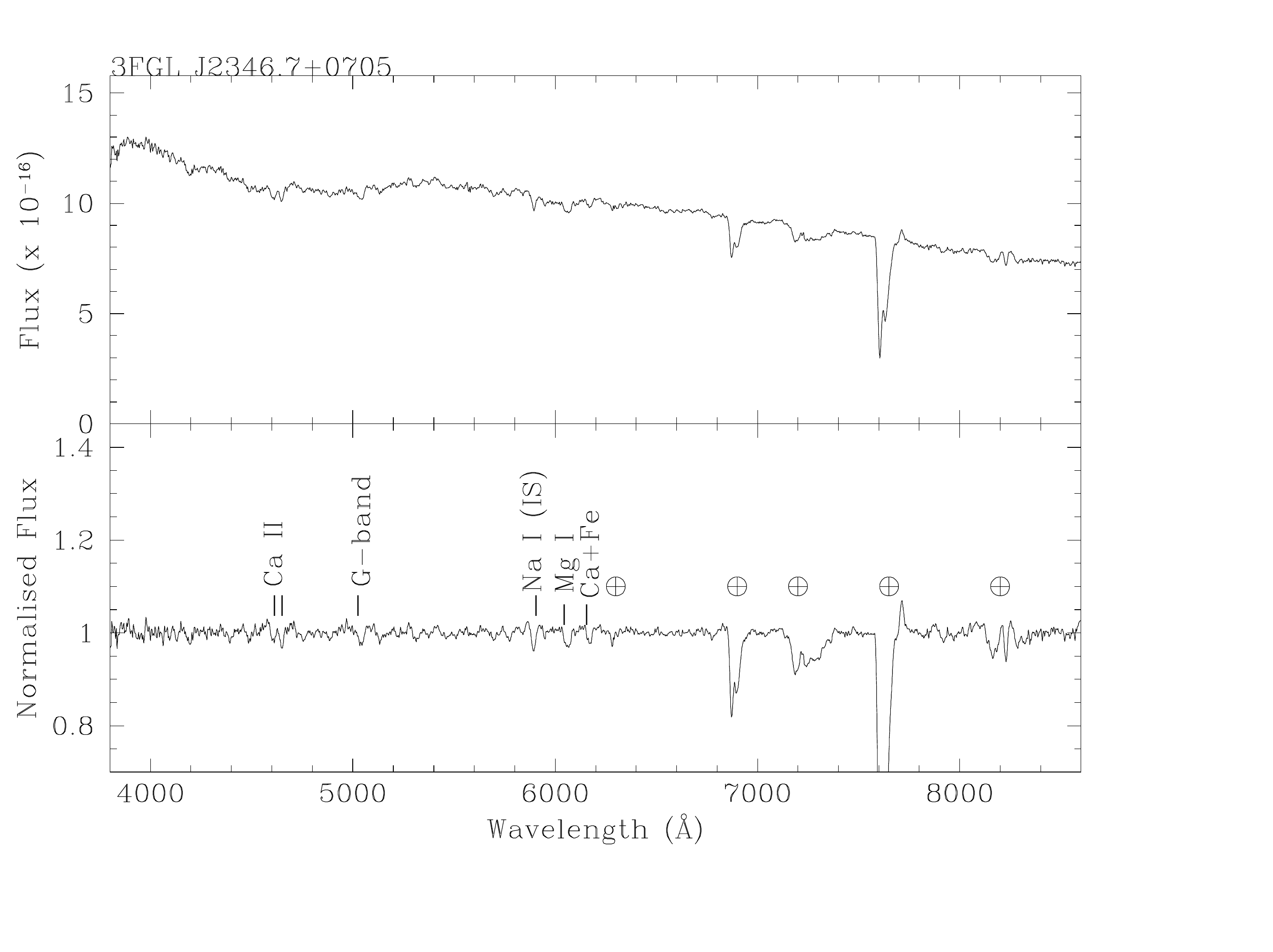}
\caption{Continued from Fig. \ref{fig:spectra}.}
\end{figure*}%[htbp]

%Close-up on spectral features
%\clearpage
%\newpage

\setcounter{figure}{2}
\begin{figure*}%[htbp]
 \includegraphics[width=0.4\textwidth, angle=-90]{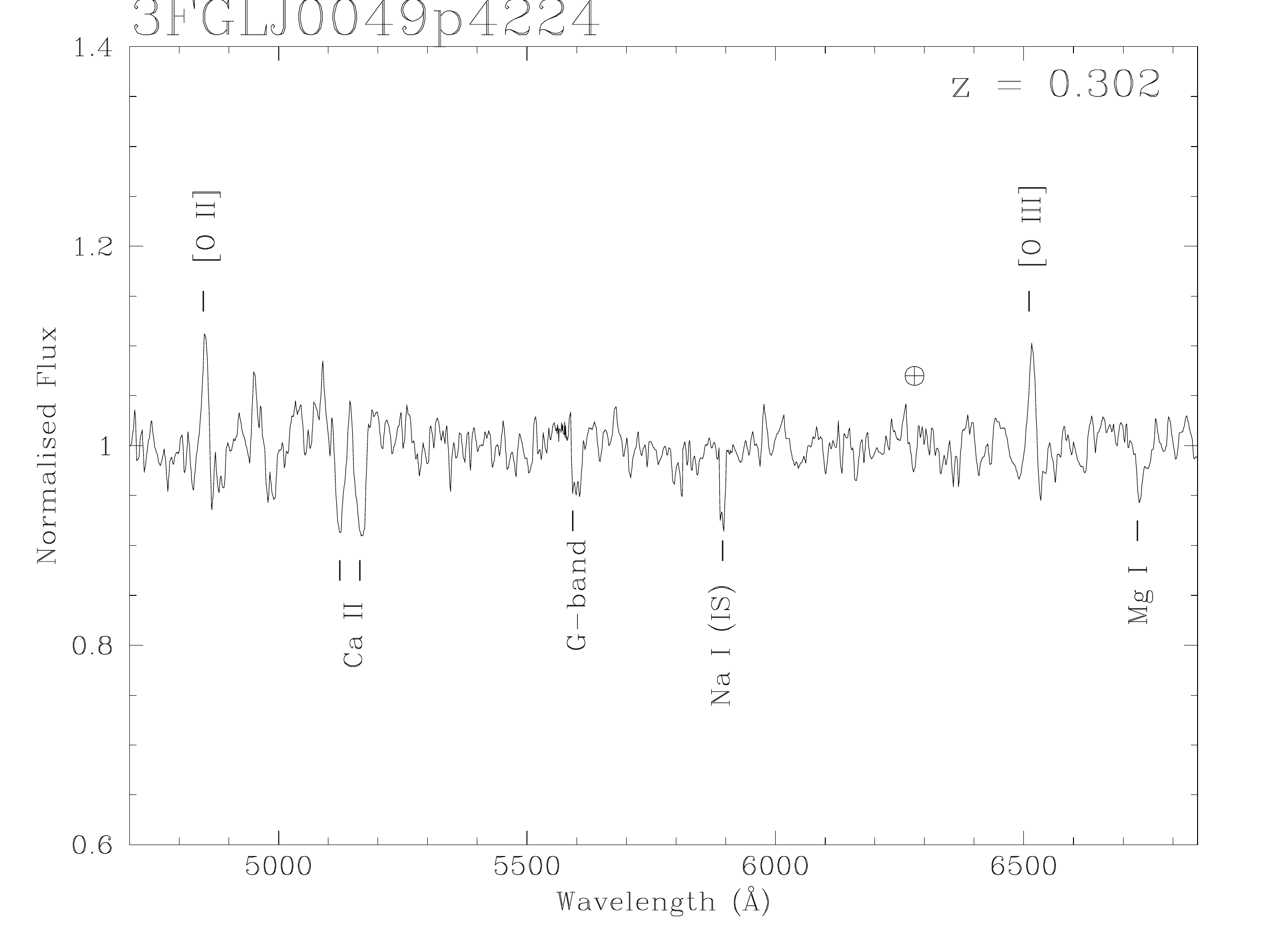}
 \includegraphics[width=0.4\textwidth, angle=-90]{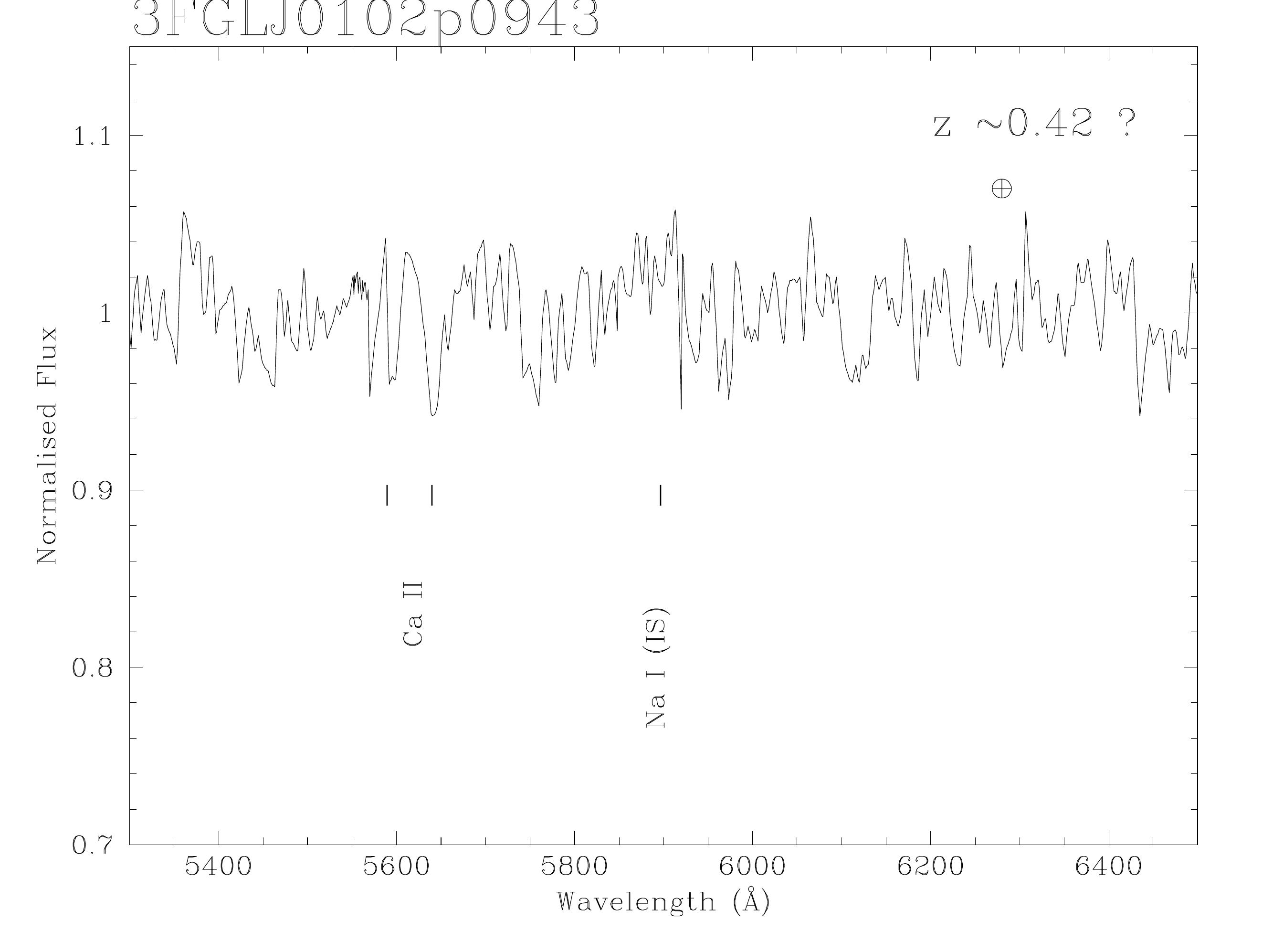}
 \includegraphics[width=0.4\textwidth, angle=-90]{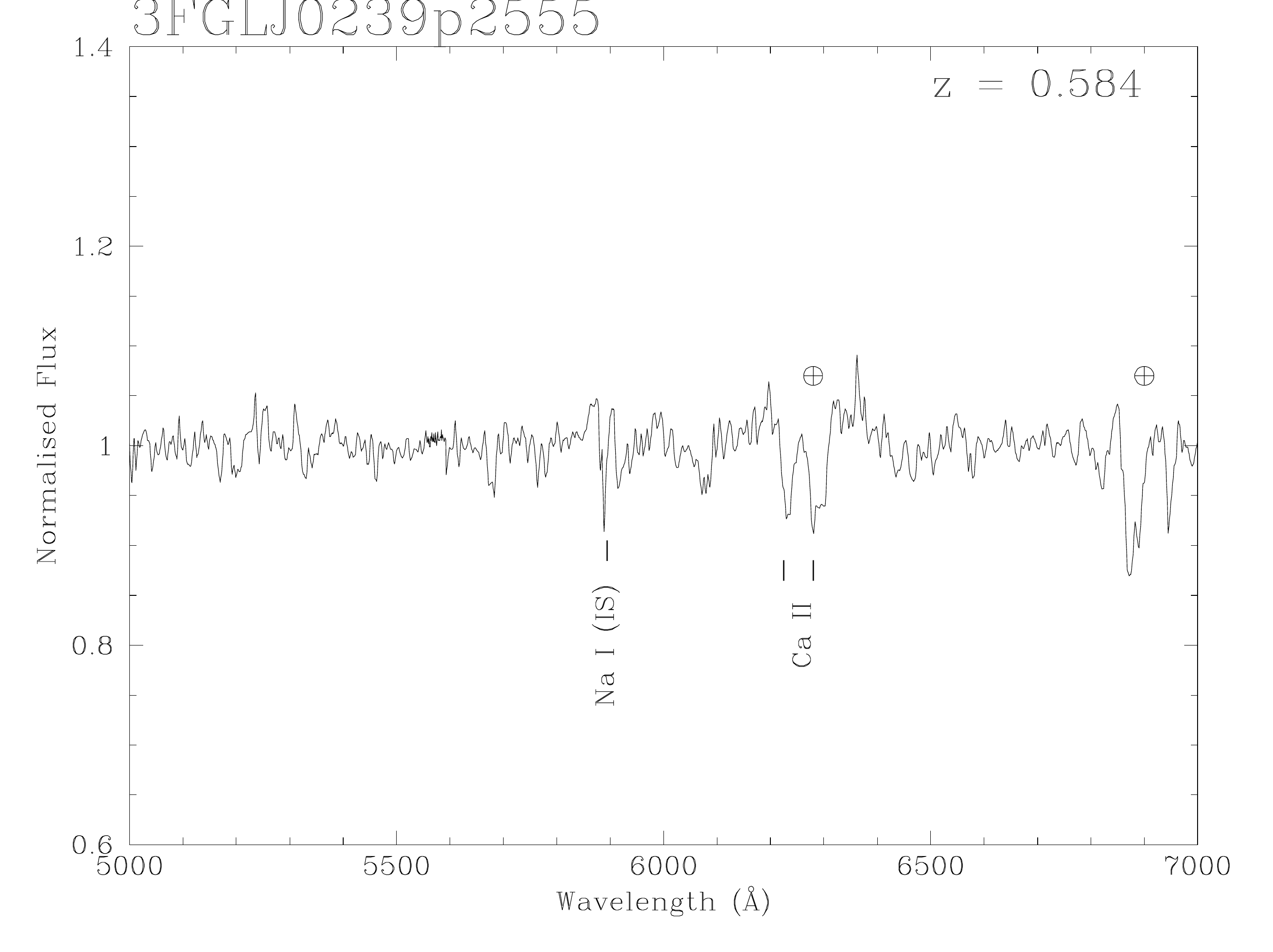}
\includegraphics[width=0.4\textwidth, angle=-90]{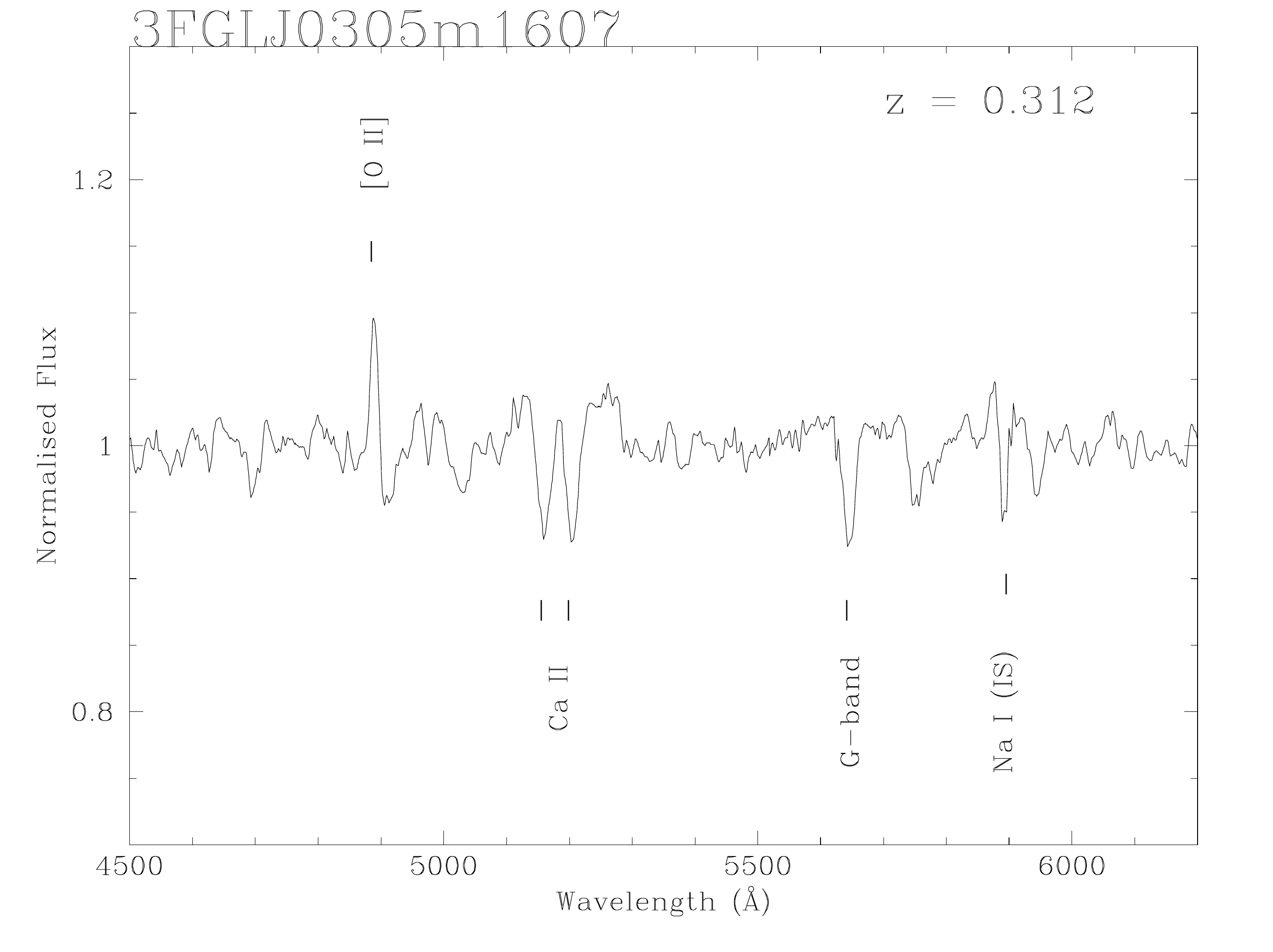}
\includegraphics[width=0.4\textwidth, angle=-90]{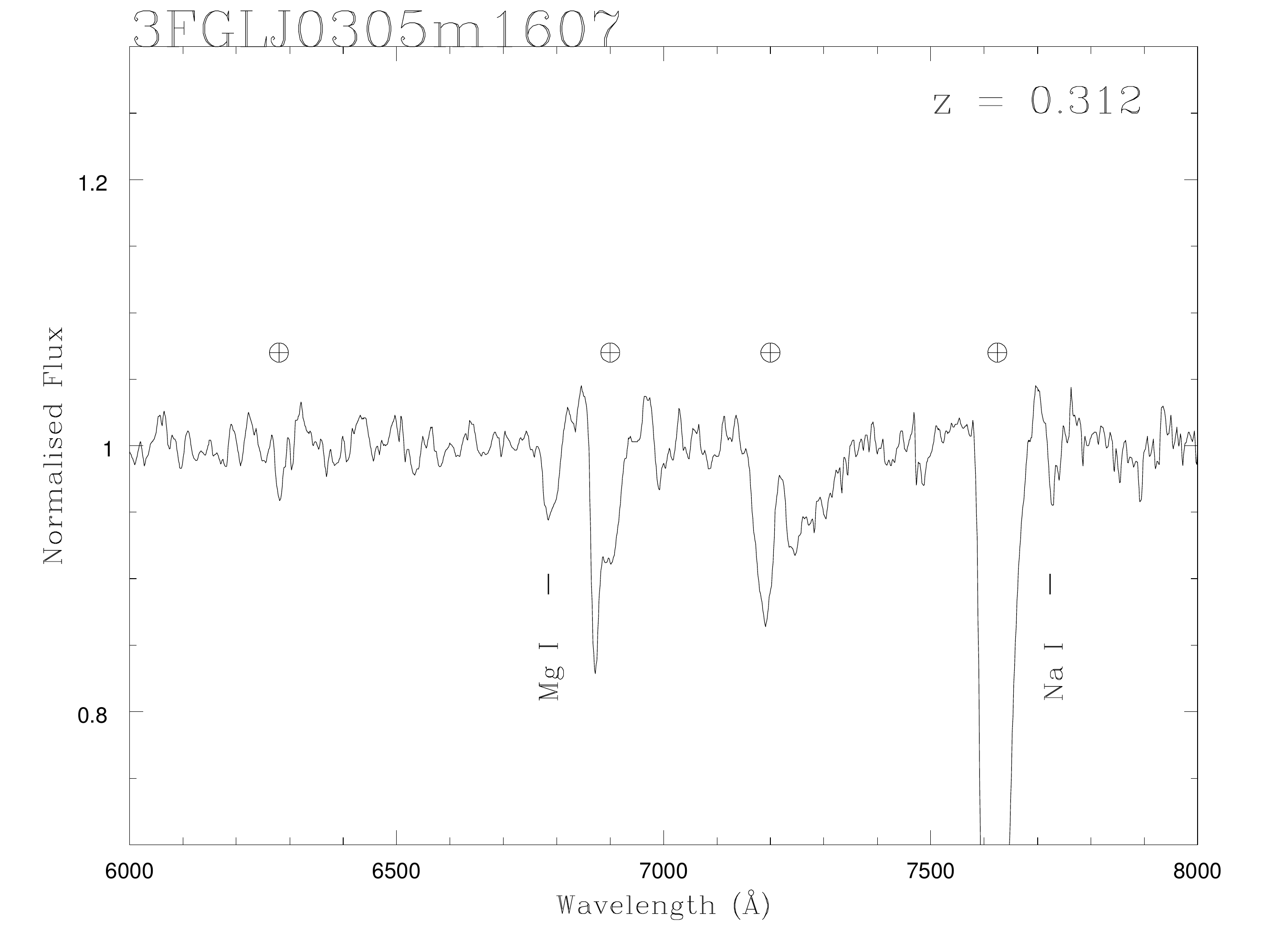}
\includegraphics[width=0.4\textwidth, angle=-90]{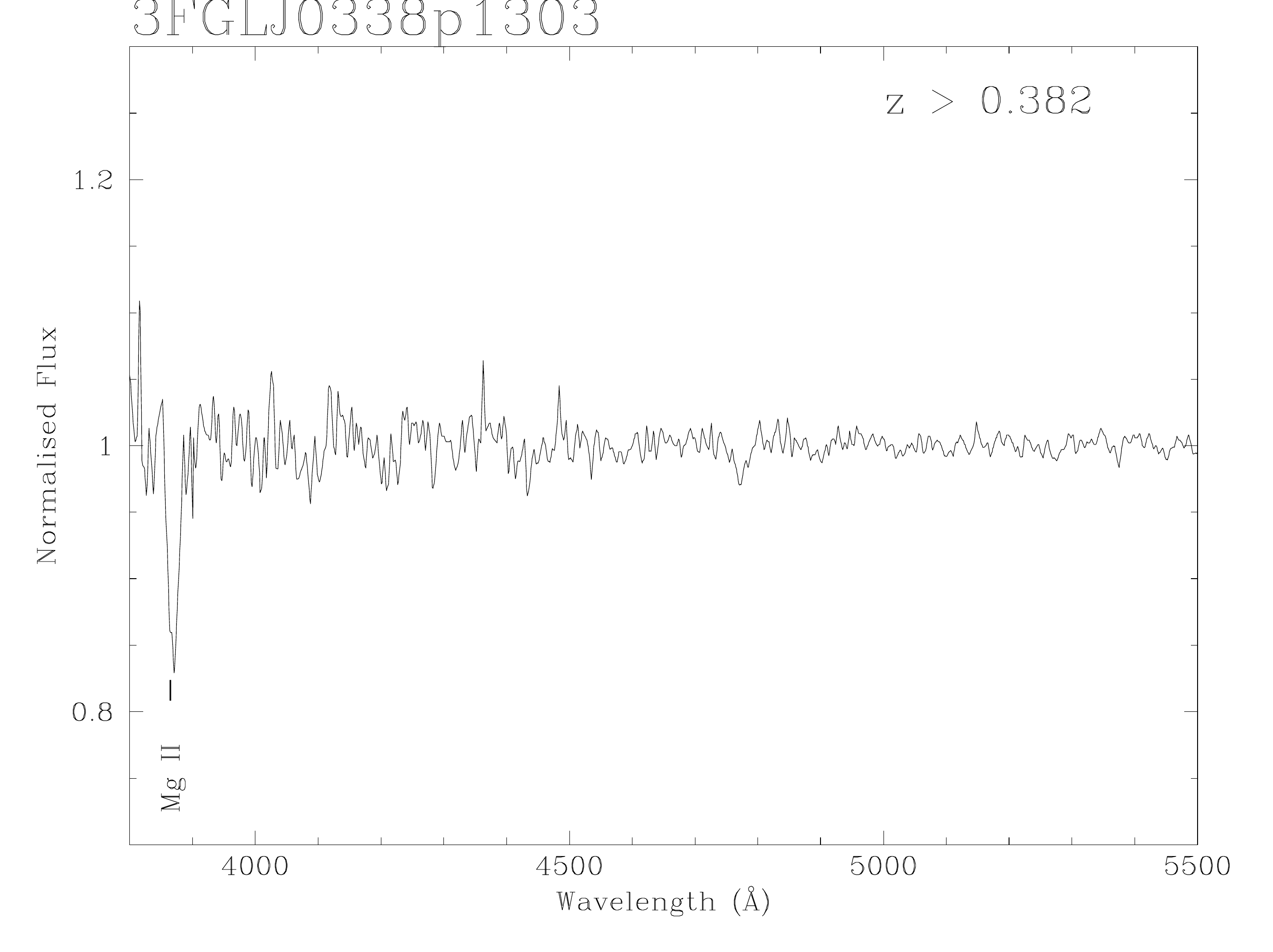}
 \caption{Close-up of the normalized spectra around the detected spectral features of the UGSs obtained at GTC. Main telluric bands are indicated as $\oplus$, spectral lines are marked by line identification. } 
   \label{fig:spectraCU}
\end{figure*}%[htbp]

\setcounter{figure}{2}
\begin{figure*}%[htbp]     
\includegraphics[width=0.4\textwidth, angle=-90]{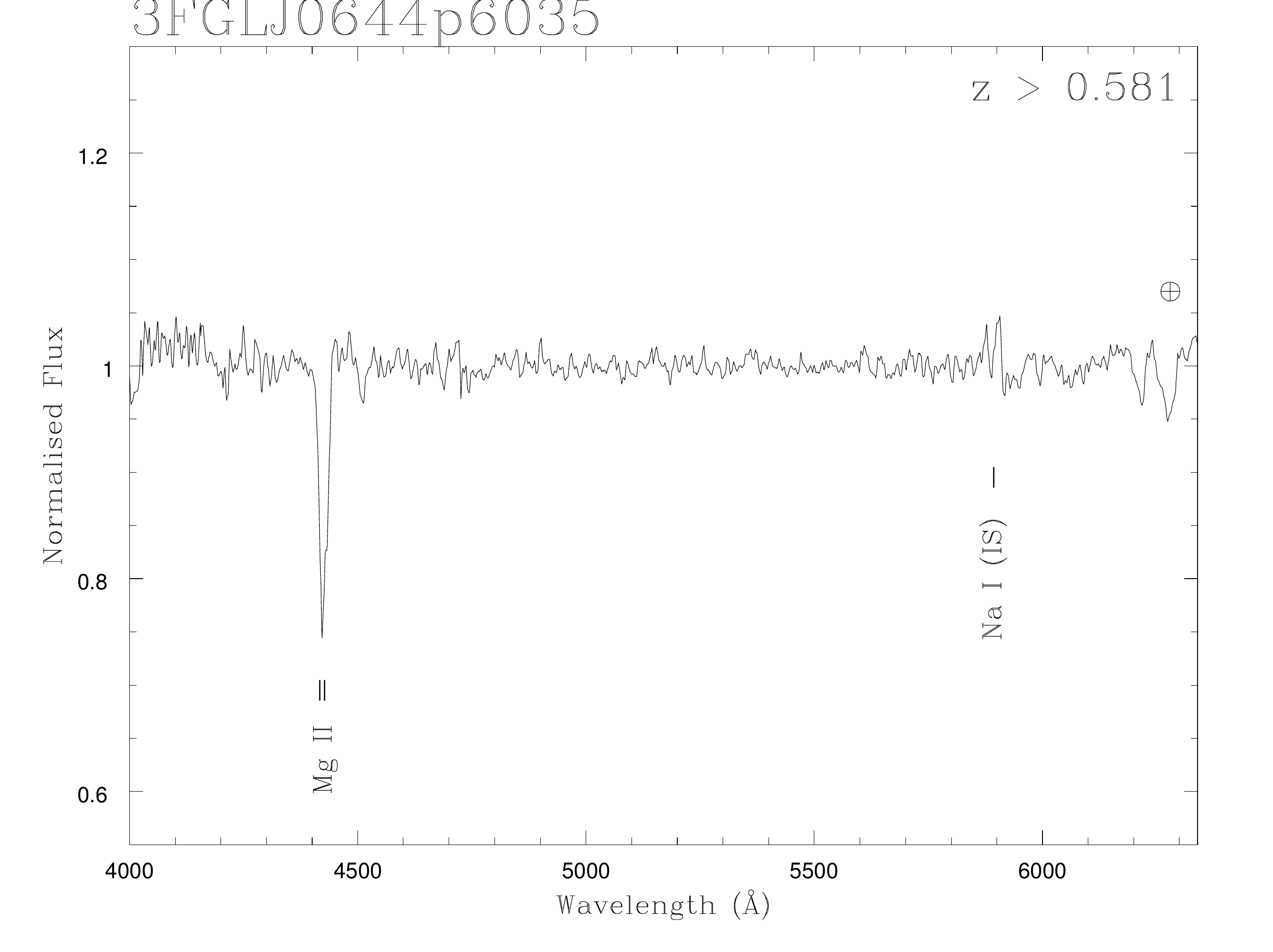}
\includegraphics[width=0.4\textwidth, angle=-90]{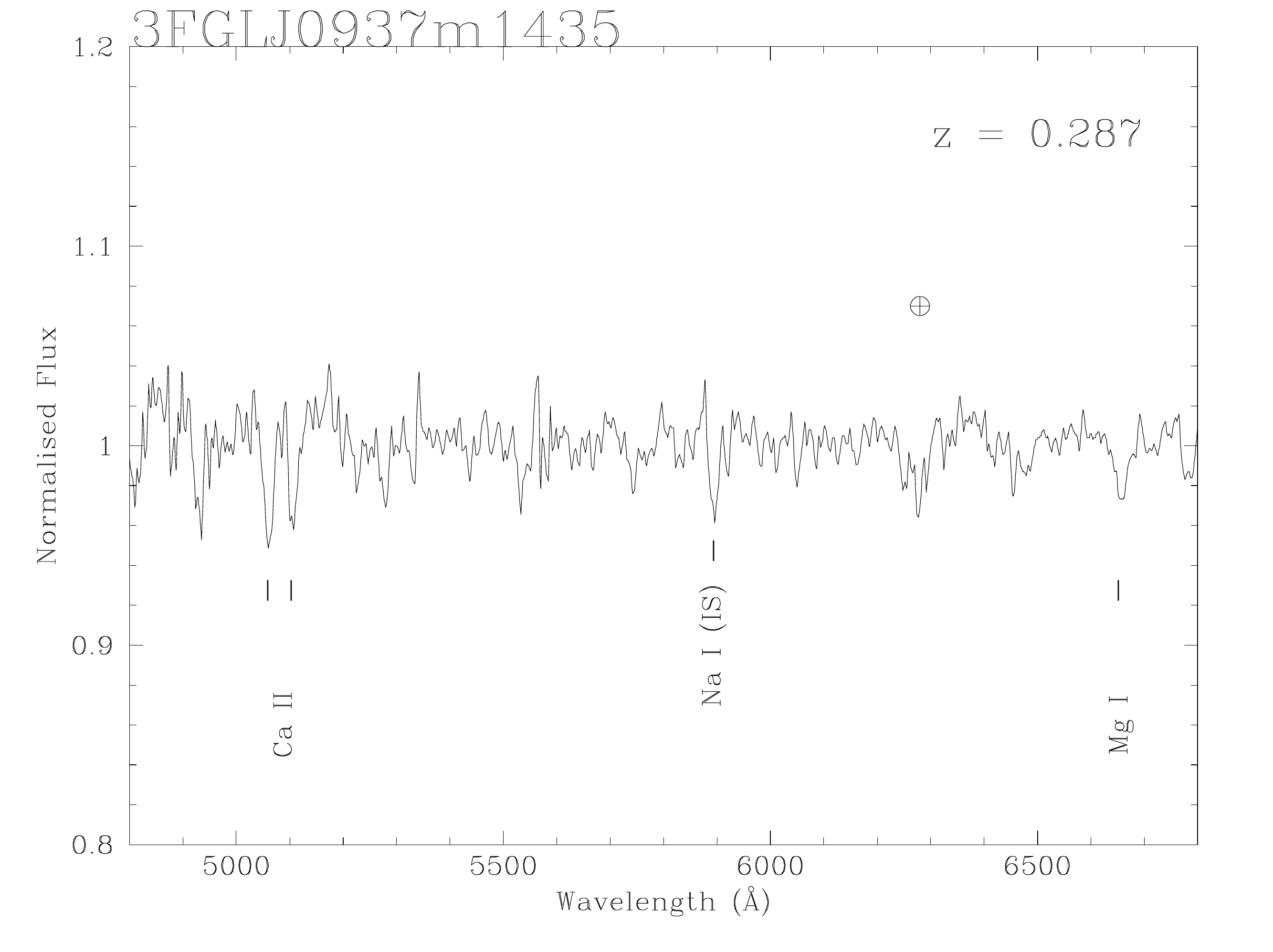}
\includegraphics[width=0.4\textwidth, angle=-90]{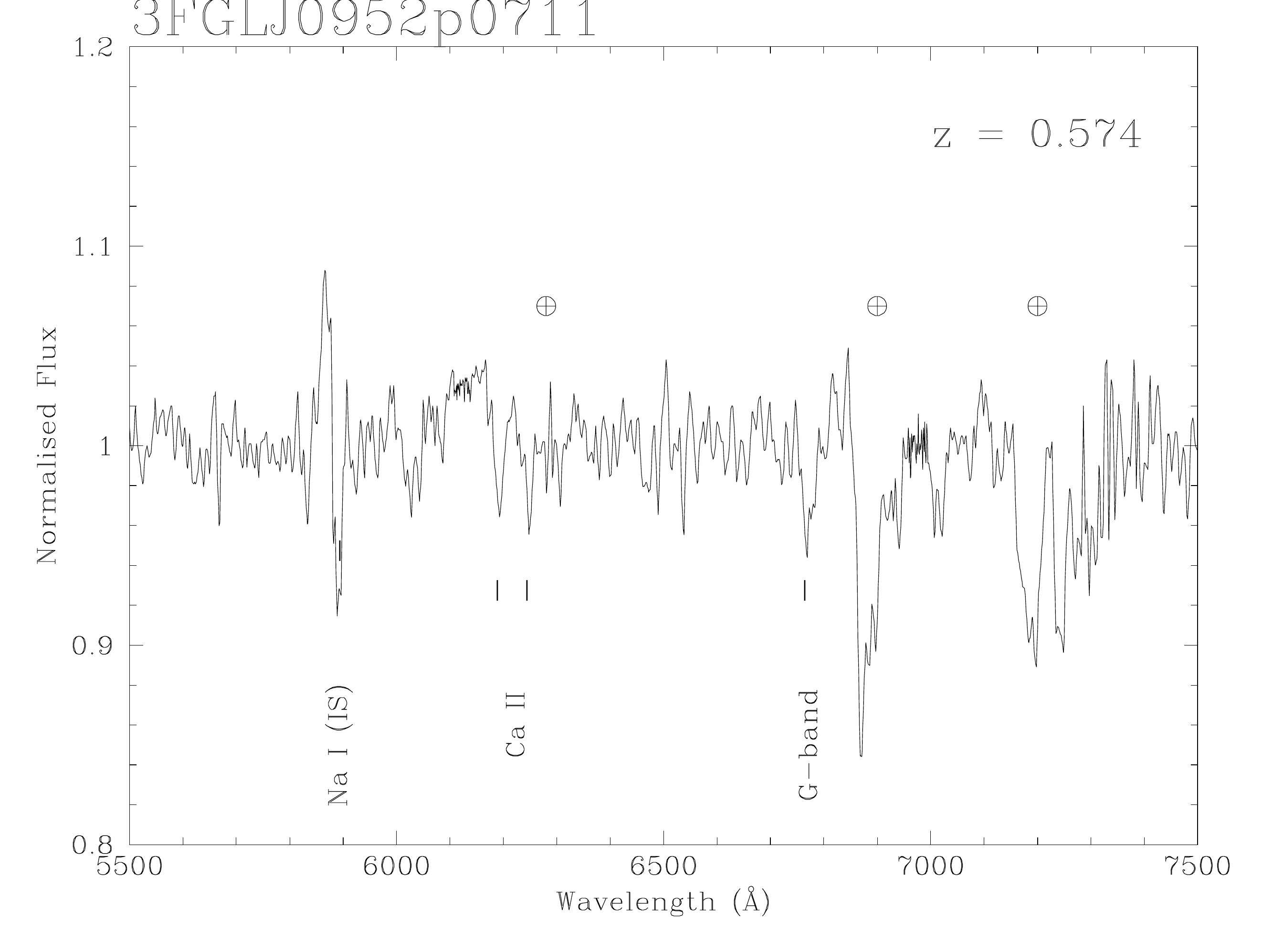}
\includegraphics[width=0.4\textwidth, angle=-90]{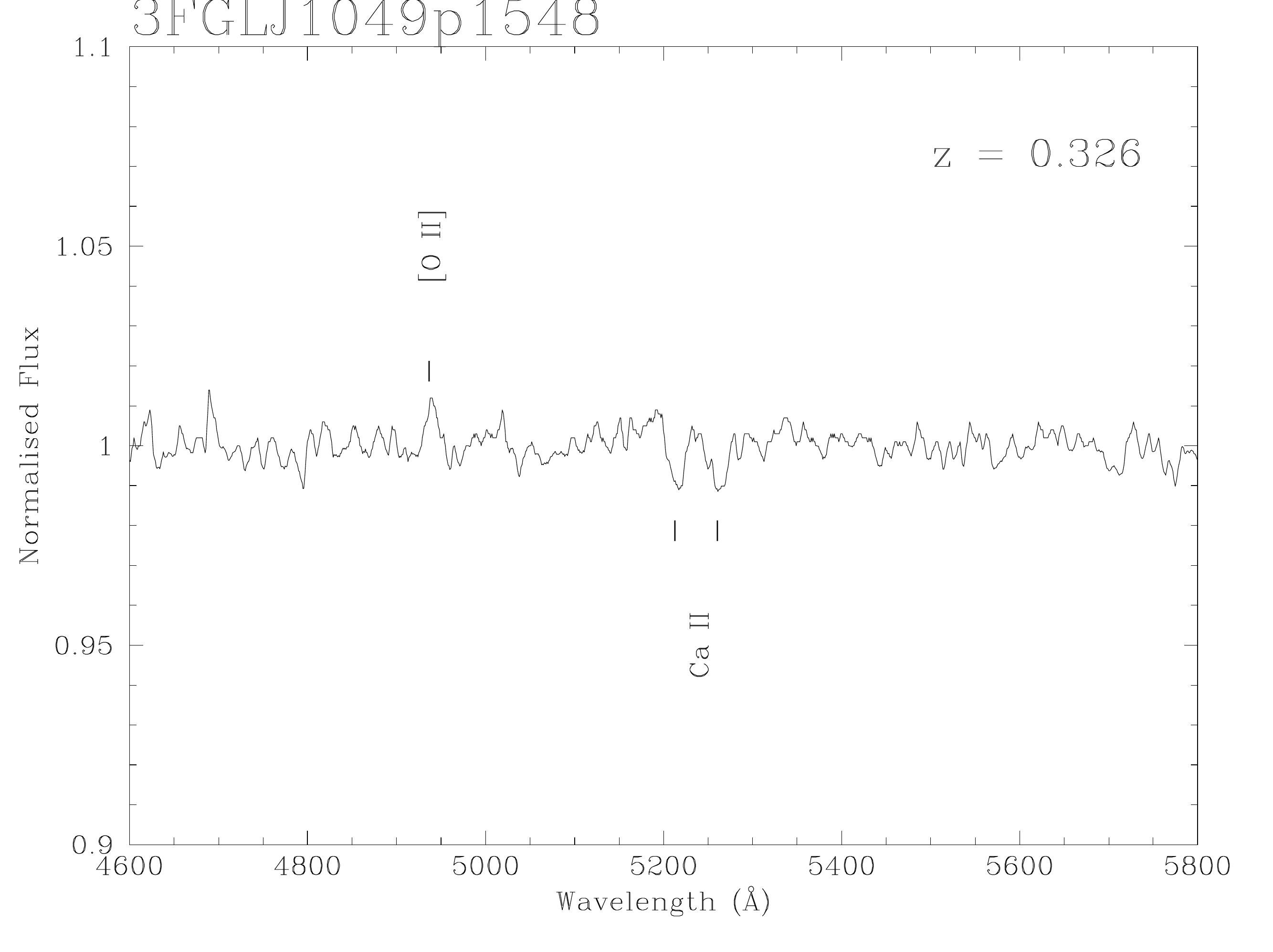}
 \includegraphics[width=0.4\textwidth, angle=-90]{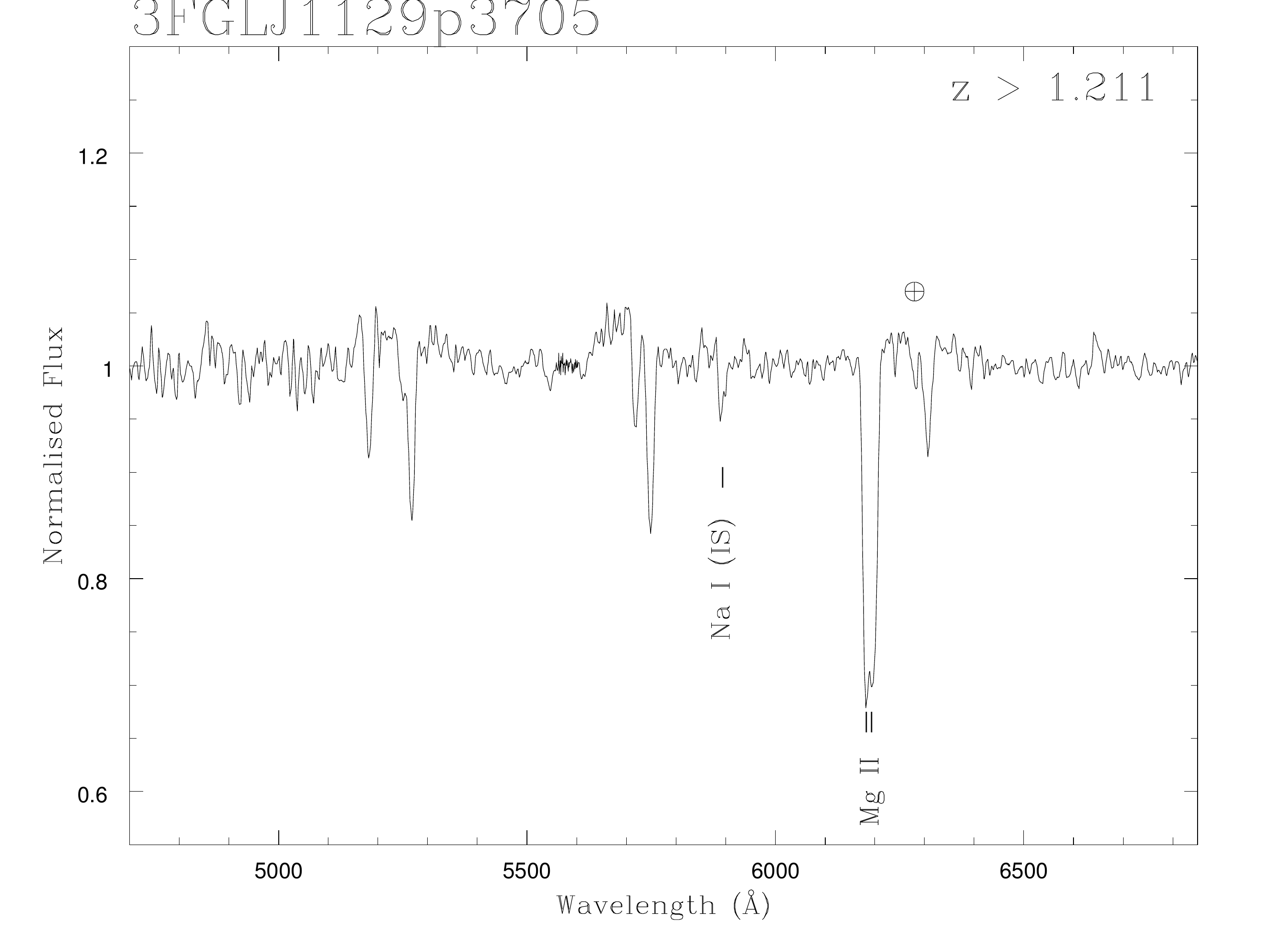}
\includegraphics[width=0.4\textwidth, angle=-90]{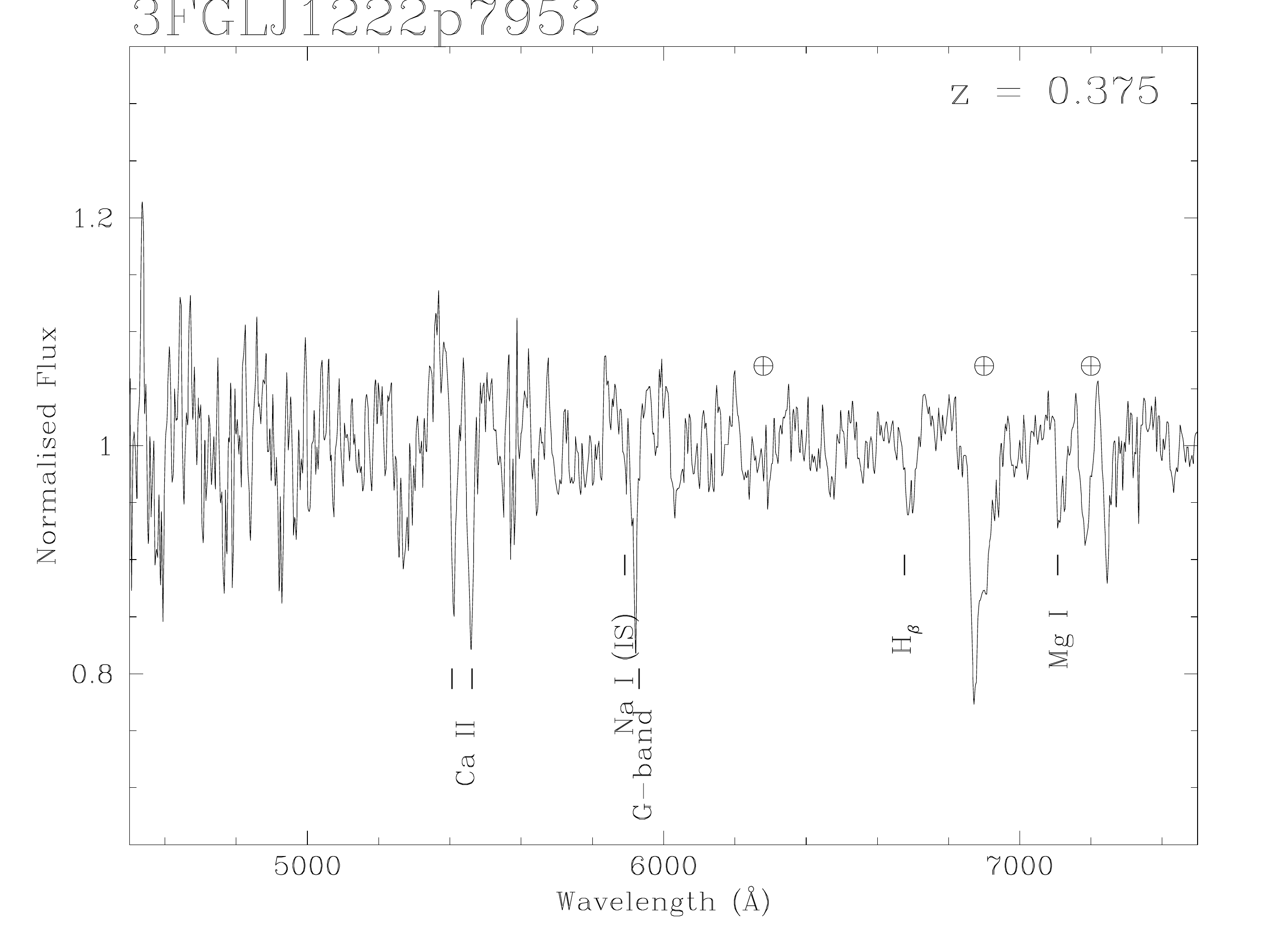}
  \caption{Continued.} 
  \end{figure*}%[htbp]

\setcounter{figure}{2}
\begin{figure*}%[htbp]     
\includegraphics[width=0.4\textwidth, angle=-90]{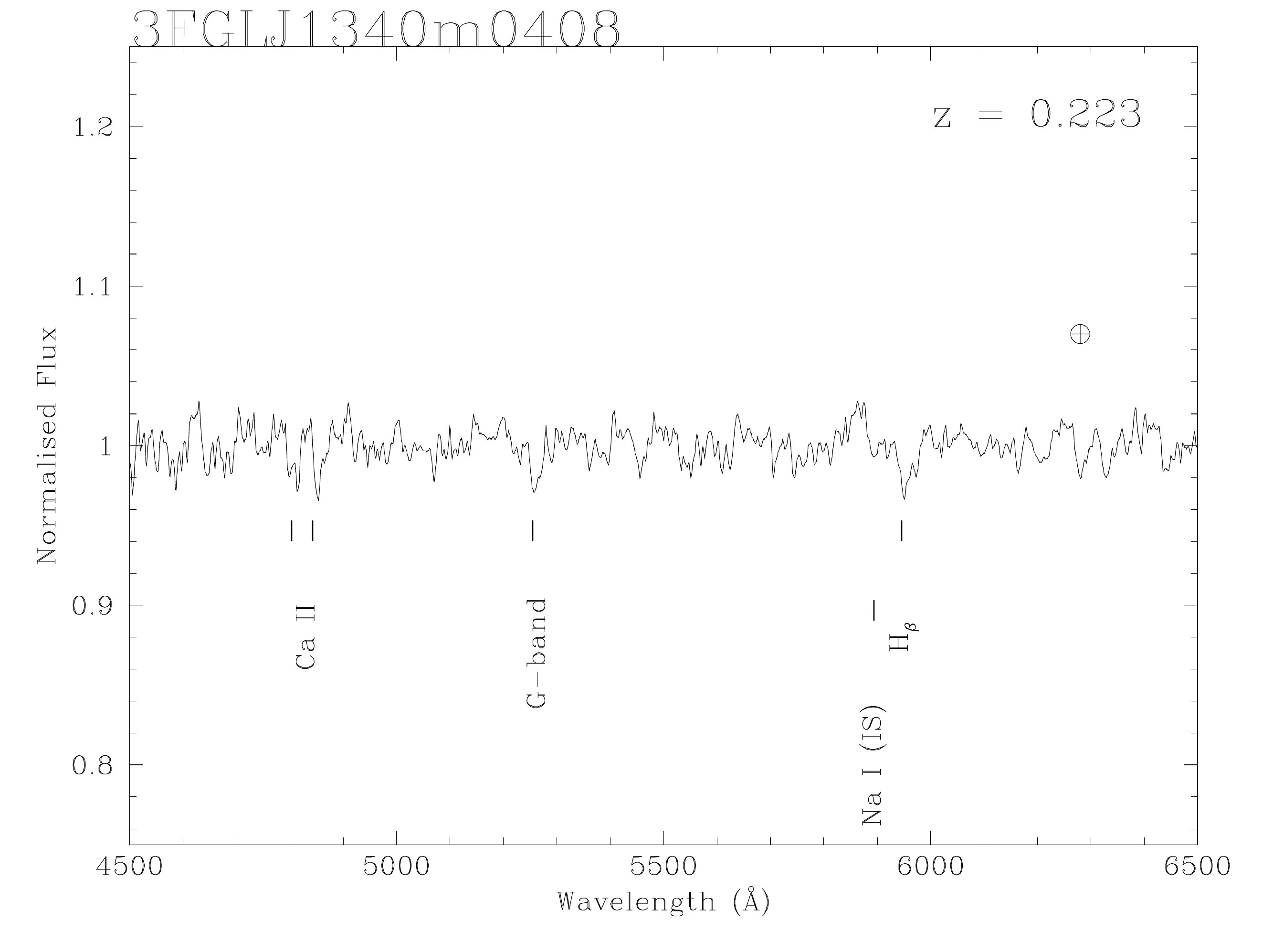}
\includegraphics[width=0.4\textwidth, angle=-90]{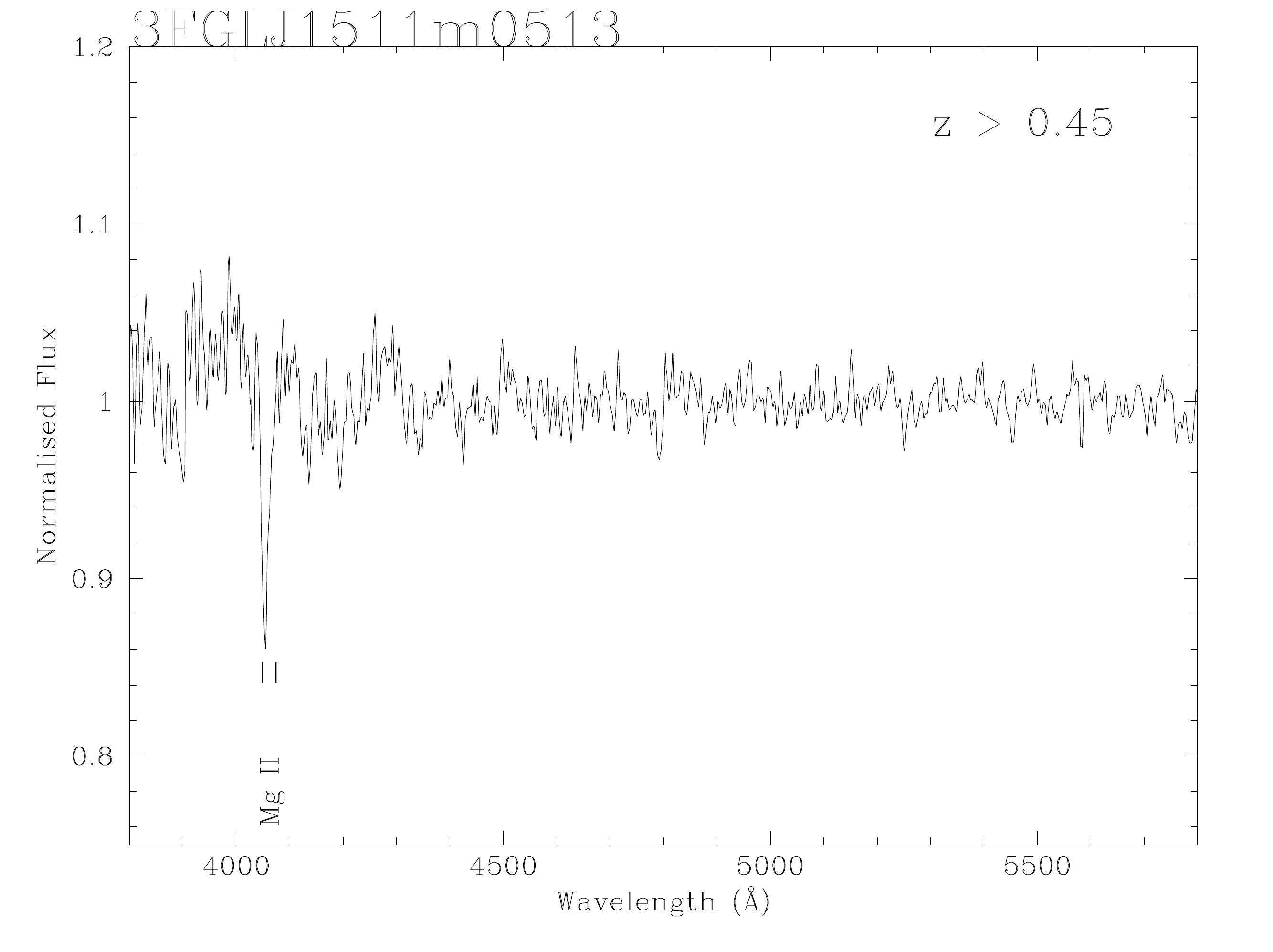}
\includegraphics[width=0.4\textwidth, angle=-90]{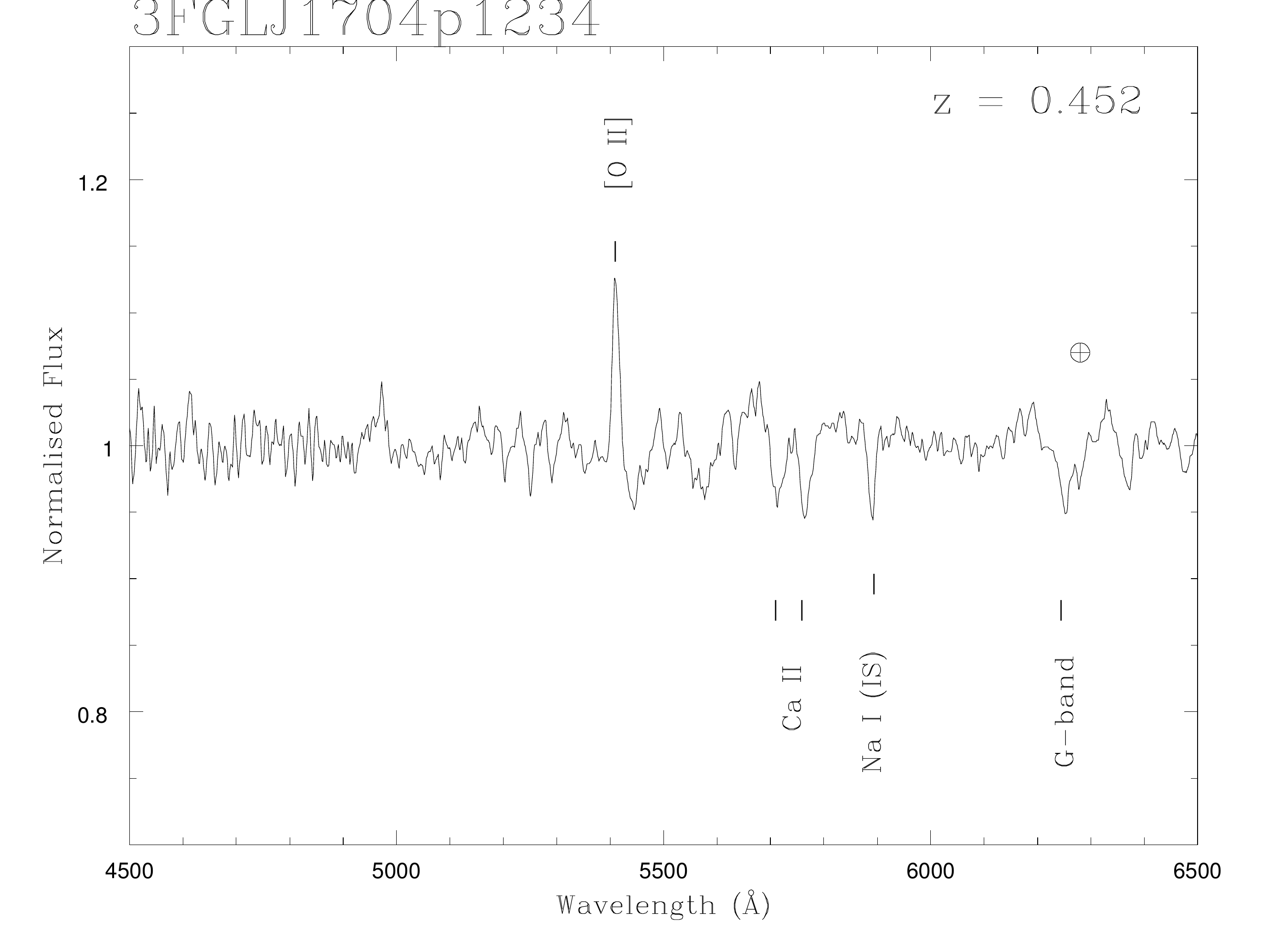}
\includegraphics[width=0.4\textwidth, angle=-90]{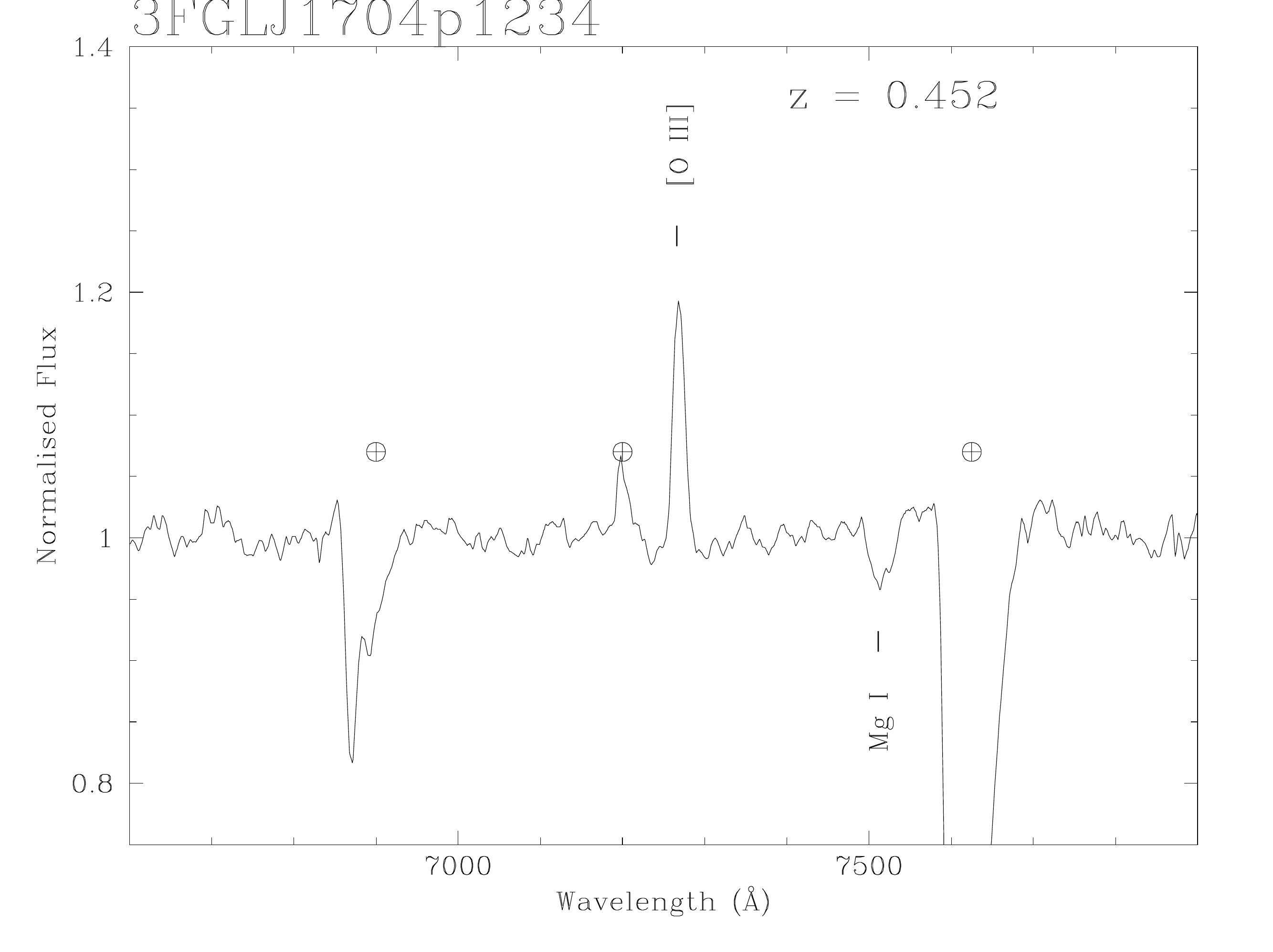}
\includegraphics[width=0.4\textwidth, angle=-90]{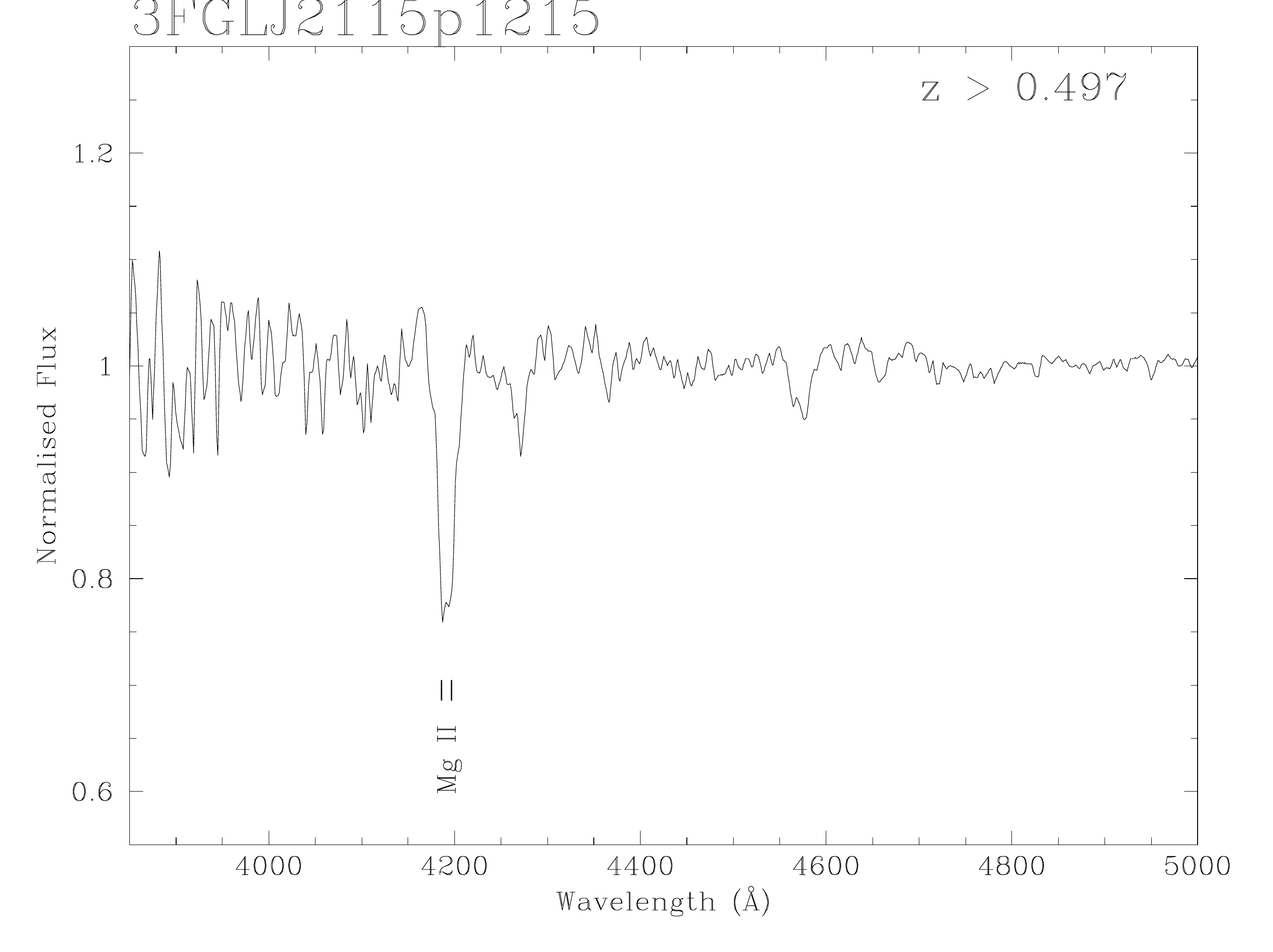}
\includegraphics[width=0.4\textwidth, angle=-90]{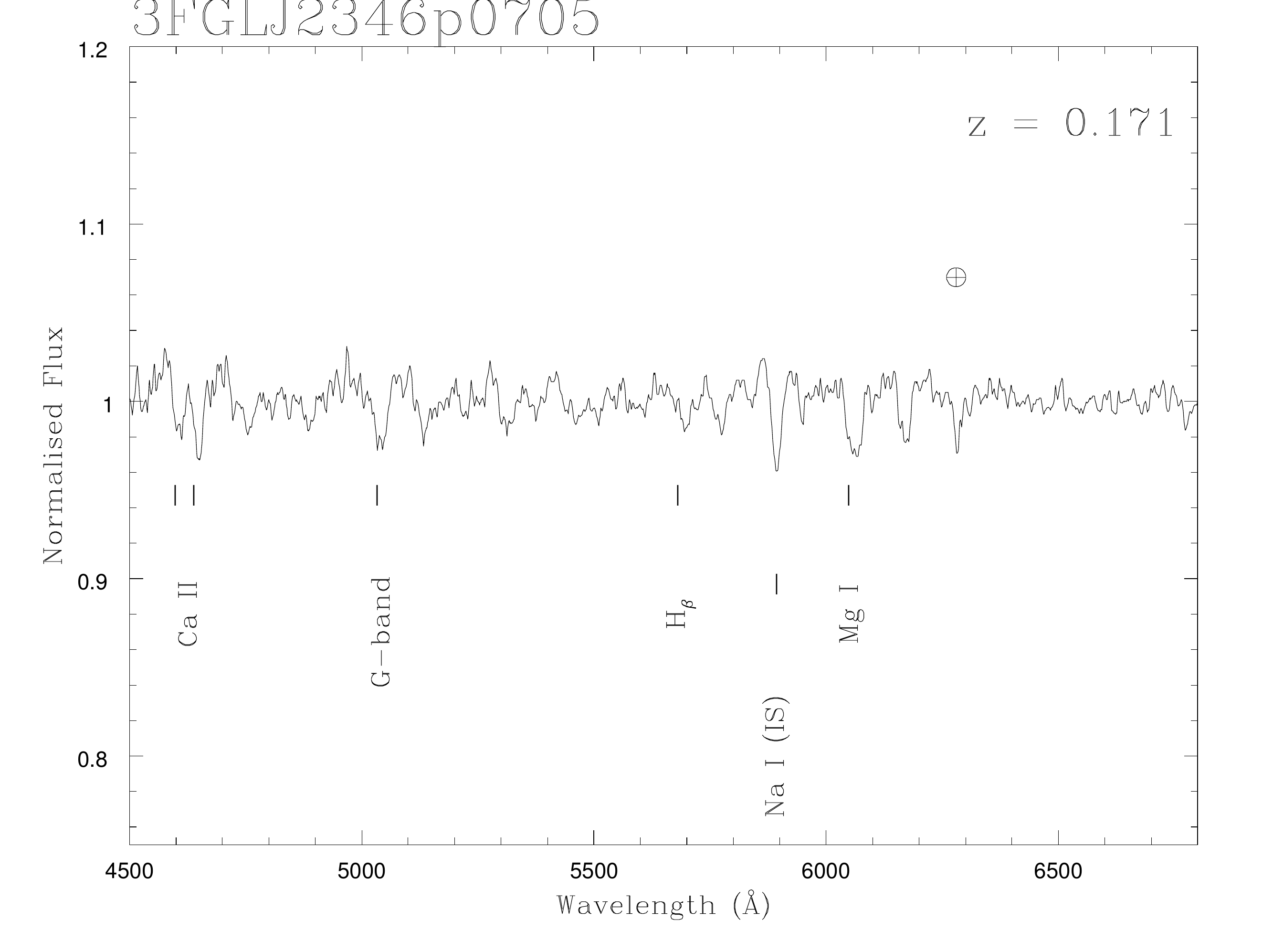}
  \caption{Continued.} 
  \end{figure*}%[htbp]

%OPTICAL IMAGES

\newpage
\setcounter{figure}{3}
\begin{figure*}%[htbp]
\centering%
\subfigure[\protect\url{3FGLJ0049.0+4224 }\label{fig:0049opt}]%
{\includegraphics[width=0.3\textwidth]{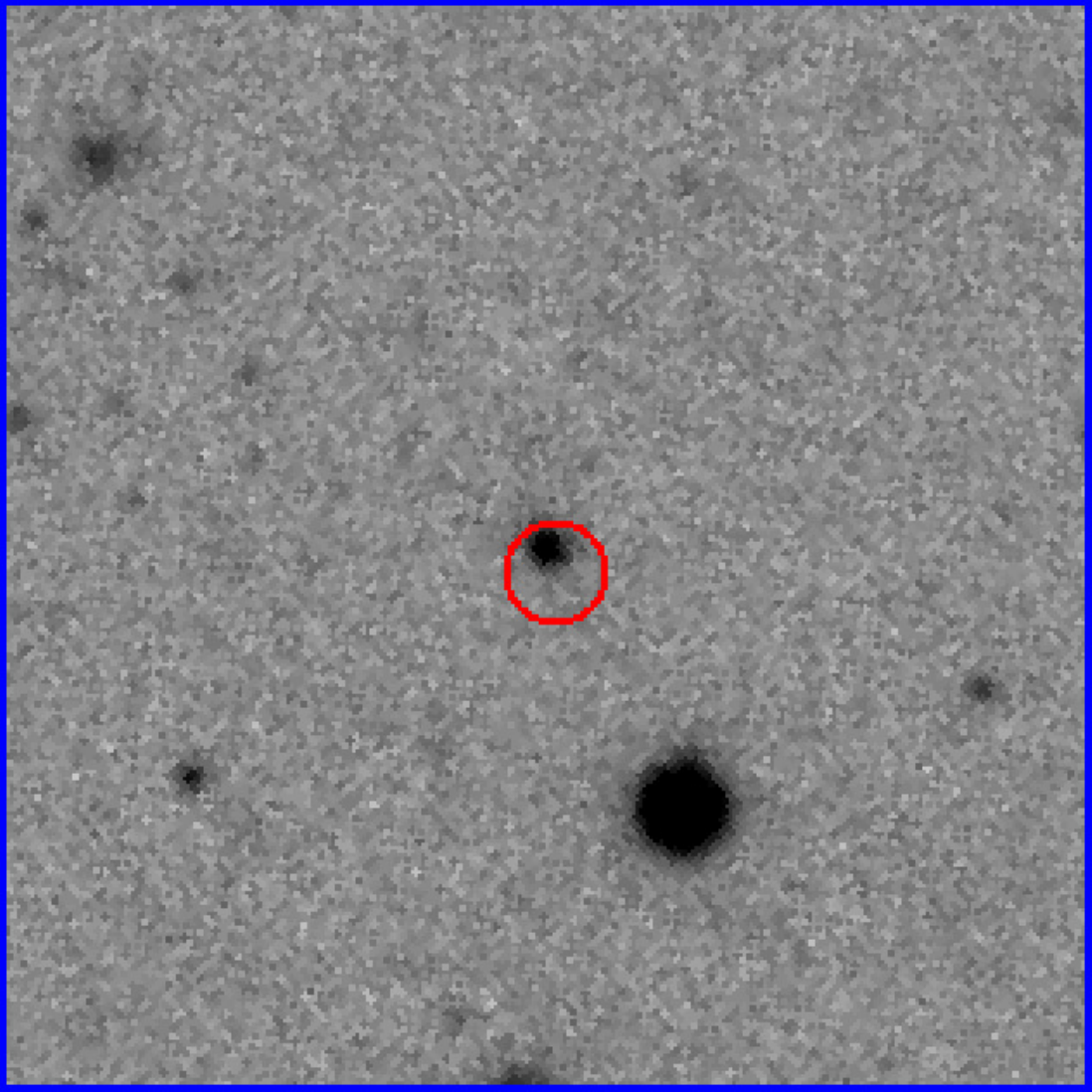}}
\subfigure[\protect\url{3FGLJ0102.1+0943}\label{fig:0102opt}]%
{\includegraphics[width=0.3\textwidth]{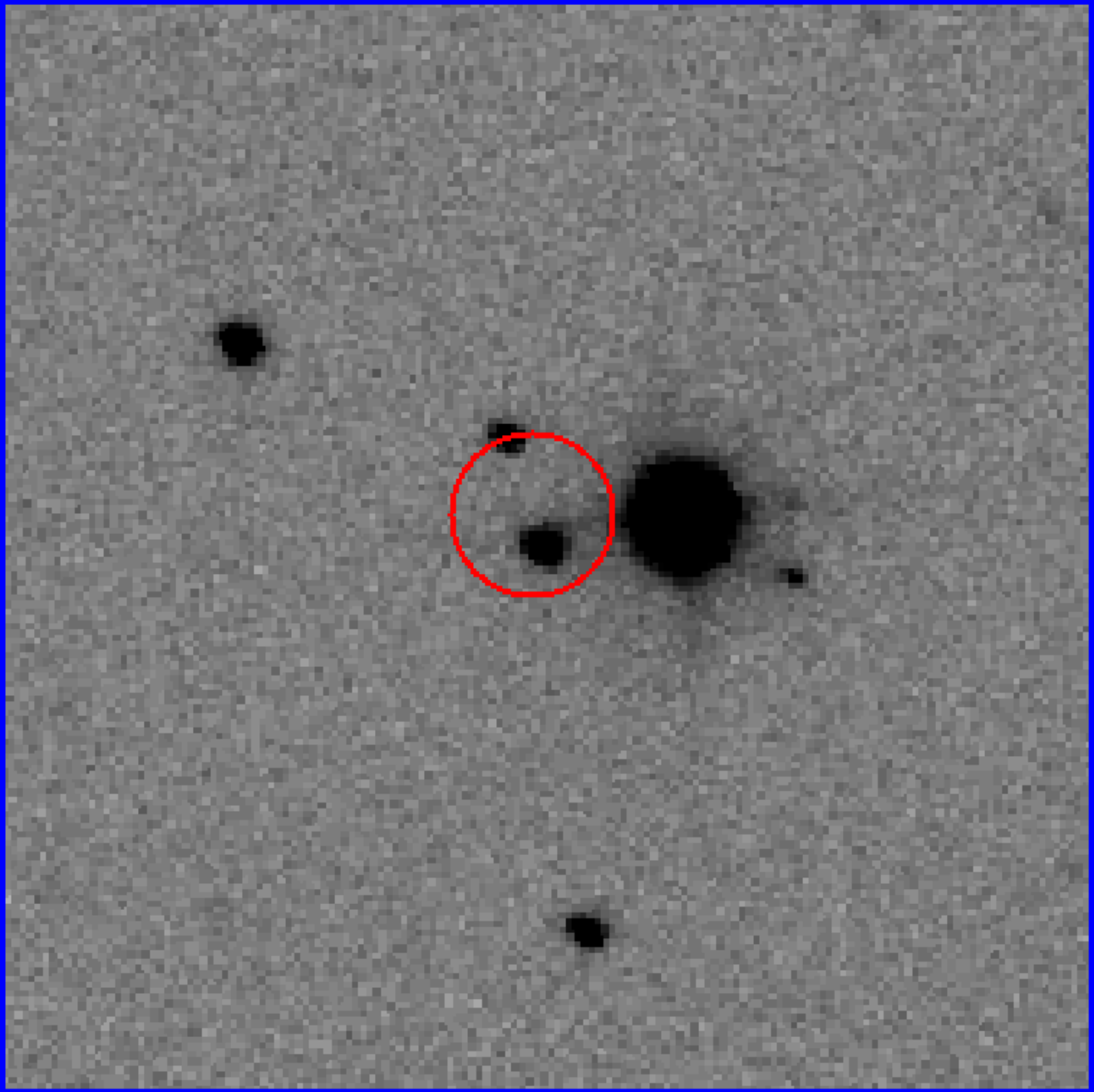}}
\subfigure[\protect\url{3FGLJ0239.0+2555}\label{fig:0239opt}]%
{\includegraphics[width=0.3\textwidth]{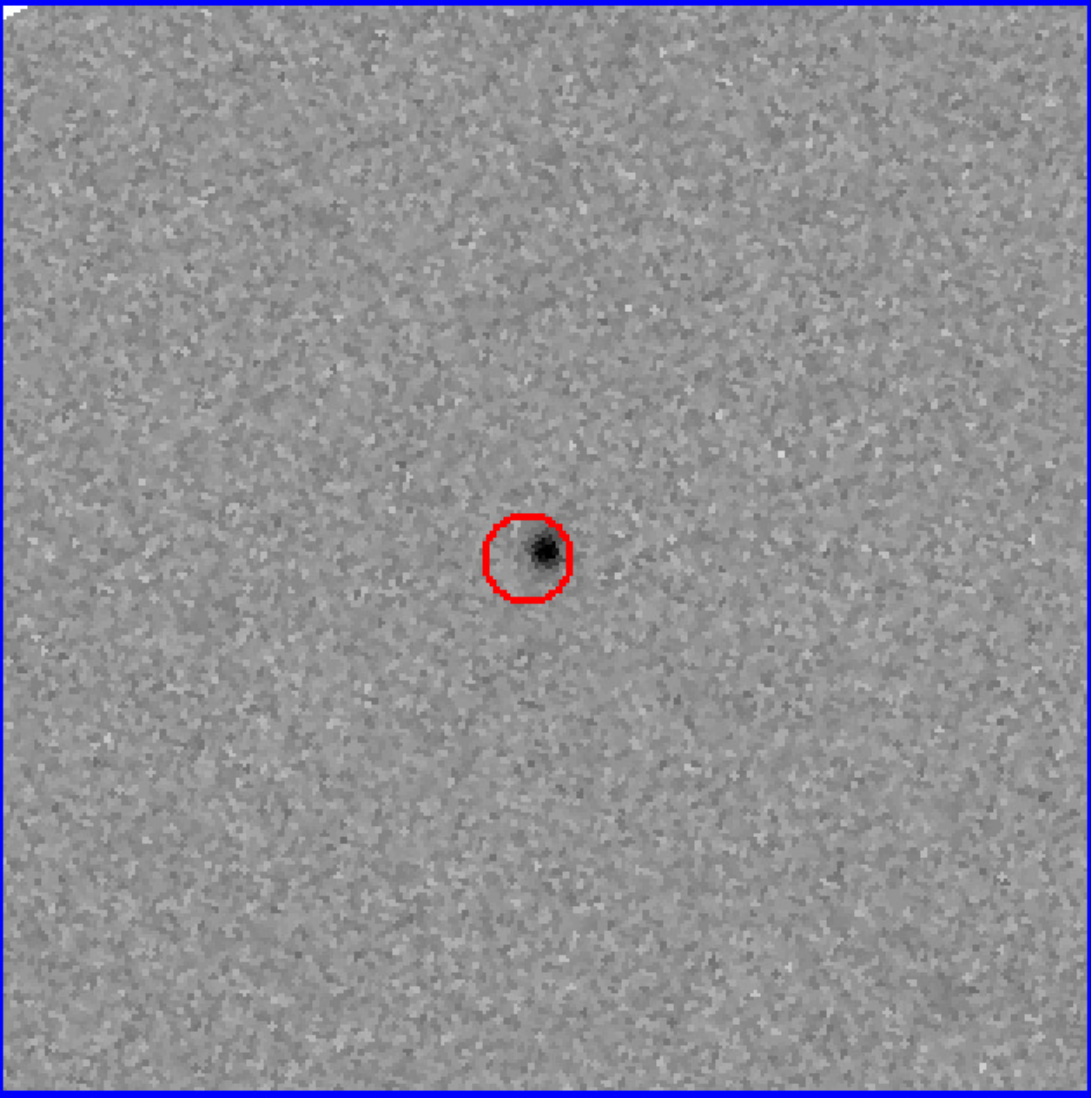}}
\subfigure[\protect\url{3FGLJ0644.6+6035}\label{fig:0644opt}]%
{\includegraphics[width=0.3\textwidth]{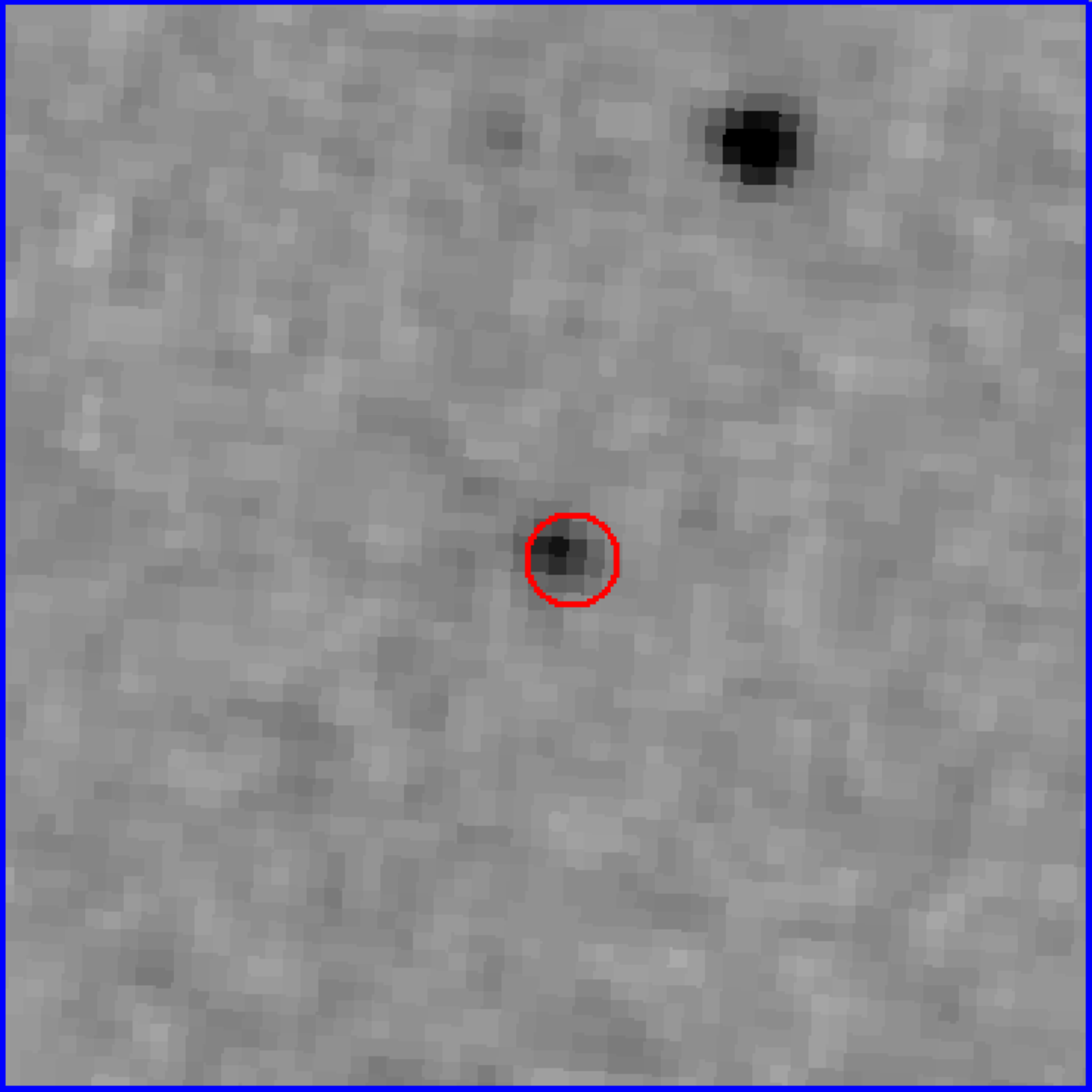}}
\subfigure[\protect\url{3FGLJ0937.9-1435}\label{fig:0937opt}]%
{\includegraphics[width=0.3\textwidth]{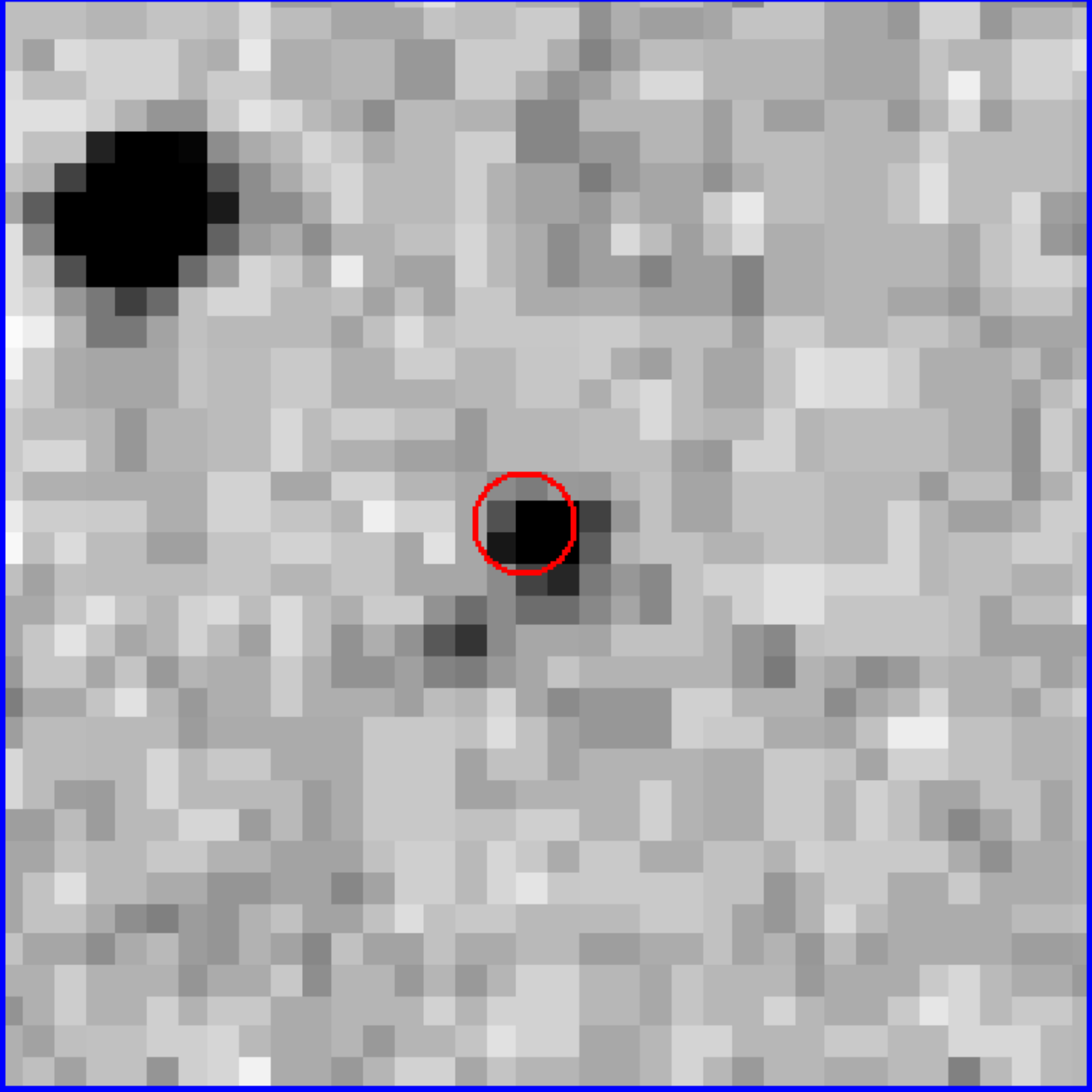}}
\subfigure[\protect\url{3FGLJ0952.8+0711}\label{fig:0952opt}]%
{\includegraphics[width=0.3\textwidth]{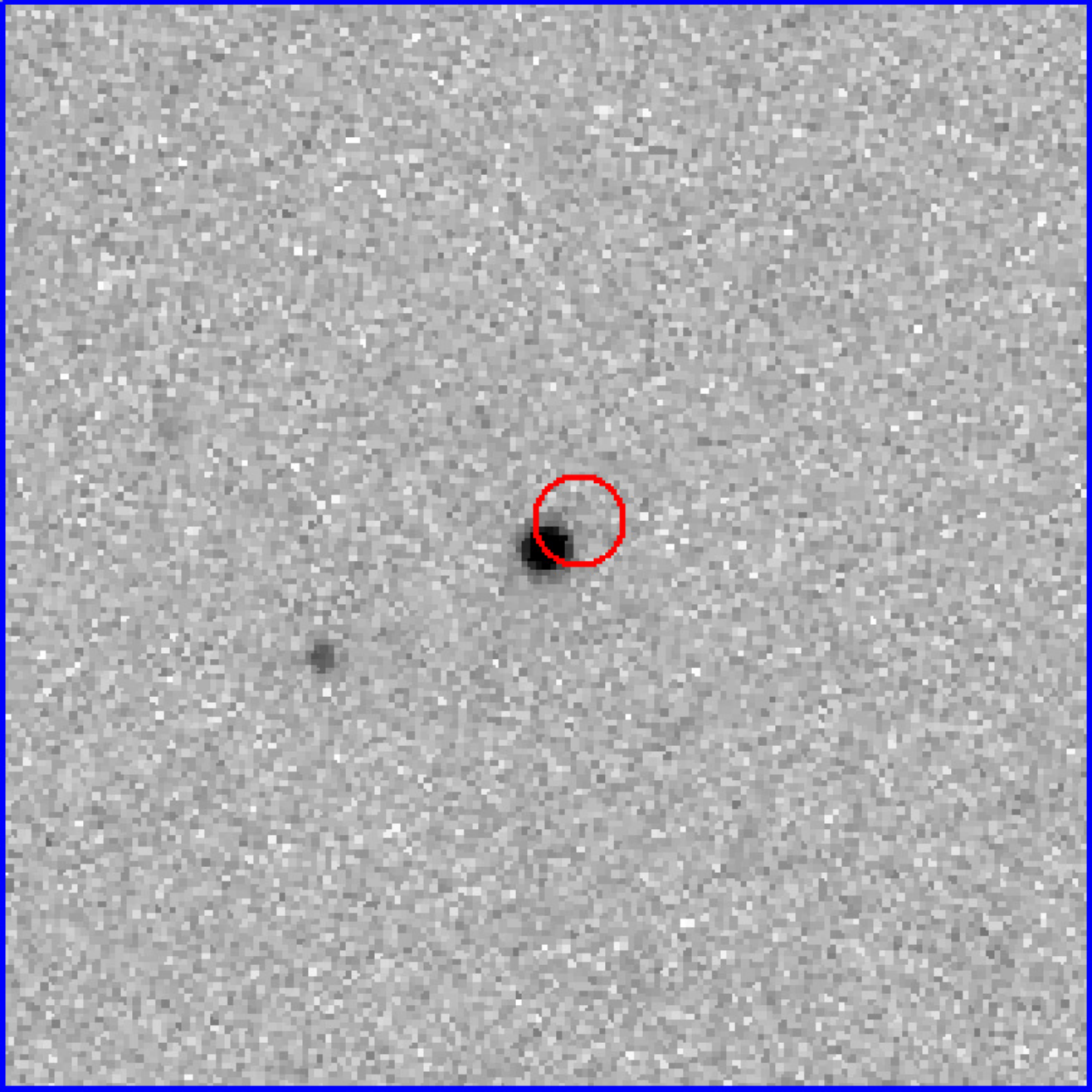}}
%\subfigure[\protect\url{3FGLJ1049.7+1548}\label{fig:1049opt}]%
%{\includegraphics[width=0.3\textwidth]{optical_skymap/1049_opt_skymap.eps}}
%\subfigure[\protect\url{3FGLJ1222.7+7952}\label{fig:1222opt}]%
%{\includegraphics[width=0.3\textwidth]{optical_skymap/1222_opt_skymap.eps}}
\subfigure[\protect\url{3FGLJ1411.4-0724}\label{fig:1411opt}]%
{\includegraphics[width=0.3\textwidth]{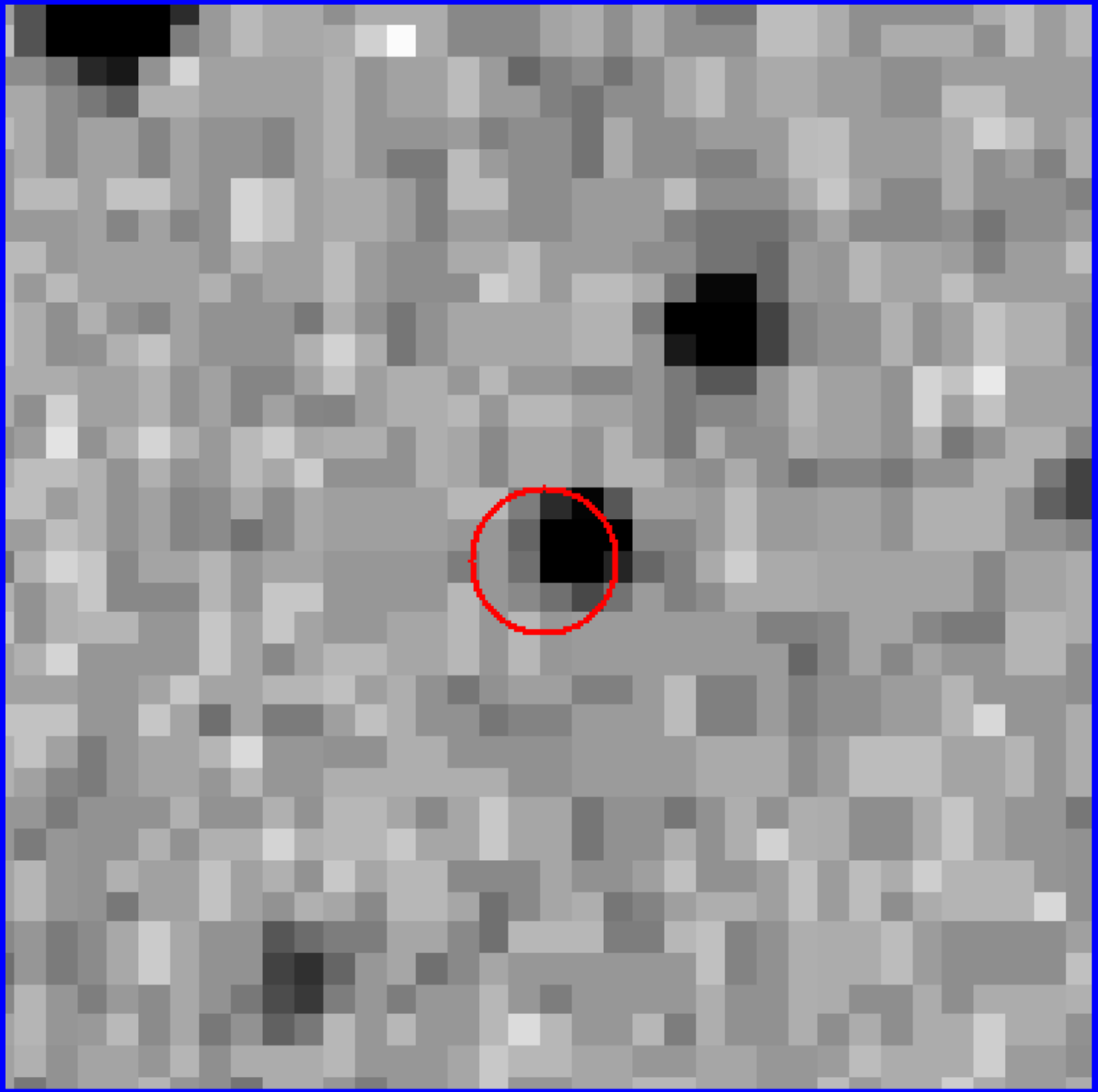}}
\subfigure[\protect\url{3FGLJ1704.4-0528}\label{fig:1704m05opt}]%
{\includegraphics[width=0.3\textwidth]{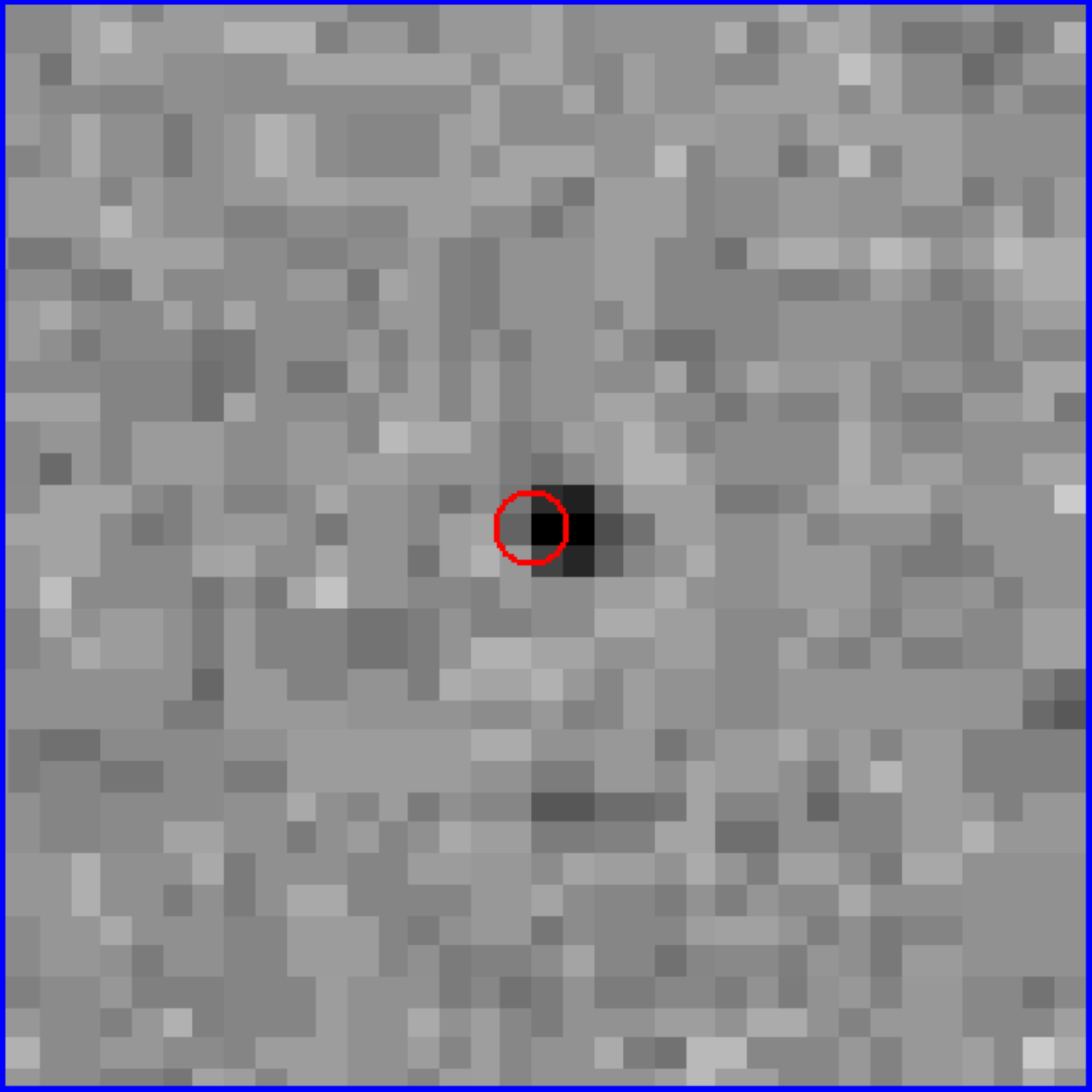}}
\subfigure[\protect\url{3FGLJ2115.2+1215}\label{fig:2115opt}]%
{\includegraphics[width=0.3\textwidth]{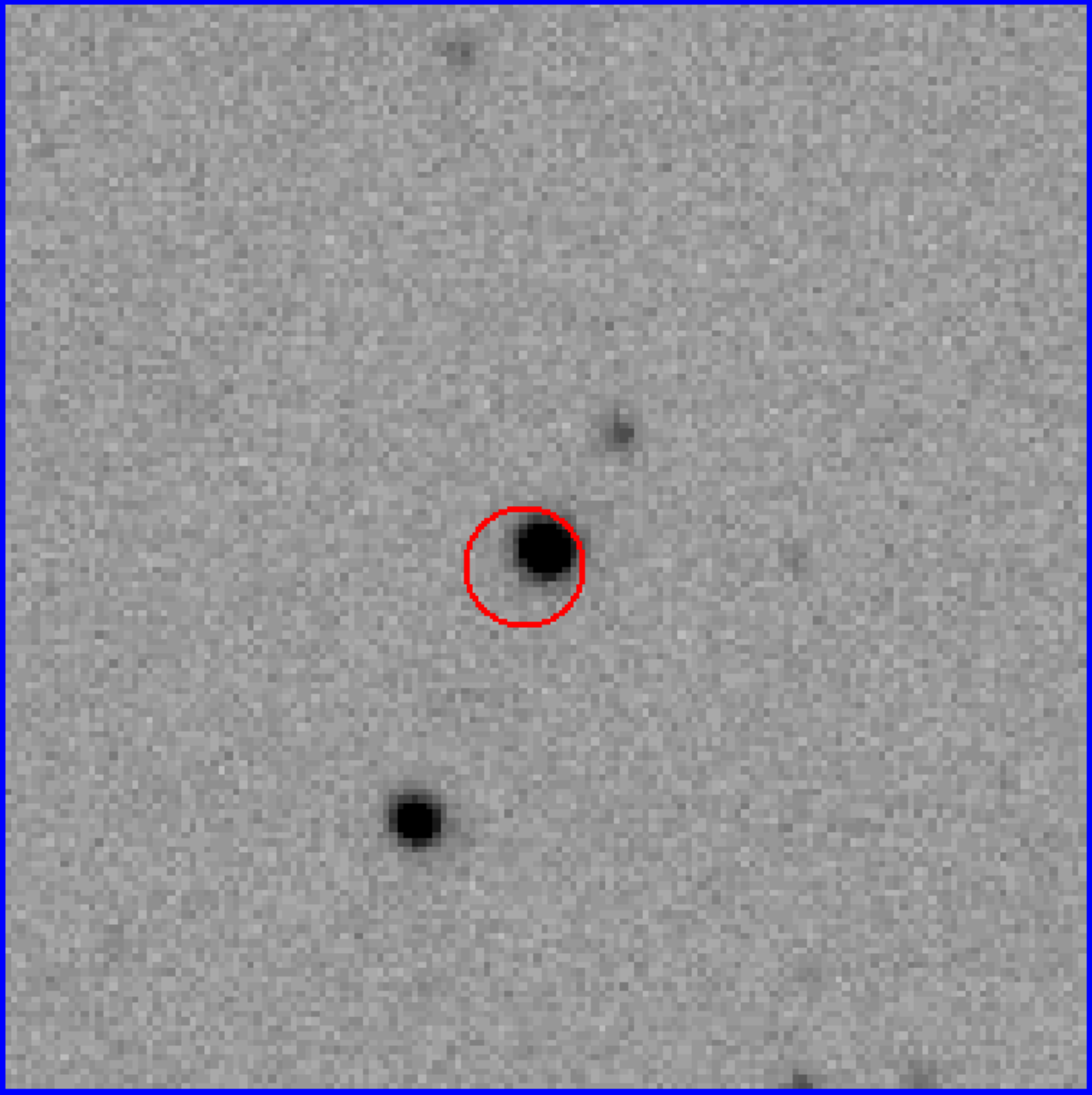}}
\caption{Optical images of nine proposed $\gamma$-ray sources. The images are taken from SDSS (g filter) or DSS survey (B filter), field= 60 arcsec x 60 arcsec, North up, East left. Details of the optical counterparts are reported in Tab. \ref{tab:table1}.  The red circles represent the error box of the proposed X-ray counterparts of the $\gamma$-ray sources (see details in Sec 2)}.
\label{fig:opt_skymap}
\end{figure*}%[htbp]

\clearpage
\newpage

\acknowledgments
%\textbf{Acknowledgments}

The financial contribution by the contract \textit{Studio e Simulazioni di Osservazioni (Immagini e Spettri) con MICADO per E-ELT} (DD 27/2016 - Ob. Fun. 1.05.02.17)  of the INAF project \textit{Micado simulazioni casi scientifici}, P.I. Dr. Renato Falomo,  is acknowledged.

%\vspace{5mm}
%\facilities{HST(STIS), Swift(XRT and UVOT), AAVSO, CTIO:1.3m,CTIO:1.5m,CXO}
\facilities{GTC-OSIRIS, \citep{cepa2003}}

%\vspace{5mm}
\software{IRAF \citep{tody1986, tody1993}}

\bibliographystyle{aasjournal}
\bibliography{UFO_GTC_biblio}

%\begin{thebibliography}{}
%\bibitem[Astropy Collaboration et al.(2013)]{2013A&A...558A..33A} Astropy Collaboration, Robitaille, T.~P., Tollerud, E.~J., et al.\ 2013, \aap, 558, A33 
%\bibitem[Bertin \& Arnouts(1996)]{1996A&AS..117..393B} Bertin, E., \& Arnouts, S.\ 1996, \aaps, 117, 393 
%\end{thebibliography}



\end{document}